\documentclass[a4,12pt]{article}
\usepackage{comment}
\usepackage{docmute}
\usepackage{amsmath,amsthm,amssymb}
\usepackage{graphicx}
\usepackage[usenames,dvipsnames]{xcolor}
\definecolor{MyGreen}{rgb}{0,0.5,0.2}
\colorlet{green}{MyGreen}
\usepackage{wrapfig}
\usepackage{enumerate}
\usepackage{mathrsfs}
\usepackage{geometry}
\usepackage{caption}
\usepackage{fancybox}
\usepackage{youngtab}

\usepackage{hyperref}
\usepackage{cite}
\usepackage{bm}
\usepackage{braket}

\allowdisplaybreaks

\usepackage{tikz}
\usetikzlibrary{calc}
\usetikzlibrary{arrows,shapes,positioning}
\usetikzlibrary{decorations.markings}
\usetikzlibrary{decorations.pathreplacing}
\usetikzlibrary{mindmap,trees,shadows}
\usetikzlibrary{patterns}
\usetikzlibrary{intersections}

\tikzstyle no arrow=[thick]
\tikzstyle end arrow=[thick,postaction={decorate,decoration={markings, mark=at position .999 with {\arrow[scale=1.4]{stealth}}}}]
\tikzstyle mid arrow=[thick,postaction={decorate,decoration={markings, mark=at position .6 with {\arrow[scale=1.4]{stealth}}}}]
\tikzstyle mid-end arrow=[thick,postaction={decorate,decoration={markings, mark=at position .8 with {\arrow[scale=1.4]{stealth}}}}]
\tikzstyle start-mid arrow=[thick,postaction={decorate,decoration={markings, mark=at position .35 with {\arrow[scale=1.4]{stealth}}}}]
\tikzstyle latter arrow=[thick,postaction={decorate,decoration={markings, mark=at position .8 with {\arrow[scale=1.4]{stealth}}}}]
\tikzstyle former arrow=[thick,postaction={decorate,decoration={markings, mark=at position .4 with {\arrow[scale=1.4]{stealth}}}}]
\tikzstyle transarrow=[white,thick,postaction={decorate,decoration={markings, mark=at position .0 with {\arrow[scale=1.4,black]{stealth}}}}]

\newcommand{\circlearrow}[6]{\draw[transarrow] ($#1+(#3:#2)+(#3+90:#4*#2)$) node[black,#5]{#6} --++ (#3+90:#2);}
\newcommand{\anticirclearrow}[6]{\draw[transarrow] ($#1+(#3:#2)+(#3-90:#4*#2)$) node[black,#5]{#6} --++ (#3-90:#2);}
\newcommand{\circlenoarrow}[5]{\node[black,#4] at ($#1+(#3:#2)$) {#5};}

\newcommand{\linenearpuncture}[6]{
\def\l{#4}
\def\w{#5}
\def\r{#6}
\coordinate (P) at (0,0);
\coordinate (U) at ($(P)+(\w,\l)$);
\coordinate (D) at ($(P)+(\w,-\l)$);
\draw[mid arrow] (D)  -- (U) node[above]{#1};
\draw[fill=white] (P) circle (1.8*\r) node[#2]{#3};
\draw[black,thick] (P) circle (1.8*\r);
\draw[fill=black] (P) circle (\r);
}

\newcommand{\linewithnoarrownearpuncture}[6]{
\def\l{#4}
\def\w{#5}
\def\r{#6}
\coordinate (P) at (0,0);
\coordinate (U) at ($(P)+(\w,\l)$);
\coordinate (D) at ($(P)+(\w,-\l)$);
\draw[no arrow] (D)  -- (U) node[above]{#1};
\draw[fill=white] (P) circle (1.8*\r) node[#2]{#3};
\draw[black,thick] (P) circle (1.8*\r);
\draw[fill=black] (P) circle (\r);
}

\newcommand{\punctureindigonPP}[8]{
\def\l{#6}
\def\r{#7}
\def\ratio{#8}
\coordinate (P) at (0,0);
\coordinate (U) at ($(P)+(0,\l)$);
\coordinate (D) at ($(P)+(0,-\l)$);
\coordinate (UU) at ($(U)+(0,\ratio*\l)$);
\coordinate (DD) at ($(D)+(0,-\ratio*\l)$);
\draw[fill=white] (P) circle (1.8*\r) node[#4]{#5};
\draw[black,thick] (P) circle (1.8*\r);
\draw[fill=black] (P) circle (\r);
\draw[mid arrow] (U)  -- (UU) node[above]{#1};
\draw[mid arrow] (DD) node[below]{#1} -- (D);
\draw[thick] (U) arc (90:-90:{\l});% node[right]{$a$};
\draw[thick] (U) arc (90:270:{\l});% node[right]{$a$};
\anticirclearrow{(P)}{\l}{180}{0.3}{left}{#2}
\circlearrow{(P)}{\l}{0}{0.3}{right}{#3}
}

\newcommand{\punctureindigonPN}[8]{
\def\l{#6}
\def\r{#7}
\def\ratio{#8}
\coordinate (P) at (0,0);
\coordinate (U) at ($(P)+(0,\l)$);
\coordinate (D) at ($(P)+(0,-\l)$);
\coordinate (UU) at ($(U)+(0,\ratio*\l)$);
\coordinate (DD) at ($(D)+(0,-\ratio*\l)$);
\draw[fill=white] (P) circle (1.8*\r) node[#4]{#5};
\draw[black,thick] (P) circle (1.8*\r);
\draw[fill=black] (P) circle (\r);
\draw[mid arrow] (U)  -- (UU) node[above]{#1};
\draw[mid arrow] (DD) node[below]{#1} -- (D);
\draw[thick] (U) arc (90:-90:{\l});% node[right]{$a$};
\draw[thick] (U) arc (90:270:{\l});% node[right]{$a$};
\anticirclearrow{(P)}{\l}{180}{0.3}{left}{#2}
\circlenoarrow{(P)}{\l}{0}{right}{#3}
}

\newcommand{\punctureindigonNP}[8]{
\def\l{#6}
\def\r{#7}
\def\ratio{#8}
\coordinate (P) at (0,0);
\coordinate (U) at ($(P)+(0,\l)$);
\coordinate (D) at ($(P)+(0,-\l)$);
\coordinate (UU) at ($(U)+(0,\ratio*\l)$);
\coordinate (DD) at ($(D)+(0,-\ratio*\l)$);
\draw[fill=white] (P) circle (1.8*\r) node[#4]{#5};
\draw[black,thick] (P) circle (1.8*\r);
\draw[fill=black] (P) circle (\r);
\draw[mid arrow] (U)  -- (UU) node[above]{#1};
\draw[mid arrow] (DD) node[below]{#1} -- (D);
\draw[thick] (U) arc (90:-90:{\l});% node[right]{$a$};
\draw[thick] (U) arc (90:270:{\l});% node[right]{$a$};
\circlenoarrow{(P)}{\l}{180}{left}{#2}
\circlearrow{(P)}{\l}{0}{0.3}{right}{#3}
}

\newcommand{\punctureindigonPPintoN}[8]{
\def\l{#6}
\def\r{#7}
\def\ratio{#8}
\coordinate (P) at (0,0);
\coordinate (U) at ($(P)+(0,\l)$);
\coordinate (D) at ($(P)+(0,-\l)$);
\coordinate (UU) at ($(U)+(0,\ratio*\l)$);
\coordinate (DD) at ($(D)+(0,-\ratio*\l)$);
\draw[fill=white] (P) circle (1.8*\r) node[#4]{#5};
\draw[black,thick] (P) circle (1.8*\r);
\draw[fill=black] (P) circle (\r);
\draw[no arrow] (U) -- (UU) node[above]{#1};
\draw[no arrow] (DD) node[below]{#1} -- (D);
\draw[thick] (U) arc (90:-90:{\l});% node[right]{$a$};
\draw[thick] (U) arc (90:270:{\l});% node[right]{$a$};
\anticirclearrow{(P)}{\l}{180}{0.3}{left}{#2}
\circlearrow{(P)}{\l}{0}{0.3}{right}{#3}
}

\usetikzlibrary{external}
\numberwithin{equation}{section}
\def\SUSYN#1{{\mathcal{N}{=}#1}}

\def\wtl#1{\widetilde{#1}}
\def\ovl#1{\overline{#1}}

\def\bbZ{\mathbb{Z}}

\def\bbR{\mathbb{R}}
\def\bbC{\mathbb{C}}
\def\bbH{\mathbb{H}}

\def\Lieg{\mathfrak{g}}

\def\Liesu{\mathfrak{su}}

\def\Lieso{\mathfrak{so}}

\def\calB{\mathcal{B}}

\def\calI{\mathcal{I}}

\def\calK{\mathcal{K}}

\def\calM{\mathcal{M}}

\def\calN{\mathcal{N}}
\def\calO{\mathcal{O}}

\def\calP{\mathcal{P}}

\def\calS{\mathcal{S}}

\def\calW{\mathcal{W}}

\def\calZ{\mathcal{Z}}

\def\qqqquad{\quad\quad\quad\quad}
\def\for{\quad \mbox{for} \;\;}
\def\ie{{\it i.e.} }

\usepackage{bigdelim}
\usepackage{color}
\usepackage{textcomp}
\usepackage{multirow}
\usepackage{blkarray}
\usepackage{arydshln}
\usepackage{ascmac}

\usepackage{classS_tikz}

\def\Lieu{\mathfrak{u}}
\def\Lieso{\mathfrak{so}}

\def\rk{\mathrm{rk}}

\def\Adj{\mathrm{Adj}}

\def\rmbf#1{\mathrm{\mathbf{#1}}}

\def\rotateL#1{\rotatebox{90}{#1}}
\def\rotateR#1{\rotatebox{-90}{#1}}
\def\YsetR#1#2{\raisebox{#1}{\rotateR{#2}}}
\def\YsetSR#1#2#3{\scalebox{#1}{\YsetR{#2}{#3}}}
\pdfoutput=1

\begin{document}

\begin{titlepage}

\begin{flushright}
IPMU 17-0017
\end{flushright}
\vskip 2cm

\begin{center}

{\Large \bfseries
Schur indices with class S line operators from networks and further skein relations
}

\vskip 1.2cm

Noriaki Watanabe$^{\mbox{\textmusicalnote}}$

\bigskip
\bigskip

\begin{tabular}{ll}
$^{\mbox{\textmusicalnote}}$  & Kavli Institute for the Physics and Mathematics of the Universe, \\
& University of Tokyo,  Kashiwa, Chiba 277-8583, Japan
\end{tabular}

\vskip 1.5cm

\textbf{Abstract}
\end{center}

\medskip
\noindent
We compute the Schur indices in the presence of some line operators based on our conjectural formula introduced in \cite{Watanabe1603}. In particular, we focus on
the rank 1 superconformal field theories with the enhanced global symmetry and the free hypermultiplets with the elementary pants networks defined on the three punctured sphere in the class S context.
From the observations on the concrete computations, we propose new kinds of the class S skein relations in the sense that they include the generic puncture non-trivially.
We also give a general formula to unify all the relations we have exhibited.

\bigskip
\vfill
\end{titlepage}
\setcounter{page}{1}

\tableofcontents

\pagebreak

\section{Introduction}

Recently, defects in field theories can be analyzed in various set-ups and in various ways.
In supersymmetric cases, we can compute them by using the localization methods \cite{Pestun07,KWY0909,GOP11,GKL1201,Okuda1412_review}.
In conformal cases, we also apply the unbroken symmetry to restrict the correlator form \cite{GLMR15,BGLM16,Gadde1602}.
In particular, there are topological defects in conformal field theories (CFTs) \cite{FFRS06,DGG10} or topological field theories (TQFTs) \cite{Witten88Jones} and they are determined by the isotropy class and, in many examples, some data such as representations of some algebra or group.

\begin{figure}[htb]
\centering
\includegraphics[width=70mm]{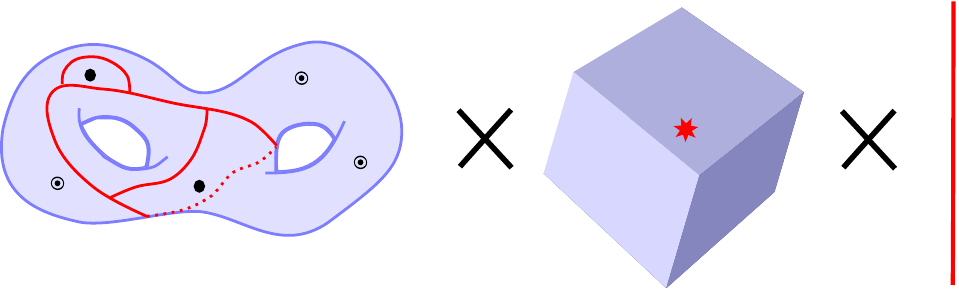}
\caption{The codimension four defects are put on the 6D bulk as the networks on the 2D punctured Riemann surface (Left), a localized point in the space direction ($S^3$ in our case) (Middle) and the straight line or the loop in the time direction ($S^1$ in our case) (Right).}
\label{fig:loop-network defect in 4D/2D}
\end{figure}

Marvellously, interesting phenomena were discovered that there are relations between the topological defects in special 2D CFTs (Liouville/Toda) or some TQFTs and BPS defects in 4D $\SUSYN{2}$ superconformal field theories (SCFTs) called ``class S theories". The class S theories are 4D $\SUSYN{2}$ systems obtained by the twisted compactifications via some Riemann surfaces of 6D $\SUSYN{(2,0)}$ SCFTs with several codimension two defects localized on the surfaces \cite{Gaiotto0904,GMN2}.
These 4D/2D duality relations were first observed in \cite{AGT09,Wyllard09,GPRR0910} in the absence of defects and, in particular, they identify the class S Schur indices with the partition functions of the 2D $q$-deformed Yang-Mills theory in \cite{GRRY1104}.
This conjecture allows us to compute the superconformal indices (SCIs) of the Lagrangian unknown theories.
This relation was extended to the correspondence between the 4D line operators and the 2D Wilson loop/network defects \cite{DMO09,AGGTV09,DGOT09,Passerini10,GomisLeFloch10,Xie1304,Bullimore13,TW1504,CGT1505,Gabella1603}.
Assuming this 4D/2D duality relation, we define the class S line operators by specifying the networks on the punctured Riemann surface even in the Lagrangian unknown theories. See Fig.~\ref{fig:loop-network defect in 4D/2D}.
In the our previous paper \cite{Watanabe1603}, we proposed the conjectural formula to compute the Schur indices with such line operators without any derivation.
\footnote{In our paper \cite{TW1504}, by using the skein relation (digon type without any puncture), we derived the formula to compute the Schur index with the elementary pants network in the $T_3$ theory (See Sec.~\ref{subsubsec:Rank 1 E_6 SCFT case}). However, the extension to higher rank cases ($N>3$) is highly non-trivial.}
To justify the formula, we checked two things there. One is the proof that they satisfy some fundamental skein relations and the other is to see that the Schur index of rank 1 SCFT with the $E_7$ global enhanced symmetry in the presence of the loop operators is written with $E_7$ characters only at the leading order in the $q$-expansion.
In Sec.~\ref{sec:Schur indices with line defects} of this paper, we give more evidences that the formula would be correct by computing the Schur indices with class S line operators in many examples.

In particular, we focus on the rank 0 SCFTs which are considered to be free hypermultiplets (Sec.~\ref{subsec:Free hypermultiplets}) and the rank 1 SCFTs with enhanced global symmetry (Sec.~\ref{subsec:Rank 1 SCFTs}).
\footnote{The rank means the complex dimension of the 4D Coulomb branch.}
In the first case, there are no non-trivial line operators but just flavor Wilson lines which are just the classical functions, namely, not operators.
This suggests that, in this case, all the Schur indices with the non-trivial networks are always factorized into the Schur indices without the line operators and the flavor Wilson line factor.
From the conjectural formula, it is almost impossible to see this factorization and this check is highly non-trivial.
In the latter case, we expect that the only one Schur index with an elementary pants network respects the global enhanced symmetry and the expression is written with the characters of the enhanced symmetry.
We check these two expectations in three cases ($E_6$,$E_7$ and $E_8$ global symmetry) by expanding the Schur indices as $q$-series.
Finally, we exhibit the superconformal QCD applications briefly.

In Sec.~\ref{sec:new kinds of skein relations}, we propose the new kinds of class S skein relations in the sense that they include the (non-degenerate) punctures. In particular, we focus on the digon-type skein relations with one puncture.
These relations account for the previous factorization in the rank 0 or 1 SCFTs in terms of the geometrical relations.
We exhibit the concrete examples in Sec.~\ref{subsec:A3 case},~\ref{subsec:A4 case} and ~\ref{subsec:A5 case} and then propose the unified form of for any puncture in Sec.~\ref{subsec:General punctures}.
In Sec.~\ref{subsec:Application}, we see one simple application of these skein relations.

\end{comment}
\section{Schur indices with line defects from 2D $q$-deformed Yang-Mills}
\label{sec:Schur indices with line defects}

As explained in the introduction briefly, the 4D/2D duality relation says that there is the map from the Riemann surfaces to the 4D SCFTs, which just means the twisted compactification of the 6D $\Liesu(N)$-type $\SUSYN{(2,0)}$ SCFT for fixed $N=2,3,\ldots$.
Notice that the Riemann surface may have (regular) punctures each of which has a type and the holonomy around it determined from the type.
We denote this map by $T^\calS_{\Lieg}$ and 
$T^\calS_{\Lieg=\Liesu(N)}[C(Y_1,Y_2,\ldots,Y_\ell)]$ denotes the resulting 4D $\SUSYN{2}$ theory in the IR after the twisted compactification on the Riemann surface $C$ with punctures whose types are specified by $Y_1,Y_2,\ldots,Y_\ell$.
We can extend this map from the pair of a punctured Riemann surface and networks on it to the 4D SCFT with line operators.

\subsubsection*{Elementary pants networks}

In this section, we analyse three cases : rank 0 SCFTs (free hypermultiplets), rank 1 SCFTs and superconformal QCDs in class S theories.

\begin{figure}[t]
\centering
\includegraphics[width=100mm]{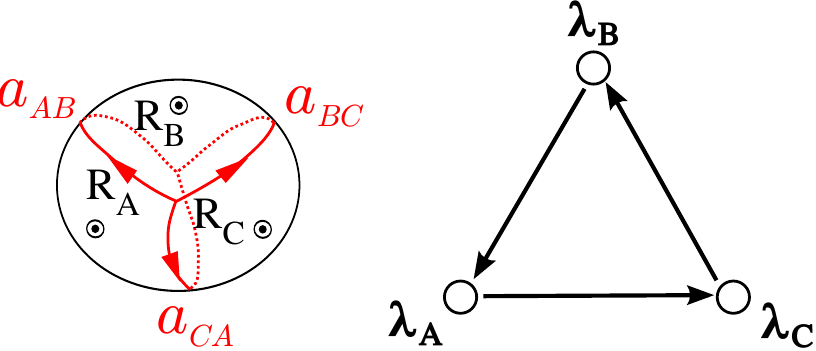}
\caption{ The $(a_{AB},a_{BC},a_{CA})$-type elementary pants network $\wp(a_{AB},a_{BC},a_{CA})$ and the associated dual decorated and oriented lattice on the two sphere.}
\label{fig:pants network again}
\end{figure}

\begin{figure}[th]
\centering
\includegraphics[width=30mm]{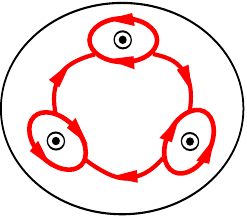}
\caption{An example of complicated networks on the trinion (three punctured sphere) which appears after the crossing resolutions in the product of two elementary pants networks. See \cite{TW1504,CGT1505}.}
\label{fig:complicated network on trinion}
\end{figure}

Among the possible networks on the trinion, there are the minimal ones depicted in Fig.~\ref{fig:pants network again}.
They were discussed explicitly at first in \cite{Xie1304} and shown to be elementary generators of the line operator algebra in \cite{CGT1505} which were called pants networks. In this paper, we refer to the network shown in Fig.~\ref{fig:pants network again} as $(a_{AB},a_{BC},a_{CA})$-type elementary pants network and denote it by $\wp(a_{AB},a_{BC},a_{CA})$.
The computation of the SCI in the simplest case, that is to say, $\wp(1,1,1)$ in the $T_3$ theory, was done in our paper \cite{TW1504}.
However, the extension to other types of pants networks had been highly non-trivial before our work \cite{Watanabe1603}. We briefly review the necessary results in Sec~\ref{subsec:Conjectural formula for elementary pants networks}.
Notice that these pants networks are expected to generate all the possible networks not touching on the punctures. See Fig~\ref{fig:complicated network on trinion}.

%The key assumption in the analysis of skein relations is that the there is the same number of independent elementary pants networks as the rank of IR charge lattice which equals to the Coulomb branch dimension.

Through this and next sections, there is an important assumption : there is the same number of independent elementary pants networks as the rank of IR charge lattice which equals to the Coulomb branch dimension.
The first example we can check readily is the $T_N$-theory \cite{Gaiotto0904}.
The Coulomb branch dimension of the $T_N$-theory equals to $\frac{(N-1)(N-2)}{2}$.
On the other hand, the number of possible junctions in the type $A_{N-1}$ case is given by the number of the possible partitions of $N$ into three parts, that is, just $\frac{(N-1)(N-2)}{2}$.
\!\footnote{In the other types, this is not true. For example, in the $\Lieg=\Lieso(2N)$ case, the Coulomb branch {\it real} dimension is given as $2N(N-2)$. In particular, when $N=4$ or $\Lieso(8)$, sixteen independent pants networks are expected if we follow the above assumption. However, the actual number of possible junctions is $10$. Here the reason why we consider the real dimension is just these networks are invariant under charge conjugation.}% For $\Lieg=\mathfrak{e}_6$ case, }
%---the above footnote %tbc

Next, let us see the rank 0 SCFTs, namely, free hypermultiplets. They have no Coulomb branch moduli and there is no dynamical gauge field.
Therefore, if all line operators given by pants networks are neutral under the background magnetic charges, they are expected to be flavor Wilson lines which are just the classical holonomies because the gauge fields are just the background fields.
At the computation level, this fact is realized as the factorization of the Schur indices with the pants networks into the no defect Schur indices and some simple factors.
%In particular, by interpreting that new kinds of skein relations happen on $C$ as discussed in Sec.~\ref{sec:new kinds of skein relations}, we assume that these factors are invariant under $q \to q^{-1}$ as explained before and check that this is true in the concrete examples.

We find that both the above assumption and the conjectural formula are consistent and support each other by the computations in this section and the analysis in the next section.

\subsection{Some formulae about 4D Schur indices and 2D $q$-deformed Yang-Mills}
\label{subsec:Conjectural formula for elementary pants networks}

In this section, we review the necessary facts to compute the $q$-deformed Yang-Mills correlators or the Schur indices.

\subsubsection*{Puncture}

The punctures are the codimension two defects in the 6D $\SUSYN{(2,0)}$ SCFTs and the important classes (regular,untwisted) are labelled by the principal embedding $\rho \;:\; \Liesu(2) \rightarrow \Liesu(N)$ \cite{GaiottoWitten0804,Gaiotto0904,GMN2}.
This classification is equivalent to the partition of $N$ when the Lie algebra type of the 6D SCFT $\SUSYN{(2,0)}$ is $\Liesu(N)$, namely, the decomposition of $N$ dimensional irreducible representations of $\Liesu(N)$ into the representation of $\Liesu(2)$. We represent this by $Y=[n_1n_2\ldots n_k]$ where $\sum_i n_i =N$.
In the language of the 4D $\SUSYN{2}$ SCFT, this corresponds to the choice of the nilpotent VEV to the $SU(2)_R$ highest part of the flavor current moment map $\mu^+$ for a $SU(N)$ flavor symmetry.
This VEV breaks $SU(2)^{UV}_R \times SU(N)$ symmetry at the original SCFT into a subgroup $G_Y \times U(1)$ and, after the flowing to another IR SCFT, we have new global symmetry $G_Y \times SU(2)^{IR}_R$. In other words, the $SU(2)^{UV}_R$ singlet matter in $\rmbf{N}$ for the $SU(N)$ symmetry are charged under $SU(2)^{IR}_R$ following the decomposition $\rmbf{N} \longrightarrow \bigoplus_i \rmbf{n}_i$.
Indeed, under the $SU(2)_R^{IR} \times G_Y$ symmetry,
the matter in $\rmbf{N}$ for $SU(N)$ symmetry at UV belongs to
\begin{align}
\rmbf{N} \longrightarrow \bigoplus_i \rmbf{n}_i = \bigoplus_{a} \rmbf{m}_a \otimes \rmbf{d}_a
\label{eq:decomposition of N by the principla embedding}
\end{align}
where $Y=[n_1n_2\ldots n_k]=[m_1^{d_1}\ldots m_t^{d_t}]$ and $G_Y=S( \prod_a U(d_a))$.
This is also true for the gauge multiplet.
Under $G \rightarrow SU(2)^{IR}_R \times G_Y$ (neither embedding nor subgroup),
\begin{align}
\Adj \longrightarrow \bigoplus_{m} \rmbf{m} \otimes R_m
\end{align}
where $R_m$ is the representation under $G_Y$ and $m$ runs over the irreducible representations of $SU(2)_R^{IR}$, namely, positive integers.

\subsubsection*{Expressions without defects}

Now, let us review the relation between the 2D $q$-deformed Yang-Mills theory correlator and the 4D Schur indices. The difference consists of the factor $\calK_Y(a;q)$ associated to each puncture $Y$ and the overall factor $\calN(q)$.
For the $[1^N]$-type puncture called full puncture, the factor from the $q$-YM to the Schur is given by
\begin{align}
\calK(a)=\calK_{[1^N]}(a)=\dfrac{1}{\calI_{\rm vector}^{\rm Schur}(a;q)^{1/2}}= \prod_{\alpha \in \Adj}\dfrac{1}{(qa^\alpha)_\infty}
\end{align}
where $\calI_{\rm vector}^{\rm Schur}(a;q)$ is the Schur index for the free vector multiplet of $SU(N)$.
This is just a basis change of the class functions, at least,  in the Schur limit.

Next, let us see the other types of punctures.
Recalling that the expression for $\calK_{[1^N]}(a;q)$ is the product over all the weights of the vector multiplet, the reduced expression should be given by the product of the  factor $(q^{1+I}a^w;q)_\infty$ ($j_2-j_1=1$) over all the weights $w$ of $R_m$ and over possible $m$. Therefore, the reduced expression is given by
\begin{align}
\calK_Y(a_{IR};q)=\prod_{m}\prod_{w \in \Pi(R_m)}\dfrac{1}{(q^{\frac{m+1}{2}}a_{IR}^w)_\infty}
\end{align}
where $a_{IR}$ is the fugacity of $G_Y$. The relation between $a_{IR}$ and $a_{UV}=a_Y$ follows from ~\eqref{eq:decomposition of N by the principla embedding}. See Table ~\ref{table:fugacity assignment} in Sec.~\ref{subsec:General punctures}. Notice that the highest of $\rmbf{m}$ for $SU(2)_R^{IR}$ has the additional $\frac{m-1}{2}$ charge compared to the UV where the gaugino (not all) contribute to SCIs as $qa^w$.
This reproduces the rule given in \cite{GRRY1104}.
See the derivations in \cite{GRR1207,Tachikawa1504} for details.

By introducing the new function for each $Y$ defined as
\begin{align}
\psi^{(Y)}_{R}(a;q):=\calK_Y(a;q)\chi_R(a_Y),
\end{align}
we can write down the complete expression of general class S Schur indices as
\begin{align}
\calI^{\rm Schur}_{T^\calS[C(Y_1,Y_2,\ldots,Y_\ell)]}(\{a\};q)=\calN(q)^{\ell-\chi_C}\sum_{\lambda \in \calP^{\Liesu(N)}_+} (\dim_q R(\lambda))^{\chi_C-\ell} \prod_{i=1}^{\ell}\psi^{(Y)}_R(a^{G_{Y_i}}_i;q).
\end{align}
where $\chi_C$ is the ordinary Euler number of $C$ on ignoring the punctures and $\dim_q R(\lambda)$ is the quantum dimension defined as ~\eqref{eq:definition of quantum dimension}.
On the other hand, the overall factor is given by
\begin{align}
\calN(q):=\calK_{[N]}(q)^{-1}=\prod_{i=2}^{N}(q^i;q).
\end{align}

\subsubsection*{Expressions in the presence of network operators}

For the loop operators in the 2D system, see \cite{CMR94,BR94,AOSV04} for example.
Since we are mainly interested in the elementary pants network operators in this paper, we recall the formula for the elementary pants network.
The claim in \cite{Watanabe1603} is that the 2D $q$-deformed Yang-Mills correlators on $C(Y_A,Y_B,Y_C)$ with the elementary pants network $\wp(a_{AB},a_{BC},a_{CA})$ is given by
\begin{align}
\calZ^{q\mathrm{YM}}_{\tiny T^\calS[C(Y_A,Y_B,Y_C)]\mbox{ w/. }\wp(a_{AB},a_{BC},a_{CA})}(a,b,c;q)=\sum_{\lambda_A,\lambda_B,\lambda_C \in \calP^{\Liesu(N)}_+}  \dfrac{\chi_{R(\lambda_A)}(a_{Y_A})\chi_{R(\lambda_B)}(b_{Y_B})\chi_{R(\lambda_C)}(c_{Y_C})}{\displaystyle \prod_{h=1}^{N-1}\prod_{\alpha=1}^{N-h}\dfrac{[(\hat{\lambda}_{ABC})_{h;\alpha}+h]_q}{[h]_q}}.
\label{eq:formula for qYM correlator with elementary pants network}
\end{align}
where $\calP_+^{\Liesu(N)}$ is the dominant weight set or all the finite dimensional irreducible representations set and $[n]_q$ is the $q$-number defined by $[n]_q:=\dfrac{q^{n/2}-q^{-n/2}}{q^{1/2}-q^{-1/2}}$. $(\hat{\lambda}_{ABC})_{h;\alpha}$ is defined as
\begin{align}
& (\hat{\lambda}_{ABC})_{h;\alpha}:=
mj\left(\sum_{s=\alpha}^{\alpha+h-1}\lambda_{A,s},\sum_{s=\alpha}^{\alpha+h-1}\lambda_{B,s},\sum_{s=\alpha}^{\alpha+h-1}\lambda_{C,s}\right) \\
& mj(a,b,c):=
\begin{cases}
a \qquad & b=a \;\textrm{or}\; c=a \\
b \qquad & a=b \;\textrm{or}\; c=b \\
c \qquad & a=c \;\textrm{or}\; b=c \\
\infty \qquad & \mbox{otherwise}
\end{cases}
\end{align}
where ``$\infty$" in this formula simply means that the contribution for the triple $(\lambda_A,\lambda_B,\lambda_C)$ in ~\eqref{eq:formula for qYM correlator with elementary pants network} vanishes and see Appendix.~\ref{app:Lie algebra convention} for $\lambda_{X,s}$.
$\lambda_{X}\in \calP_+^{\Liesu(N)}$ ($X=A,B,C$) just means that each $\lambda_{X,s}$ runs over non-negative integers for $s=1,2,\ldots,N-1$.
In the absence of the elementary pants networks ($a_{AB}=N$ and $a_{BC}=a_{CA}=0$ for example), this reduces to the well-known 2D topological $q$-deformed Yang-Mills partition function.
The corresponding Schur index is obtained by replacing $\chi_R(x_Y)$ by $\psi^{(Y)}_R(x^{G_Y};q)$ in the above expression and multiplying $\calN(q)$.

\subsection{Good theories for $q$-series expansion}
\label{subsec:Good theories for q-series expansion}

In the actual computations, we expand the above expression ~\eqref{eq:formula for qYM correlator with elementary pants network} in order of ascending powers of $q$.
The important point is that the expansions are well-defined only when we choose the types of punctures appropriately. For simplicity, let us consider the no network defect cases.

We can expand the inverse of $q$-dimensions as
\begin{align}
\dfrac{1}{\dim_q R(\lambda)}=& q^{\rho^\alpha \lambda_\alpha}(1- \# \{ \alpha \;|\; \lambda_\alpha \neq 0 \} + \calO(q^2))
\end{align}
where $\rho^\alpha=\frac{1}{2} \alpha(N-\alpha)$.
Therefore, the leading exponent for fixed $\lambda$ is given by
\begin{align}
g_{R}:=\dfrac{1}{2} \sum_{\alpha=1}^{N-1} \alpha(N-\alpha) \lambda_{\alpha}.
\end{align}
On the other hand, the characters is expanded as
\begin{align}
\chi_{R(\lambda)}(a_Y)=:& q^{\rho_{L;Y}^\alpha \lambda_\alpha} (f_{R,Y}(a^{G_Y})+\calO(q^{1/2}))
\end{align}
where $\rho^\alpha_{L,Y} \le 0$ for $\alpha=1,2,\ldots,N-1$.
For $Y=[1^N]$, $\rho^\alpha_{L,Y}=0$.
For $Y=[N]$, $\rho^\alpha_{L,Y}=-\frac{1}{2}\alpha(N-\alpha)$.
See Table.~\ref{tab:A3 examples of leading exponent},~\ref{tab:A4 examples of leading exponent} and ~\ref{tab:A5 examples of leading exponent} for example.

In the total expressions,
\begin{align}
\sum_{\lambda \in \calP^{\Liesu(N)}_+} \dfrac{\prod_{i=1}^{n}\chi_{R(\lambda)}(a_{i,Y_i})}{(\dim_q R(\lambda))^{\ell-\chi}}
=
\sum_{\lambda \in \calP^{\Liesu(N)}_+}
q^{(-\chi \rho^\alpha+\sum \rho_{L;Y_i}^\alpha)\lambda_a}\left( f_R(\{a\})+\calO(q^{1/2}) \right)
\end{align}
and we can see that the leading exponent is linear in the dominant weight.
To have a well-defined $q$-series, we must require that $L^\alpha:=(-\chi \rho^\alpha+\sum \rho_{L;Y_i}^\alpha) > 0$ for all $\alpha$.
We conjecture that this is always true for any good or ugly class S theories whose formal number of the Coulomb branch operators at each scaling dimension is non-negative.
\!\footnote{The author thanks Y.Tachikawa for a few discussions on these subjects.}
It is expected that this is also equivalent to the condition that the explicit symmetry at UV is unbroken at the IR SCFT \cite{GMT11}.
In particular, for bad theories with vanishing $\rho_{L,Y}^\alpha$, we can sometimes compute the Schur indices but some $\delta$ functions may appear. It causes the symmetry breaking.
We do not consider this ugly case in this paper.
When we add some network defects, it just shifts the leading order by some constant and the $q$-expansion is still valid.

\subsection{Breakdown of unitarity bound for defects}

Finally, we make a comment on the breakdown of unitarity bound.
As we see in the later examples, some Schur indices have the negative exponents of $q$.
In the absence of loop operators, it is guaranteed that the exponents of $q$ are always non-negative because there are three (Poincar\'{e}) supercharges anti-commuting the supercharges to define the Schur indices.
However, the insertion of line defects breaks the full superconformal algebra into the subalgebra and there left no supercharges anti-commuting the defining supercharge.
See \cite{CGS1606} for the subalgebras.
This no longer ensures the non-negativity of the exponents of $q$.

Let us consider a simple example of the breakdown recalling the discussion in \cite{GMN3,IOT11}.
If we have a purely electric line operator and another purely magnetic line operator, they classically generate the Poynting vector around the axis through two on ignoring the Euclidean time direction.
This contributes to the Schur indices as some negative power of $q$ when we put the line operators in the appropriate order.
\!\footnote{If we exchange the positions of two line operators, the sign of the contribution in the exponent of $q$ is reversed.}%, that is to say, $q$ coming from the classical bulk electromagnetic field is replaced by $q^{-1}$.}

Notice also that the $\SUSYN{(2,2)}$ type surface operators also break this bound because they create angular momentum around them when we realize them as vortex strings \cite{GRR1207}.
At mathematical level, they act on the superconformal indices as difference operators and many times applications lead to the negative power of $q$.

\subsection{Free hypermultiplets}
\label{subsec:Free hypermultiplets}

In our analysis, we assume that all the SCFTs without any Coulomb branch are free hypermultiplets specified by the representation of the flavor symmetry.
Under the condition that only untwisted regular punctures are allowed, we can classify them.
Using the formula ~\eqref{eq:dimension formula for Coulomb branch with each scaling dimension}
and the condition $d_{C,k}=0$ leads to the constraint $p^A_k+p^B_k+p^C_k=2k-1$ for $k=2,3,\ldots,N$.
We use the property that the regular punctures must satisfy $p_{k+1}-p_k=0 \mbox{ or }1$ in addition.
These constraints determine all the possibility of $p^{A,B,C}_k$ and all the three punctures combinations are listed in Table.~\ref{table:classification of free hypermultiplet}.
We categorize them into three classes.
From the analysis later, we call each bi-fundamental type, second rank anti-symmetric type and exceptional type.

To determine what representations of the explicit global symmetry given by the three punctured sphere, we use the information about the dimension of the Higgs branch and flavor central charges for non-Abelian simple group.

\begin{table}[htb]
\centering
\begin{tabular}{|c||c|c|c|}
\hline
theory & \!\!explicit flavor symmetry\!\! & $d_H$ & \!\!flavor central charges\!\! \\
\hline
\hline
\scalebox{0.9}{$T^\calS[C([1^N],[1^N],[N-1,1])]$} & \scalebox{0.8}{$SU(N)_1 \times SU(N)_2 \times U(1)$} & $N^2$ & \scalebox{0.9}{$k_{SU(N)_1}\!=\!k_{SU(N)_2}\!=\!2N$} \\
\hline
\scalebox{0.75}{$T^\calS[C([1^N],[n^2],[n,n\!-\!1,1])]$} \scalebox{0.6}{($N\!=\!2n \ge 4$)}\!\! & \scalebox{0.9}{$SU(N) \times SU(2) \times U(1)^2$} & \!\!\scalebox{0.75}{$\frac{1}{2}N(N\!\!+\!3)\!$}\! & \scalebox{0.85}{$k_{SU(N)}\!=\!N$, $k_{SU(2)}\!=\!2N$} \\
\scalebox{0.75}{$T^\calS[C([1^N],[n^21],[n\!+\!1,n])]$} \scalebox{0.5}{($N\!=\!2n\!+\!1 \ge 5$)}\!\! & & & \\
\hline
\scalebox{0.9}{$T^\calS[C([1^6],[2^3],[42])]$} & \scalebox{0.8}{$SU(6) \times SU(3) \times U(1)$} & $28$ & \scalebox{0.85}{$k_{SU(6)}\!=\!12$, $k_{SU(3)}\!=12\!$} \\
\hline
\scalebox{0.9}{$T^\calS[C([21^4],[2^3],[3^2])]$} & \scalebox{0.75}{$SU(4) \!\times\! SU(3) \!\times\! SU(2) \!\times\! U(1)$} & $24$ & \scalebox{0.65}{$k_{SU(4)}=10$, $k_{SU(3)}\!=\!k_{SU(2)}\!=\!12$} \\
\hline
\end{tabular}
\caption{Classification of the class S free hypermultiplets and their data}
\label{table:classification of free hypermultiplet}
\end{table}

\subsubsection*{Bi-fundamental type : $\bf F^{\rm (bf)}_{A_{N-1}}$}

The first case $F^{\rm (bf)}_{A_{N-1}}=T^\calS[C([1^N],[1^N],[N-1,1])]$ is well-known to be the bi-fundamental free hypermultiplet \cite{Gaiotto0904}.
In fact, the bi-fundamental free hypermultiplets have $SU(N)_1 \times SU(N)_2 \times U(1)$-symmetry where $U(1)$ is the baryon symmetry.
In term of $\SUSYN{1}$ chiral multiplets, the representation is given as $(\rmbf{N},\rmbf{N},1) \oplus (\rmbf{\ovl{N}},\rmbf{\ovl{N}},-1)$ and this means $k_{SU(N)_i} = 2N$ ($i=1,2$).
The degree of freedom in $\bbH$ also equals to $d_H=N^2$.
We can also see this fact from the 4D/2D duality relation conjecture stated above by computing it concretely.

\subsubsection*{Second rank anti-symmetric type : $\bf F^{\rm (as)}_{A_{N-1}}$}

In the case that $N=2n$ ($N \ge 4$ or $n \ge 2$) is even, the theory $F^{\rm (as)}_{A_{N-1}}=T^\calS[C([1^N],[n^2],[n,n-1,1])]$ is rank 0.
In the case that $N=2n+1$ ($N \ge 5$ or $n \ge 2$) is odd, the theory $F^{\rm (as)}_{A_{N-1}}=T^\calS[C([1^N],[n^2,1],[n+1,n])]$ is also rank 0.
Although the class S realizations of the even and odd cases look different, we can discuss both cases at the same time because the 4D physical properties are uniformly treated except $U(1)$ charges.
We can read off explicit flavor symmetry
as $SU(N) \times SU(2) \times U(1)^2$ for $N \ge 5$ and $SU(4) \times SU(2)^2 \times U(1)$ when $N=4$.
In both cases, $k_{SU(N)}=2N$ and $k_{SU(2)}=2N$ ($k_{SU(2)_1}=8$ and $k_{SU(2)_2}=6$ when $N=4$) holds true.
The dimension of Higgs branch is given by $d_H=\frac{1}{2}N(N+3)$.

The matter content, namely, the representation in terms of the chiral multiplets, satisfying these conditions is following :
\begin{align}
(\rmbf{N},\rmbf{2},\ast)\oplus (\rmbf{\ovl{N}},\rmbf{2},\ast) \oplus (\wedge^2 \rmbf{N},\rmbf{1},\ast)\oplus (\wedge^2 \rmbf{\ovl{N}},\rmbf{1},\ast) \quad \mbox{ when } N \ge 5
\label{eq:matter contents in hyper coming from type II trinions} \\
(\rmbf{4},\rmbf{2},\rmbf{1},\ast)\oplus (\rmbf{\ovl{4}},\rmbf{2},\rmbf{1},\ast) \oplus (\wedge^2 \rmbf{6},\rmbf{1},\rmbf{2},\ast) \quad\mbox{ when } N=4
\label{eq:matter contents in hyper coming from type II trinions when N=4}
\end{align}
where $\ast$ represents an undetermined $U(1)$ charge.

By computing the Schur indices from the $q$-deformed Yang-Mills partition function, we can fix $U(1)$ charges as follows.
\begin{align}
& \nonumber N=4 \\
& (\rmbf{4},\mathbf{2},\rmbf{1},1)\oplus (\rmbf{\ovl{4}},\mathbf{2},\rmbf{1},-1) \oplus (\rmbf{6},\mathbf{1},\rmbf{2},0)
\label{eq:matter contents in hyper coming from type II trinions when N=4 from 4D/2D} \\
& \nonumber N\mbox{ : even} \\
& (\rmbf{N},\mathbf{2},(1,0))\oplus (\rmbf{\ovl{N}},\mathbf{2},(-1,0)) \oplus (\wedge^2 \rmbf{N},\mathbf{1},(1,1))\oplus (\wedge^2 \rmbf{\ovl{N}},\mathbf{1},(-1,-1))
\label{eq:matter contents in hyper coming from type II trinions when even from 4D/2D} \\
& \nonumber N\mbox{ : odd} \\
& (\rmbf{N},(\mathbf{2},1),1)\oplus (\rmbf{\ovl{N}},(\mathbf{2},-1),-1) \oplus (\wedge^2 \rmbf{N},(\mathbf{1},2),-\tfrac{2}{N-1})\oplus (\wedge^2 \rmbf{\ovl{N}},(\mathbf{1}-2),\tfrac{2}{N-1}).
\label{eq:matter contents in hyper coming from type II trinions when odd from 4D/2D}
\end{align}

\subsubsection*{Exceptional case 1 : $\bf F^{\rm (ex1)}_{A_5}=T^\calS[C([1^6],[2^3],[42])]$}

In this case, the flavor symmetry is
$SU(6) \times SU(3) \times U(1)$, the flavor central charges are $k_{SU(6)}=12$ and $k_{SU(3)}=12$ and the Higgs branch dimension is $d_H=28$. The candidates satisfying the above conditions are $(\wedge^3 \mathbf{6}=\mathbf{20} , \mathbf{1}, *)\oplus (\rmbf{6},\rmbf{3},*)\oplus(\rmbf{\ovl{6}},\rmbf{\ovl{3}},*)$ or $(\wedge^3 \mathbf{6}=\mathbf{20} , \mathbf{1}, *) \oplus (\wedge^3 \mathbf{6}=\mathbf{20} , \mathbf{1}, *) \oplus (\mathbf{1},\rmbf{Adj}=\mathbf{8},*) \oplus (\mathbf{1},\rmbf{Adj}=\mathbf{8},*)$.
To the best of the author's knowledge, we cannot determine which candidate is actually true.
\!\footnote{Naively speaking, if the latter case is true, there may exists a $SU(2)$ symmetry but such symmetry does not appear.}
However, by the 4D/2D computation, we find that
\begin{align}
(\mathbf{20} , \mathbf{1}, 0)\oplus (\rmbf{6},\rmbf{3},1)\oplus(\rmbf{\ovl{6}},\rmbf{\ovl{3}},-1)
\end{align}
is the correct answer.

\subsubsection*{Exceptional case 2 : $\bf F^{\rm (ex2)}_{A_5}=T^\calS[C([21^4],[2^3],[3^2])]$}

In this case, the flavor symmetry is
$SU(4) \times SU(3) \times SU(2) \times U(1)$, the flavor central charges are $k_{SU(4)}=10$, $k_{SU(3)}=12$ and $k_{SU(2)}=12$ and the Higgs branch dimension is $d_H=24$.
Indeed, the matter content is given by
\begin{align}
(\wedge^2 \mathbf{4}=\mathbf{6},\mathbf{1},\mathbf{2},0)\oplus (\mathbf{4},\mathbf{\ovl{3}},\mathbf{1},\tfrac{1}{2})\oplus
(\mathbf{\ovl{4}},\mathbf{3},\mathbf{1},-\tfrac{1}{2})\oplus
(\mathbf{1},\mathbf{3},\mathbf{2},1)\oplus (\mathbf{1},\mathbf{\ovl{3}},\mathbf{2},-1).
\end{align}

\subsubsection{Bi-fundamental type}

\paragraph{$A_2$ bi-fundamental}

Let us see the simple case at first.
This is the $A_2$ bi-fundamental hypermultiplet whose flavor symmetry is given by $SU(3) \times SU(3) \times U(1)$. There, we consider the $(a_{AB},a_{BC},a_{CA})=(1,1,1)$-type pants network on $T^\calS[C([1^3],[1^3],[2,1])]=F^{\rm (bf)}_{A_2}=\mathrm{Hyper}(\rmbf{3},\rmbf{3},1)$.
Let $a,b$ and $c$ be the holonomies of $SU(3) \times SU(3) \times U(1)$. There are six sectors.
\!\footnote{The sum in ~\eqref{eq:formula for qYM correlator with elementary pants network} splits into the possible pairs of $\lambda_B-\lambda_A$ and $\lambda_C-\lambda_A$ (finite sum) and $\lambda_A$ for the fixed pair (infinite sum). We refer to the former pair as ``sector".}
Up to $q^{1/2}$-order, we can see that the $q$-deformed Yang-Mills correlators receive the contributions from all the triple of dominant weights at punctures listed in Table.~\ref{table: contributing dominant configuration up to q^1/2 in A2 bifundamental}.

% a-(1)->c-(1)->b-(1)->a anticlockwise

\begin{table}[htb]
\centering
\begin{tabular}{c|c|c|c|}
\cline{2-4} &
$((\lambda_{A})_{1},(\lambda_{A})_{2})$ & $((\lambda_{B})_{1},(\lambda_{B})_{2})$ & $((\lambda_{C})_{1},(\lambda_{C})_{2})$ \\
\cline{2-4}
\cline{2-4}
\ldelim\{{2}{1.3cm}[$q^{0},q^{1/2}$]
& $(0,0)$ & $(0,1)$ & $(1,0)$ \\
& $(1,0)$ & $(0,0)$ & $(0,1)$ \\
\cline{2-4}
\ldelim\{{6}{0.4cm}[$q^{1/2}$]
& $(2,0)$ & $(1,0)$ & $(1,1)$ \\
& $(1,1)$ & $(0,1)$ & $(0,2)$ \\
& $(1,0)$ & $(1,1)$ & $(0,2)$ \\
& $(0,1)$ & $(0,2)$ & $(1,1)$ \\
& $(0,1)$ & $(1,0)$ & $(0,0)$ \\
& $(0,1)$ & $(1,0)$ & $(1,1)$ \\
\cline{2-4}
\end{tabular}
\caption{The dominant configurations contributing to $q^{0}$ and $q^{1/2}$ terms.}
\label{table: contributing dominant configuration up to q^1/2 in A2 bifundamental}
\end{table}

The result is given by
\begin{align}
\nonumber & \calI^{\rm Schur}_{\tiny F^{\rm (bf)}_{A_2}\mbox{ w/. }\wp(1,1,1)}(a,b,c)
=\left[
\chi_{\rmbf{3}}(a)c^{-1}+\chi_{\rmbf{\ovl{3}}}(b)c
\right] +
q^{1/2}
\left[
2\chi_{\rmbf{\ovl{3}}}(a)\chi_{\rmbf{3}}(b)
+
\chi_{\rmbf{\ovl{3}}}(b)c^{-2}
+
\chi_{\rmbf{3}}(a)c^2 \right. \\
\nonumber & \left. +
\chi_{\rmbf{8}}(a)\chi_{\rmbf{\ovl{3}}}(b)c^{-2}
+
\chi_{\rmbf{3}}(a)\chi_{\rmbf{8}}(b)c^{2}
+
\chi_{\rmbf{6}}(a)\chi_{\rmbf{3}}(b)
+
\chi_{\rmbf{\ovl{3}}}(a)\chi_{\rmbf{\ovl{6}}}(b) \right] +q \left[
c^{3}\chi_{\rmbf{\ovl{3}}}(a)\chi_{\rmbf{3}}(b) \right. \\
\nonumber & +c^{-1}\chi_{\rmbf{15}'}(a)\chi_{\rmbf{8}}(b)+2c^{-1}\chi_{\rmbf{\ovl{6}}}(a)\chi_{\rmbf{8}}(b)+c^{-3}\chi_{\rmbf{\ovl{15}}'}(a)\chi_{\rmbf{\ovl{6}}}(b)+c^{-1}\chi_{\rmbf{\ovl{6}}}(a)\chi_{\rmbf{\ovl{10}}}(b)+c\chi_{\rmbf{10}}(a)\chi_{\rmbf{6}}(b)\\
\nonumber & +c^{-3}\chi_{\rmbf{6}}(a)\chi_{\rmbf{3}}(b)+c^{3}\chi_{\rmbf{6}}(a)\chi_{\rmbf{15}'}(b)+3c\chi_{\rmbf{\ovl{3}}}(b)+3c^{-1}\chi_{\rmbf{3}}(a)+c\chi_{\rmbf{\ovl{15}}'}(b)+3c^{-1}\chi_{\rmbf{3}}(a)\chi_{\rmbf{8}}(b)\\
\nonumber & +c^{-3}\chi_{\rmbf{\ovl{3}}}(a)\chi_{\rmbf{\ovl{6}}}(b)+c^{3}\chi_{\rmbf{\ovl{3}}}(a)\chi_{\rmbf{\ovl{6}}}(b)+2c\chi_{\rmbf{8}}(a)\chi_{\rmbf{6}}(b)+c\chi_{\rmbf{8}}(a)\chi_{\rmbf{\ovl{15}}'}(b)+3c\chi_{\rmbf{8}}(a)\chi_{\rmbf{\ovl{3}}}(b)\\
\nonumber & \left. +c^{-1}\chi_{\rmbf{15}'}(a)+c^{-1}\chi_{\rmbf{\ovl{6}}}(a)+c^{-3}\chi_{\rmbf{\ovl{3}}}(a)\chi_{\rmbf{3}}(b)+c^{3}\chi_{\rmbf{6}}(a)\chi_{\rmbf{3}}(b)+c\chi_{\rmbf{6}}(b)\right]+\calO(q^{3/2}) \\
&=\left[
\chi_{\rmbf{3}}(a)c^{-1}+\chi_{\rmbf{\ovl{3}}}(b)c
\right] \calI^{\rm Schur}_{F^{\rm (bf)}_{A_2}}(a,b,c)
\end{align}
where
\begin{align}
\nonumber \calI^{\rm Schur}_{F^{\rm (bf)}_{A_2}}(a,b,c)&=
1+q^{1/2}\left[c\chi_{\rmbf{3}}(a)\chi_{\rmbf{3}}(b)+c^{-1}\chi_{\rmbf{\ovl{3}}}(a)\chi_{\rmbf{\ovl{3}}}(b)\right]+q \left[1+\chi_{\rmbf{8}}(a)+\chi_{\rmbf{8}}(b)+\chi_{\rmbf{8}}(a)\chi_{\rmbf{8}}(b) \right. \\
\nonumber & \left. +c^{2}\chi_{\rmbf{\ovl{3}}}(a)\chi_{\rmbf{\ovl{3}}}(b)+c^{-2}\chi_{\rmbf{\ovl{6}}}(a)\chi_{\rmbf{\ovl{6}}}(b)+c^{2}\chi_{\rmbf{6}}(a)\chi_{\rmbf{6}}(b)+c^{-2}\chi_{\rmbf{3}}(a)\chi_{\rmbf{3}}(b)\right]+\calO(q^{3/2}) \\
\end{align}
and $\rmbf{15}'=R(2\omega_1+\omega_2)$.
Then, we find the flavor Wilson line factor.
In terms of the representations of the global symmetry, it is given by
\begin{align}
W_{\wp(1,1,1)}^{A_2\; \rm bi-fund}=(\rmbf{3},\rmbf{1},-1) \oplus (\rmbf{1},\rmbf{\ovl{3}},1).
\end{align}

\paragraph{$A_5$ bi-fundamental}

We can check that all the elementary pants networks in the $A_5=\Liesu(6)$ bi-fundamental free hypermultiplet are factorized as the SCI in the absence of lines and the following factors :
\begin{align}
& W_{\wp(4,1,1)}^{A_5\; \rm bi-fund}=c^{-1}\chi_{\rmbf{\ovl{15}}}(a)+c\chi_{\rmbf{15}}(b) \\
& W_{\wp(3,2,1)}^{A_5\; \rm bi-fund}=[2]_q c\chi_{\rmbf{20}}(b)+c^{-2}\chi_{\rmbf{20}}(a) \\
& W_{\wp(3,1,2)}^{A_5\; \rm bi-fund}=[2]_qc^{-1}\chi_{\rmbf{20}}(a)+
c^{2}\chi_{\rmbf{20}}(b) \\
& W_{\wp(2,3,1)}^{A_5\; \rm bi-fund}=[3]_qc\chi_{\rmbf{\ovl{15}}}(b)+c^{-3}\chi_{\rmbf{15}}(a) \\
& W_{\wp(2,2,2)}^{A_5\; \rm bi-fund}=[3]_q
\left(c^{-2}\chi_{\rmbf{15}}(a)+c^{2}\chi_{\rmbf{\ovl{15}}}(b)\right) \\
& W_{\wp(2,1,3)}^{A_5\; \rm bi-fund}=[3]_q
c^{-1}\chi_{\rmbf{15}}(a)+c^{3}\chi_{\rmbf{\ovl{15}}}(b) \\
& W_{\wp(1,4,1)}^{A_5\; \rm bi-fund}=[4]_qc\chi_{\rmbf{\ovl{6}}}(b)+c^{-4}\chi_{\rmbf{6}}(a) \\
& W_{\wp(1,3,2)}^{A_5\; \rm bi-fund}=([5]_q+1)c^{2}\chi_{\rmbf{\ovl{6}}}(b)+[4]_qc^{-3}\chi_{\rmbf{6}}(a) \\
& W_{\wp(1,2,3)}^{A_5\; \rm bi-fund}=([5]_q+1)c^{-2}\chi_{\rmbf{6}}(a)+[4]_qc^{3}\chi_{\rmbf{\ovl{6}}}(b) \\
& W_{\wp(1,1,4)}^{A_5\; \rm bi-fund}=[4]_q
c^{-1}\chi_{\rmbf{6}}(a)+c^{4}\chi_{\rmbf{\ovl{6}}}(b)
\end{align}

\paragraph{$A_{N-1}$ bi-fundamental}

In the general $A_{N-1}=\Liesu(N)$ case ( $T_{\Liesu(N)}^\calS[C([1^N],[1^N],[N-1,1])]=F^{\rm(bf)}_{A_{N-1}}=\mathrm{Hyper}(\rmbf{N},\rmbf{N},1)=\frac{1}{2}\mathrm{Hyper}(\rmbf{N},\rmbf{N},1)\oplus \frac{1}{2}\mathrm{Hyper}(\rmbf{\ovl{N}},\rmbf{\ovl{N}},-1)$ ), we conjecture that the $(p,r,s)$-type ($N=p+r+s$) pants network gives the flavor Wilson lines
\begin{align}
W_{(p,r,s)}^{A_{N-1}\; \rm bi-fund}=
\left[\begin{array}{c} r+s-1 \\ r \end{array}\right]_q
W^{\rm flavor}(\wedge^p \rmbf{N},\rmbf{1},-r) \oplus
\left[\begin{array}{c} r+s-1 \\ s \end{array}\right]_q
W^{\rm flavor}(\rmbf{1},\wedge^p \rmbf{\ovl{N}},s)
\label{eq:flavor Wilson of bi-fundamental type}
\end{align}
where $W^{\rm flavor}(R_A,R_B,q_C)$ is the flavor line in the representation $R_A \otimes R_B \otimes \bbC_{q_C}$ under $SU(N)_A \times SU(N)_B \times U(1)_C$ flavor symmetry of the bi-fundamental hypermultiplet. We have also introduced the q-binomial coefficient
\begin{align}
\left[\begin{array}{c} x+y \\ y \end{array}\right]_q
:=
\dfrac{[x+y]_q!}{[x]_q![y]_q!}
=\dfrac{\prod_{i=1}^{x+y} [i]_q}{\prod_{i=1}^{x} [i]_q\prod_{i=1}^{y} [i]_q}.
\end{align}

\subsubsection{Second rank anti-symmetric type}

Here we focus on two cases $\Lieg=\Liesu(4)$ and $\Liesu(5)$. The higher rank computations are totally similar.

\paragraph{$A_3$ second rank anti-symmetric}

We focus on $F^{\rm (as)}_{A_3}=T^\calS[C([1^4],[2^2],[21^2])]$ which consists of $(\rmbf{4},\rmbf{2},\rmbf{1},1)\oplus (\rmbf{\ovl{4}},\rmbf{2},\rmbf{1},-1) \oplus (\rmbf{6},\rmbf{1},\rmbf{2},0)$.
The global symmetry is $SU(4) \times SU(2)_B \times U(1) \times SU(2)_C$ and the associated fugacities are chosen as $a_{[1^4]}=(a_1,a_2,a_3,a_4)$ ($a_1a_2a_3a_4=1$), $b_{[2^2]}=(q^{1/2}b,q^{-1/2}b,q^{1/2}b^{-1},q^{-1/2}b^{-1})$ and $c_{[21^2]}=(q^{1/2}c,q^{-1/2}c,c^{-1}c'_1,c^{-1}c'_1\,\!^{-1})$.

\begin{align}
W_{\wp(2,1,1)}^{A_3\; \rm as}&=
[2]_qc+
c^{-1}\chi^{SU(2)}_{\rmbf{2}}(c')+\chi_{\rmbf{\ovl{4}}}(a)\chi^{SU(2)}_{\rmbf{2}}(b) \\
W_{\wp(1,2,1)}^{A_3\; \rm as}&=
[2]_qc\chi^{SU(2)}_{\rmbf{2}}(b)+c^{-1}\chi^{SU(2)}_{\rmbf{2}}(b)\chi^{SU(2)}_{\rmbf{2}}(c')+\chi_{\rmbf{\ovl{4}}}(a) \\
W_{\wp(1,1,2)}^{A_3\; \rm as}&=
c^{-1}\chi_{\rmbf{4}}(a)+\chi^{SU(2)}_{\rmbf{2}}(b)\chi^{SU(2)}_{\rmbf{2}}(c')+c\chi_{\rmbf{\ovl{4}}}(a)
\label{eq:flavor Wilson for A3 as-type free hypers}
\end{align}

\paragraph{$A_4$ second rank anti-symmetric}

Next, we focus on $F^{\rm (as)}_{A_4}=T^\calS[C([1^5],[2^21],[32])]$ which consists of $(\rmbf{5},\rmbf{2}_1,1)\oplus (\rmbf{\ovl{5}},\rmbf{2}_{-1},-1) \oplus (\mathrm{c.c.})$.
The global symmetry is $SU(5) \times SU(2) \times U(1)_B \times U(1)_C$ and the associated fugacities are chosen as $a_{[1^5]}=(a_1,a_2,a_3,a_4,a_5)$ ($a_1a_2a_3a_4a_5=1$), $b_{[2^21]}=(q^{1/2}bb',q^{-1/2}bb',q^{1/2}bb'^{-1},q^{-1/2}bb'^{-1},b^{-4})$ and\\ $c_{[32]}=(qc,c,q^{-1}c,q^{1/2}c^{-3/2},q^{-1/2}c^{-3/2})$.
The flavor Wilson lines are given by
\begin{align}
W_{\wp(3,1,1)}^{A_4\; \rm as}&=
[2]_qb^{2}c+b^{-3}c\chi^{SU(2)}_{\rmbf{2}}(b')+b^{2}c^{-3/2}+b\chi_{\rmbf{\ovl{5}}}(a)\chi^{SU(2)}_{\rmbf{2}}(b')+b^{-2}c^{-1/2}\chi_{\rmbf{5}}(a) \\
\nonumber W_{\wp(2,2,1)}^{A_4\; \rm as}&=
[2]_q^2 b^{-2}c+[2]_q \left( b^{-2}c^{-3/2}+b^{3}c\chi^{SU(2)}_{\rmbf{2}}(b') \right) +
b^{-2}c\chi^{SU(2)}_{\rmbf{3}}(b')+b^{2}\chi_{\rmbf{\ovl{5}}}(a) \\
&+b^{3}c^{-3/2}\chi^{SU(2)}_{\rmbf{2}}(b')+b^{-1}c^{-1/2}\chi_{\rmbf{5}}(a)\chi^{SU(2)}_{\rmbf{2}}(b') \\
\nonumber W_{\wp(2,1,2)}^{A_4\; \rm as}&=
[2]_qb^{-2}c^{-1/2}+c^{-1}\chi_{\rmbf{10}}(a)+b^{2}c\chi_{\rmbf{\ovl{5}}}(a)+b^{3}c^{-1/2}\chi^{SU(2)}_{\rmbf{2}}(b')+b^{-2}c^{2} \\
&+b^{-1}c^{1/2}\chi_{\rmbf{5}}(a)\chi^{SU(2)}_{\rmbf{2}}(b') \\
W_{\wp(1,3,1)}^{A_4\; \rm as}&=
[2]_q^2b^{-1}c\chi^{SU(2)}_{\rmbf{2}}(b')+[2]_q\left( b^{-1}c^{-3/2}\chi^{SU(2)}_{\rmbf{2}}(b')+b^{4}c \right) +b^{4}c^{-3/2}+c^{-1/2}\chi_{\rmbf{5}}(a) \\
\nonumber W_{\wp(1,2,2)}^{A_4\; \rm as}&=
[2]_q^2b^{-1}c^{-1/2}\chi^{SU(2)}_{\rmbf{2}}(b')+[2]_q \left( c^{1/2}\chi_{\rmbf{5}}(a)+b^{-1}c^{2}\chi^{SU(2)}_{\rmbf{2}}(b')+b^{4}c^{-1/2} \right) \\
&+c^{-2}\chi_{\rmbf{5}}(a)+b^{4}c^{2} \\
W_{\wp(1,1,3)}^{A_4\; \rm as}&=
[2]_q \left( b^{-1}c^{1/2}\chi^{SU(2)}_{\rmbf{2}}(b')+c^{-1}\chi_{\rmbf{5}}(a) \right) +c^{3/2}\chi_{\rmbf{5}}(a) +b^{4}c^{1/2}.
\label{eq:flavor Wilson for A4 as-type free hypers}
\end{align}

\subsubsection{Exceptional type}

For the two exceptional type free hypermultiplets, we can also check the factorizations at order-by-order in $q$.
The factorized factors are the flavor Wilson lines and the results are following.

\subsubsection*{$A_5$ exceptional type 1}

The first case is given by $F^{\rm (ex1)}_{A_5}=T^\calS[C([1^6],[2^3],[42])]$.
The global symmetry is given by $SU(6) \times SU(3) \times U(1)$ and the associated fugacities are chosen as $a_{[1^6]}=(a_1,a_2,a_3,a_4,a_5,a_6)$ ($a_1a_2a_3a_4a_5a_6=1$), $b_{[2^3]}=(q^{1/2}b_1,q^{-1/2}b_1,q^{1/2}b_2,q^{-1/2}b_2,q^{1/2}b_3,q^{-1/2}b_3)$ ($b_1b_2b_3=1$) and $c_{[42]}=(q^{3/2}c,q^{1/2}c,q^{-1/2}c,q^{-3/2}c,q^{1/2}c^{-2},q^{-1/2}c^{-2})$.
The flavor Wilson lines are given by
\begin{align}
W_{\wp(4,1,1)}^{A_5\; \rm ex1}&=[2]_q c\chi^{SU(3)}_{\rmbf{\ovl{3}}}(b)+c^{-1}\chi_{\rmbf{6}}(a)+\chi_{\rmbf{\ovl{6}}}(a)\chi^{SU(3)}_{\rmbf{3}}(b) \\
W_{\wp(3,2,1)}^{A_5\; \rm ex1}&=[3]_q[2]_q c+[2]_q\left( c\chi^{SU(3)}_{\rmbf{8}}(b)+c^{-2} \right)+\chi_{\rmbf{\ovl{6}}}(a)\chi^{SU(3)}_{\rmbf{\ovl{3}}}(b)+c^{-1}\chi_{\rmbf{6}}(a)\chi^{SU(3)}_{\rmbf{3}}(b) \\
W_{\wp(3,1,2)}^{A_5\; \rm ex1}&=[2]_q \left( c^{-1}+c^{2} \right) +c^{-1}\chi_{\rmbf{20}}(a)+\chi_{\rmbf{6}}(a)\chi^{SU(3)}_{\rmbf{3}}(b)+c\chi_{\rmbf{\ovl{6}}}(a)\chi^{SU(3)}_{\rmbf{\ovl{3}}}(b) \\
\nonumber W_{\wp(2,3,1)}^{A_5\; \rm ex1}&=[3]_q[2]_q c\chi^{SU(3)}_{\rmbf{3}}(b)+[2]_q\left( c\chi^{SU(3)}_{\rmbf{\ovl{6}}}(b)+c^{-2}\chi^{SU(3)}_{\rmbf{3}}(b)\right) \\
&+c^{-1}\chi_{\rmbf{6}}(a)\chi^{SU(3)}_{\rmbf{\ovl{3}}}(b)+\chi_{\rmbf{\ovl{6}}}(a) \\
\nonumber W_{\wp(2,2,2)}^{A_5\; \rm ex1}&=[3]_q\left( c^{2}+c^{-1} \right)\chi^{SU(3)}_{\rmbf{3}}(b)+[2]_q\left( \chi_{\rmbf{6}}(a)\chi^{SU(3)}_{\rmbf{\ovl{3}}}(b)+c\chi_{\rmbf{\ovl{6}}}(a) \right) \\
&+c^{2}\chi^{SU(3)}_{\rmbf{\ovl{6}}}(b)+c^{-2}\chi_{\rmbf{15}}(a)+c^{-1}\chi^{SU(3)}_{\rmbf{3}}(b) \\
W_{\wp(2,1,3)}^{A_5\; \rm ex1}&=[2]_q \left( \chi^{SU(3)}_{\rmbf{3}}(b)+c^{-1}\chi_{\rmbf{15}}(a) \right)+c\chi_{\rmbf{6}}(a)\chi^{SU(3)}_{\rmbf{\ovl{3}}}(b)+c^{2}\chi_{\rmbf{\ovl{6}}}(a) \\
W_{\wp(1,4,1)}^{A_5\; \rm ex1}&=[3]_q[2]_qc\chi^{SU(3)}_{\rmbf{\ovl{3}}}(b)+[2]_qc^{-2}\chi^{SU(3)}_{\rmbf{\ovl{3}}}(b)+c^{-1}\chi_{\rmbf{6}}(a) \\
W_{\wp(1,3,2)}^{A_5\; \rm ex1}&=[3]_q[2]_q \left( c^{2}+c^{-1} \right)\chi^{SU(3)}_{\rmbf{\ovl{3}}}(b)+[3]_q\chi_{\rmbf{6}}(a)+c^{-3}\chi_{\rmbf{6}}(a) \\
W_{\wp(1,2,3)}^{A_5\; \rm ex1}&=[3]_q[2]_q\chi^{SU(3)}_{\rmbf{\ovl{3}}}(b)+[3]_q\left( c+c^{-2} \right)\chi_{\rmbf{6}}(a)+[2]_qc^{3}\chi^{SU(3)}_{\rmbf{\ovl{3}}}(b) \\
W_{\wp(1,1,4)}^{A_5\; \rm ex1}&=[3]_qc^{-1}\chi_{\rmbf{6}}(a)+[2]_qc\chi^{SU(3)}_{\rmbf{\ovl{3}}}(b)+c^{2}\chi_{\rmbf{6}}(a).
\label{eq:flavor Wilson for A5 ex1-type free hypers}
\end{align}

\subsubsection*{$A_5$ exceptional type 2}

The second case is given by $F^{\rm (ex2)}_{A_5}=T^\calS[C([21^4],[2^3],[3^2])]$.
The global symmetry is given by $U(1) \times SU(4) \times SU(3) \times SU(2)$ and the associated fugacities are chosen as $a_{[21^4]}=(q^{1/2}a,q^{-1/2}a,a^{-1/2}a'_1,a^{-1/2}a'_2,a^{-1/2}a'_3,a^{-1/2}a'_4)$ ($a'_1a'_2a'_3a'_4=1$),\\ $b_{[2^3]}=(q^{1/2}b_1,q^{-1/2}b_1,q^{1/2}b_2,q^{-1/2}b_2,q^{1/2}b_3,q^{-1/2}b_3)$ ($b_1b_2b_3=1$) and\\
$c_{[3^2]}=(qc,c,q^{-1}c,qc^{-1},c^{-1},q^{-1}c^{-1})$.
The flavor Wilson lines are given by
\begin{align}
W_{\wp(4,1,1)}^{A_5\; \rm ex2}&=[2]_q \left( a^{-1}+a^{1/2}\chi^{SU(4)}_{\rmbf{\ovl{4}}}(a')\right)\chi^{SU(3)}_{\rmbf{3}}(b)+[2]_qa +\chi^{SU(3)}_{\rmbf{\ovl{3}}}(b)\chi^{SU(2)}_{\rmbf{2}}(c)+a^{-1/2}\chi^{SU(4)}_{\rmbf{4}}(a') \\
\nonumber W_{\wp(3,2,1)}^{A_5\; \rm ex2}&=[2]_q^2\chi^{SU(2)}_{\rmbf{2}}(c)+
[2]_q \left( c\chi^{SU(3)}_{\rmbf{3}}(b)+a^{-1}\chi^{SU(3)}_{\rmbf{\ovl{3}}}(b) \right)+a^{-1/2}\chi^{SU(4)}_{\rmbf{4}}(a')\chi^{SU(3)}_{\rmbf{3}}(b) \\
& +a^{1/2}\chi^{SU(4)}_{\rmbf{\ovl{4}}}(a')\chi^{SU(3)}_{\rmbf{\ovl{3}}}(b)+\chi^{SU(3)}_{\rmbf{8}}(b)\chi^{SU(2)}_{\rmbf{2}}(c) \\
\nonumber W_{\wp(3,1,2)}^{A_5\; \rm ex2}&=2[2]_q+a^{1/2}\chi^{SU(4)}_{\rmbf{4}}(a')\chi^{SU(3)}_{\rmbf{\ovl{3}}}(b)+\chi^{SU(4)}_{\rmbf{6}}(a')\chi^{SU(2)}_{\rmbf{2}}(c)+a^{-3/2}\chi^{SU(4)}_{\rmbf{4}}(a') \\
& \!\!\!\! \!\!\!\! \!\!\!\! \!\!\!\! +a^{-1}\chi^{SU(3)}_{\rmbf{\ovl{3}}}(b)\chi^{SU(2)}_{\rmbf{2}}(c)+\left(a^{3/2}+a^{-1/2}\chi^{SU(3)}_{\rmbf{3}}(b)\right)\chi^{SU(4)}_{\rmbf{\ovl{4}}}(a')+a \chi^{SU(3)}_{\rmbf{3}}(b)\chi^{SU(2)}_{\rmbf{2}}(c) \\
\nonumber W_{\wp(2,3,1)}^{A_5\; \rm ex2}&=[2]_q^2 \chi^{SU(3)}_{\rmbf{3}}(b)\chi^{SU(2)}_{\rmbf{2}}(c)+[2]_q \left( a\chi^{SU(3)}_{\rmbf{\ovl{3}}}(b)+a^{-1} \right) \\
& +a^{-1/2}\chi^{SU(4)}_{\rmbf{4}}(a')\chi^{SU(3)}_{\rmbf{\ovl{3}}}(b)+\chi^{SU(3)}_{\rmbf{\ovl{6}}}(b)\chi^{SU(2)}_{\rmbf{2}}(c)+a^{1/2}\chi^{SU(4)}_{\rmbf{\ovl{4}}}(a') \\
\nonumber W_{\wp(2,2,2)}^{A_5\; \rm ex2}&=[2]_q^2 \chi^{SU(3)}_{\rmbf{3}}(b)+[2]_q\left( a^{1/2}\chi^{SU(4)}_{\rmbf{4}}(a')+a^{-1}\chi^{SU(2)}_{\rmbf{2}}(c)+a\chi^{SU(3)}_{\rmbf{\ovl{3}}}(b)\chi^{SU(2)}_{\rmbf{2}}(c) \right) \\
\nonumber & +\chi^{SU(3)}_{\rmbf{\ovl{6}}}(b)+\chi^{SU(3)}_{\rmbf{3}}(b)\chi^{SU(2)}_{\rmbf{3}}(c)+a^{-1/2}\chi^{SU(4)}_{\rmbf{4}}(a')\chi^{SU(3)}_{\rmbf{\ovl{3}}}(b)\chi^{SU(2)}_{\rmbf{2}}(c)\\ 
& +a^{1/2}\chi^{SU(4)}_{\rmbf{\ovl{4}}}(a')\chi^{SU(2)}_{\rmbf{2}}(c)+a^{2}+a^{-1}\chi^{SU(4)}_{\rmbf{6}}(a') \\
\nonumber W_{\wp(2,1,3)}^{A_5\; \rm ex2}&=[2]_q\left( a\chi^{SU(3)}_{\rmbf{\ovl{3}}}(b)+a^{1/2}\chi^{SU(4)}_{\rmbf{4}}(a')\chi^{SU(2)}_{\rmbf{2}}(c)+a^{-1}\right)+a^{-1/2}\chi^{SU(4)}_{\rmbf{4}}(a')\chi^{SU(3)}_{\rmbf{\ovl{3}}}(b) \\
& +\chi^{SU(3)}_{\rmbf{3}}(b)\chi^{SU(2)}_{\rmbf{2}}(c)+a^{-1}\chi^{SU(4)}_{\rmbf{6}}(a')\chi^{SU(2)}_{\rmbf{2}}(c)+a^{2}\chi^{SU(2)}_{\rmbf{2}}(c)+a^{1/2}\chi^{SU(4)}_{\rmbf{\ovl{4}}}(a')
\\
W_{\wp(1,4,1)}^{A_5\; \rm ex2}&=[2]_q^2\chi^{SU(3)}_{\rmbf{\ovl{3}}}(b)\chi^{SU(2)}_{\rmbf{2}}(c)+[2]_q a+a^{-1/2}\chi^{SU(4)}_{\rmbf{4}}(a') \\
W_{\wp(1,3,2)}^{A_5\; \rm ex2}&=[2]_q\chi^{SU(3)}_{\rmbf{\ovl{3}}}(b)\left([3]_q+\chi^{SU(2)}_{\rmbf{3}}(c)\right)+[2]_q\left([2]_qa+a^{-1/2}\chi^{SU(4)}_{\rmbf{4}}(a')\right) \chi^{SU(2)}_{\rmbf{2}}(c) \\
\nonumber W_{\wp(1,2,3)}^{A_5\; \rm ex2}&=[3]_q[2]_qa+[3]_q \left( a^{-1/2}\chi^{SU(4)}_{\rmbf{4}}(a')+\chi^{SU(3)}_{\rmbf{\ovl{3}}}(b)\chi^{SU(2)}_{\rmbf{2}}(c) \right) +[2]_qa\chi^{SU(2)}_{\rmbf{3}}(c) \\
& +a^{-1/2}\chi^{SU(4)}_{\rmbf{4}}(a')\chi^{SU(2)}_{\rmbf{3}}(c)+\chi^{SU(3)}_{\rmbf{\ovl{3}}}(b)\chi^{SU(2)}_{\rmbf{2}}(c) \\
W_{\wp(1,1,4)}^{A_5\; \rm ex2}&=[3]_q a\chi^{SU(2)}_{\rmbf{2}}(c)+[2]_q \left( a^{-1/2}\chi^{SU(4)}_{\rmbf{4}}(a')\chi^{SU(2)}_{\rmbf{2}}(c)+\chi^{SU(3)}_{\rmbf{\ovl{3}}}(b)\right)+
a\chi^{SU(2)}_{\rmbf{2}}(c).
\label{eq:flavor Wilson for A5 ex2-type free hypers}
\end{align}

\subsection{Rank 1 SCFTs}
\label{subsec:Rank 1 SCFTs}

In this section, we focus on the special rank 1 SCFTs with enhanced global symmetries, namely, $E_{6,7,8}$ symmetries.

\subsubsection{Rank 1 $E_6$ SCFT case}
\label{subsubsec:Rank 1 E_6 SCFT case}

The first result is already discussed in \cite{TW1504} in a direct way.
There is only one type elementary pants network specified by $(1,1,1)$.
The Schur index with the line defect corresponding to the $\wp(1,1,1)$ is given by
\begin{align}
\nonumber & \calI^{\rm Schur}_{\tiny \mbox{rank 1 $E_6$ SCFT w/. } \wp(1,1,1)}(a,b,c)
=
q^{1/2} \chi_{\rmbf{27}}^{E_6}(a,b,c)
+
q^{3/2} \chi_{\rmbf{1728}}^{E_6}(a,b,c)\\
\nonumber & \qqqquad+
q^{5/2} \left( \chi_{\rmbf{46332}}^{E_6}(a,b,c)+\chi_{\rmbf{1728}}^{E_6}(a,b,c)+\chi_{\rmbf{351}}^{E_6}(a,b,c)+\chi_{\rmbf{27}}^{E_6}(a,b,c) \right)\\
\nonumber & \qqqquad+
q^{7/2} \left(
\chi_{\rmbf{741312}}^{E_6}(a,b,c)+\chi_{\rmbf{51975}}^{E_6}(a,b,c)+\chi_{\rmbf{46332}}^{E_6}(a,b,c)+\chi_{\rmbf{17550}}^{E_6}(a,b,c) \right. \\
& \left. \qqqquad+2\chi_{\rmbf{1728}}^{E_6}(a,b,c)+\chi_{\rmbf{351}}^{E_6}(a,b,c)+2\chi_{\rmbf{27}}^{E_6}(a,b,c) \right)+\calO(q^{9/2}).
\end{align}
Here all the irreducible representations appearing in characters are non-trivially charged under its center group $\bbZ_3$.

Notice that the Schur indices in the absence of defects up to $q^3$ order is given as
\begin{align}
\nonumber \calI^{\rm Schur}_{\tiny \mbox{rank 1 $E_6$ SCFT}}&(a,b,c)=1+q\chi_{\rmbf{78}}^{E_6}(a,b,c)+q^2 \left( \chi_{\rmbf{2430}}^{E_6}(a,b,c)+\chi_{\rmbf{78}}^{E_6}(a,b,c)+1 \right)\\
&+
q^{3} \left(
\chi_{\rmbf{43758}}^{E_6}(a,b,c)+\chi_{\rmbf{2925}}^{E_6}(a,b,c)
+\chi_{\rmbf{2430}}^{E_6}(a,b,c)+2\chi_{\rmbf{78}}^{E_6}(a,b,c)+1
\right)
+\calO(q^4).
\end{align}

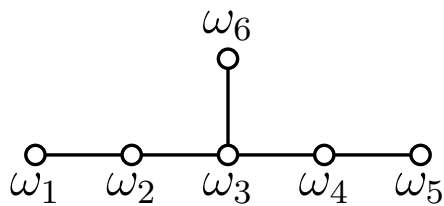
\begin{figure}[thb]
\centering
\scalebox{1.6}{\tikzsetnextfilename{-figure-computation-E6}
\begin{tikzpicture}
\def\l{0.8}
\def\r{0.08}

\coordinate (N3) at (0,0);
\coordinate (N2) at ($(N3)+(-\l,0)$);
\coordinate (N1) at ($(N2)+(-\l,0)$);
\coordinate (N4) at ($(N3)+(\l,0)$);
\coordinate (N5) at ($(N4)+(\l,0)$);
\coordinate (N6) at ($(N3)+(0,\l)$);

\draw[thick] (N1) -- (N2) -- (N3) -- (N4) -- (N5);
\draw[thick] (N3) -- (N6);

\draw[fill=white,thick] (N1) circle (\r) node[below]{$\omega_1$};
\draw[fill=white,thick] (N2) circle (\r) node[below]{$\omega_2$};	
\draw[fill=white,thick] (N3) circle (\r) node[below]{$\omega_3$};
\draw[fill=white,thick] (N4) circle (\r) node[below]{$\omega_4$};
\draw[fill=white,thick] (N5) circle (\r) node[below]{$\omega_5$};
\draw[fill=white,thick] (N6) circle (\r) node[above]{$\omega_6$};
\end{tikzpicture}}
\caption{$E_6$ Dynkin diagram. $\omega_i$ gives $i \mod 3$ charge under the center group $\bbZ_3$. See \cite{LieART} and its \emph{Mathematica} package.}
\end{figure}

\begin{table}[thb]
\begin{tabular}{|c|cccccc|}
\hline
dimension & $\rmbf{27}$ & $\rmbf{351}$ & $\rmbf{1728}$ & $\rmbf{17550}$ & $\rmbf{46332}$ & $\rmbf{51975}$
\\
Dynkin labels & (100000) & (000100) & (100001) & (000101) & (100002) & (101000) \\
\hline
dimension & $\rmbf{741312}$ & $\rmbf{78}$ & $\rmbf{2430}$ & $\rmbf{2925}$ & $\rmbf{43758}$ & \\
Dynkin labels & (100003) & (000001) & (000002) & (001000) & (000003) & \\
\hline
\end{tabular}
\caption{The dimensions of irreducible representations and their Dynkin labels in $E_6$.}
\end{table}

\subsubsection{Rank 1 $E_7$ SCFT case}
\label{subsubsec:Rank 1 E_7 SCFT case}

In this theory, we find that the $(2,1,1)$-type pants network gives the non-trivial 4D line operator respecting the $E_7$-symmetry.
We computed this case in \cite{Watanabe1603} at the leading order of $q$.
Indeed, the computation up to $q^{7/2}$ gives
\!\footnote{We do not exactly check the terms at $q^{7/2}$ but only see the match of the values at the trivial fugacities $a=b=c=1$. Notice also that there is the common structure between the previous $E_6$ SCFT and this $E_7$ SCFT on replacing the fundamental weights $\omega_6^{E_6},\omega_3^{E_6},\omega_1^{E_6},\omega_4^{E_6}$ by $\omega_1^{E_7},\omega_2^{E_7},\omega_6^{E_7},\omega_7^{E_7}$ respectively. The similar structure also appears in the $SU(2)$ $N_f=4$ SQCD, namely, the $SO(8)$ SCFT with or without the fundamental Wilson line, for example. However, it seems to be not true in the $E_8$ SCFT as we see later.}
\begin{align}
\nonumber & \calI^{\rm Schur}_{\tiny \mbox{rank 1 $E_7$ SCFT w/. } \wp(2,1,1)}(a,b,c)
=
q^{1/2} \chi_{\rmbf{56}}^{E_7}(a,b,c)
+
q^{3/2} \chi_{\rmbf{6480}}^{E_7}(a,b,c)\\
\nonumber & \qqqquad+
q^{5/2} \left( \chi_{\rmbf{320112}}^{E_7}(a,b,c)+\chi_{\rmbf{6480}}^{E_7}(a,b,c)+\chi_{\rmbf{912}}^{E_7}(a,b,c)+\chi_{\rmbf{56}}^{E_7}(a,b,c) \right) \\
\nonumber & \qqqquad+q^{7/2} \left(
\chi_{\rmbf{9405760}}^{E_7}(a,b,c)+\chi_{\rmbf{362880}}^{E_7}(a,b,c)+\chi_{\rmbf{320112}}^{E_7}(a,b,c)+\chi_{\rmbf{86184}}^{E_7}(a,b,c)\right. \\
& \left.\qqqquad+2\chi_{\rmbf{6480}}^{E_7}(a,b,c)+\chi_{\rmbf{912}}^{E_7}(a,b,c)+2\chi_{\rmbf{56}}^{E_7}(a,b,c) \right)+\calO(q^{9/2}).
% % q^{5/2} term is conjectural but check the dimension i.e. evaluated at a=b=c=1
\end{align}
Notice that all the irreducible representations appearing in characters are non-trivially charged under its center group $\bbZ_2$.

The no defect Schur index of this SCFT is given by
\begin{align}
\nonumber & \calI^{\rm Schur}_{\tiny \mbox{rank 1 $E_7$ SCFT}}(a,b,c)=1+q\chi_{\rmbf{133}}^{E_7}(a,b,c)+q^2 \left( \chi_{\rmbf{7371}}^{E_7}(a,b,c)+\chi_{\rmbf{133}}^{E_7}(a,b,c)+1 \right) \\
& \qqqquad+q^{3} \left(\chi_{\rmbf{238602}}^{E_7}(a,b,c)+\chi_{\rmbf{8645}}^{E_7}(a,b,c)+\chi_{\rmbf{7371}}^{E_7}(a,b,c)+2\chi_{\rmbf{133}}^{E_7}(a,b,c)+1\right)
+\calO(q^4).
\end{align}

%$q^{5/2}$ term for both E_6 and E_7
%\begin{align}
%\chi_{\lambda+2\theta}+\chi_{\lambda} \chi_{Adj}
%\end{align}

The other two pants networks are just flavor Wilson lines, that is to say, factorized into the Schur index without any networks and the following factors.
\begin{align}
\nonumber & \wp(1,2,1) : \chi_{\rmbf{\ovl{4}}}(b)\chi^{SU(2)}_{\rmbf{2}}(c)+\chi_{\rmbf{4}}(a) \\
\nonumber & \wp(1,1,2) : \chi_{\rmbf{4}}(a)\chi^{SU(2)}_{\rmbf{2}}(c)+\chi_{\rmbf{\ovl{4}}}(b)
\label{eq:flavor Wilson for rank 1 E7 SCFT}
\end{align}

\begin{figure}[thb]
\centering
\scalebox{1.6}{\tikzsetnextfilename{-figure-computation-E7}
\begin{tikzpicture}
\def\l{0.8}
\def\r{0.08}

\coordinate (N3) at (0,0);
\coordinate (N2) at ($(N3)+(-\l,0)$);
\coordinate (N1) at ($(N2)+(-\l,0)$);
\coordinate (N4) at ($(N3)+(\l,0)$);
\coordinate (N5) at ($(N4)+(\l,0)$);
\coordinate (N6) at ($(N5)+(\l,0)$);
\coordinate (N7) at ($(N3)+(0,\l)$);

\draw[thick] (N1) -- (N2) -- (N3) -- (N4) -- (N5) -- (N6);
\draw[thick] (N3) -- (N7);

\draw[fill=white,thick] (N1) circle (\r) node[below]{$\omega_1$};
\draw[fill=white,thick] (N2) circle (\r) node[below]{$\omega_2$};	
\draw[fill=white,thick] (N3) circle (\r) node[below]{$\omega_3$};
\draw[fill=white,thick] (N4) circle (\r) node[below]{$\omega_4$};
\draw[fill=white,thick] (N5) circle (\r) node[below]{$\omega_5$};
\draw[fill=white,thick] (N6) circle (\r) node[below]{$\omega_6$};
\draw[fill=white,thick] (N7) circle (\r) node[above]{$\omega_7$};
\end{tikzpicture}}
\caption{$E_7$ Dynkin diagram. $\omega_{4,6,7}$ gives non-trivial charge under the center group $\bbZ_2$.}
\end{figure}
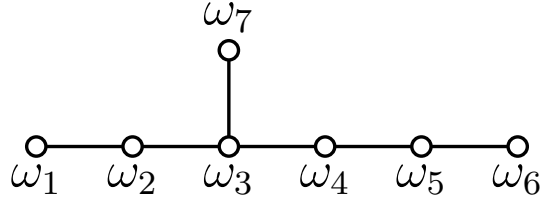

\begin{table}[thb]
\begin{tabular}{|c|cccccc|}
\hline
dimension & $\rmbf{56}$ & $\rmbf{912}$ & $\rmbf{6480}$ & $\rmbf{86184}$ & $\rmbf{320112}$ & $\rmbf{362880}$ \\
Dynkin labels & (0000010) & (0000001) & (1000010) & (1000001) & (2000010) & (0100010) \\
\hline
dimension & $\rmbf{9405760}$ & $\rmbf{133}$ & $\rmbf{7371}$ & $\rmbf{8645}$ & $\rmbf{238602}$ & \\
Dynkin labels & (3000010) & (1000000) & (2000000) & (0100000) & (3000000) & \\
\hline
\end{tabular}
\caption{The dimensions of irreducible representations and their Dynkin labels in $E_7$.}
\end{table}

\subsubsection{Rank 1 $E_8$ SCFT case}
\label{subsubsec:Rank 1 E_8 SCFT case}

In this SCFT, only the $(3,1,2)$-type pants network is the non-trivial network respecting the $E_8$-symmetry. The result is following.
\begin{align}
\nonumber & \calI^{\rm Schur}_{\tiny \mbox{rank 1 $E_8$ SCFT w/. } \wp(3,1,2)}(a,b,c)-[2]_q \calI^{\rm Schur}_{\tiny \mbox{rank 1 $E_8$ SCFT}}(a,b,c)
=
q^{1/2} \left(\chi_{\rmbf{248}}^{E_8}(a,b,c)+1\right) \\
\nonumber &\qqqquad+
q^{3/2} \left(\chi_{\rmbf{30380}}^{E_8}(a,b,c)+\chi_{\rmbf{27000}}^{E_8}(a,b,c)+\chi_{\rmbf{248}}^{E_8}(a,b,c)+1 \right) \\
\nonumber &\qqqquad +
q^{5/2} \left( \chi_{\rmbf{4096000}}^{E_8}(a,b,c)+\chi_{\rmbf{1763125}}^{E_8}(a,b,c)+\chi_{\rmbf{30380}}^{E_8}(a,b,c)+2\chi_{\rmbf{27000}}^{E_8}(a,b,c) \right. \\
& \qqqquad\qquad \left.+\chi_{\rmbf{3875}}^{E_8}(a,b,c)+3\chi_{\rmbf{248}}^{E_8}(a,b,c)+1 \right)
+\calO(q^{7/2})
\end{align}
where, to express this as possible as simply, we have subtracted the Schur index without the defects expressed as
\begin{align}
\nonumber & \calI^{\rm Schur}_{\tiny \mbox{rank 1 $E_8$ SCFT}}(a,b,c)
=
1+
q \chi_{\rmbf{248}}^{E_8}(a,b,c)
+
q^{2} \left( \chi_{\rmbf{27000}}^{E_8}(a,b,c)+\chi_{\rmbf{248}}^{E_8}(a,b,c)+1 \right) \\
& \qqqquad+
q^{3} \left( \chi_{\rmbf{1763125}}^{E_8}(a,b,c)+\chi_{\rmbf{30380}}^{E_8}(a,b,c)+\chi_{\rmbf{27000}}^{E_8}(a,b,c)+2\chi_{\rmbf{248}}^{E_8}(a,b,c)+1 \right)
+\calO(q^{4}).
\end{align}
Notice that in the presence of $(3,1,2)$-type pants network, the lowest exponent of $q$ is $-\frac{1}{2}$. The other pants networks are written as the flavor Wilson lines as follows :
\begin{align}
\nonumber & \wp(1,1,4) : [2]_q\chi_{\rmbf{6}}(a)\chi^{SU(2)}_{\rmbf{2}}(c)+[2]_q\chi^{SU(3)}_{\rmbf{\ovl{3}}}(b) \\
\nonumber & \wp(1,2,3) : [3]_q\chi_{\rmbf{6}}(a)+\chi_{\rmbf{6}}(a)\chi^{SU(2)}_{\rmbf{3}}(c)+([2]_q)^2\chi^{SU(3)}_{\rmbf{\ovl{3}}}(b)\chi^{SU(2)}_{\rmbf{2}}(c) \\
\nonumber & \wp(1,3,2) : [3]_q[2]_q\chi^{SU(3)}_{\rmbf{\ovl{3}}}(b)+[2]_q\chi^{SU(3)}_{\rmbf{\ovl{3}}}(b)\chi^{SU(2)}_{\rmbf{3}}(c)+[2]_q\chi_{\rmbf{6}}(a)\chi^{SU(2)}_{\rmbf{2}}(c) \\
\nonumber & \wp(1,4,1) :  ([2]_q)^2\chi^{SU(3)}_{\rmbf{\ovl{3}}}(b)\chi^{SU(2)}_{\rmbf{2}}(c)+\chi_{\rmbf{6}}(a) \\
\nonumber & \wp(2,1,3) : \chi_{\rmbf{15}}(a)\chi^{SU(2)}_{\rmbf{2}}(c)+\chi_{\rmbf{6}}(a)\chi^{SU(3)}_{\rmbf{\ovl{3}}}(b)+\chi_{\rmbf{\ovl{6}}}(a)+\chi^{SU(3)}_{\rmbf{3}}(b)\chi^{SU(2)}_{\rmbf{2}}(c) \\
\nonumber & \wp(2,2,2) : ([2]_q)^2\chi^{SU(3)}_{\rmbf{3}}(b)+\chi_{\rmbf{15}}(a)+\chi_{\rmbf{6}}(a)\chi^{SU(3)}_{\rmbf{\ovl{3}}}(b)\chi^{SU(2)}_{\rmbf{2}}(c)+\chi_{\rmbf{\ovl{6}}}(a)\chi^{SU(2)}_{\rmbf{2}}(c)\\
\nonumber & \qquad\qquad +\chi^{SU(3)}_{\rmbf{3}}(b)\chi^{SU(2)}_{\rmbf{3}}(c)+\chi^{SU(3)}_{\rmbf{\ovl{6}}}(b) \\
\nonumber & \wp(2,3,1) : ([2]_q)^2\chi^{SU(3)}_{\rmbf{3}}(b)\chi^{SU(2)}_{\rmbf{2}}(c)+\chi_{\rmbf{6}}(a)\chi^{SU(3)}_{\rmbf{\ovl{3}}}(b)+\chi_{\rmbf{\ovl{6}}}(a)+\chi^{SU(3)}_{\rmbf{\ovl{6}}}(b)\chi^{SU(2)}_{\rmbf{2}}(c) \\
\nonumber & \wp(3,2,1) : ([2]_q)^2\chi^{SU(2)}_{\rmbf{2}}(c)+\chi_{\rmbf{6}}(a)\chi^{SU(3)}_{\rmbf{3}}(b)+\chi_{\rmbf{\ovl{6}}}(a)\chi^{SU(3)}_{\rmbf{\ovl{3}}}(b)+\chi^{SU(3)}_{\rmbf{8}}(b)\chi^{SU(2)}_{\rmbf{2}}(c) \\
\nonumber & \wp(4,1,1) : \chi_{\rmbf{6}}(a)+\chi_{\rmbf{\ovl{6}}}(a)\chi^{SU(3)}_{\rmbf{3}}(b)+\chi^{SU(3)}_{\rmbf{\ovl{3}}}(b)\chi^{SU(2)}_{\rmbf{2}}(c)
\label{eq:flavor Wilson for rank 1 E8 SCFT}
\end{align}

\begin{figure}[thb]
\centering
\scalebox{1.6}{\tikzsetnextfilename{-figure-computation-E8}
\begin{tikzpicture}
\def\l{0.8}
\def\r{0.08}

\coordinate (N3) at (0,0);
\coordinate (N2) at ($(N3)+(-\l,0)$);
\coordinate (N1) at ($(N2)+(-\l,0)$);
\coordinate (N4) at ($(N3)+(\l,0)$);
\coordinate (N5) at ($(N4)+(\l,0)$);
\coordinate (N6) at ($(N5)+(\l,0)$);
\coordinate (N7) at ($(N6)+(\l,0)$);
\coordinate (N8) at ($(N3)+(0,\l)$);

\draw[thick] (N1) -- (N2) -- (N3) -- (N4) -- (N5) -- (N6) -- (N7);
\draw[thick] (N3) -- (N8);

\draw[fill=white,thick] (N1) circle (\r) node[below]{$\omega_1$};
\draw[fill=white,thick] (N2) circle (\r) node[below]{$\omega_2$};	
\draw[fill=white,thick] (N3) circle (\r) node[below]{$\omega_3$};
\draw[fill=white,thick] (N4) circle (\r) node[below]{$\omega_4$};
\draw[fill=white,thick] (N5) circle (\r) node[below]{$\omega_5$};
\draw[fill=white,thick] (N6) circle (\r) node[below]{$\omega_6$};
\draw[fill=white,thick] (N7) circle (\r) node[below]{$\omega_7$};
\draw[fill=white,thick] (N8) circle (\r) node[above]{$\omega_8$};
\end{tikzpicture}}
\caption{$E_8$ Dynkin diagram. The center group is trivial.}
\end{figure}
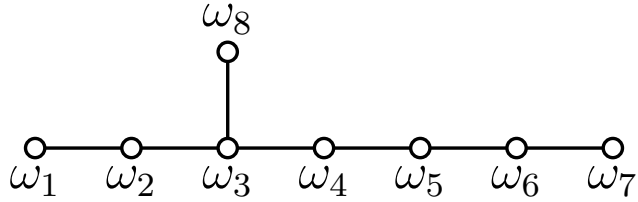

\begin{table}[h]
\centering
\begin{tabular}{|c|cccccc|}
\hline
dimension & $\rmbf{248}$ & $\rmbf{3875}$ & $\rmbf{27000}$ & $\rmbf{30380}$ & $\rmbf{1763125}$ & $\rmbf{4096000}$
\\
Dynkin labels & \!(00000010)\!\! & \!\!(10000000)\!\! & \!\!(00000020)\!\! & \!\!(00000100)\!\! & \!\!(00000030)\!\! & \!\!(000000110)\! \\
\hline
\end{tabular}
\caption{The dimensions of irreducible representations and their Dynkin labels in $E_8$.}
\end{table}

In each case of the rank 1 SCFTs, it is interesting problem to rewrite them in terms of the affine characters in the admissible representations of $(\mathfrak{e}_6)_{k=-3},(\mathfrak{e}_7)_{k=-4}$ and $(\mathfrak{e}_8)_{k=-6}$ as conjectured in \cite{BLLPRR13,CGS1606}.
The other non-trivial checks are the ``IR computations" discussed in \cite{CordovaShao1506,CGS1606} by using the techniques and results on the spectral networks in the Minahan-Nemeschansky theory in \cite{GMN5,HollandsNeitzke1607} and on the quantum traces in \cite{CGT1505,Gabella1603}.

\subsection{Superconformal QCDs}

\subsubsection*{$SU(2)$ $N_f=4$ SQCD}

\begin{figure}[th]
\centering
\begin{tabular}{rcl}
\begin{minipage}{0.3\hsize}
\includegraphics[width=35mm]{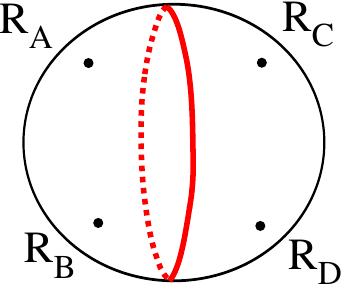}
\end{minipage}
&
\begin{minipage}{0.3\hsize}
\includegraphics[width=35mm]{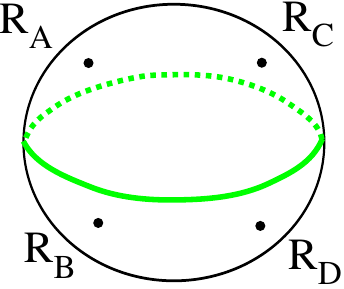}
\end{minipage}
&
\begin{minipage}{0.3\hsize}
\includegraphics[width=35mm]{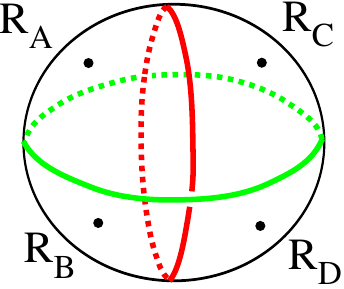}
\end{minipage}
\end{tabular}
\caption{For the $A_1$ or $SU(2)$ SCQCD cases obtained by the four punctured sphere, we show a 4D Wilson loop in $\Box=R(\omega_1)=\rmbf{2}$ (Left), a 4D 't Hooft loop labelled by $\Box=R(\omega_1)=\rmbf{2}$ (Middle) and both Wilson and 't Hooft loops (Right).}
\label{fig:loops in SQCD for A1}
\end{figure}

Here we introduce the following symbol.
\begin{align}
\chi(R_A,R_B,R_C,R_D)=\chi^{SU(2)}_{R_A}(a)\chi^{SU(2)}_{R_B}(b)\chi^{SU(2)}_{R_C}(c)\chi^{SU(2)}_{R_D}(d).
\end{align}

We also written down the Schur index expression of $SU(2)$ $N_f=4$ SQCD for comparison.
\begin{align}
\nonumber \calI^{\rm Schur}_{\tiny SU(2)\;N_f=4 \mbox{ SQCD}}(a,b,c,d)&=1+q\chi^{SO(8)}_{\rmbf{28}}+q^2\left[ \chi^{SO(8)}_{\rmbf{300}}+\chi^{SO(8)}_{\rmbf{28}}+1 \right] \\
& \quad +q^3\left[ \chi^{SO(8)}_{\rmbf{1925}}+\chi^{SO(8)}_{\rmbf{350}}+\chi^{SO(8)}_{\rmbf{300}}+2\chi^{SO(8)}_{\rmbf{28}}+1 \right]
+\calO(q^4).
\end{align}
In this system described by the Lagrangian, $\chi^{SO(8)}_{\rmbf{28}}$ corresponds to the meson operator $M_{AB}=\epsilon^{ab}Q_{aA}Q_{bB}$ ($a,b$ : $SU(2)$ gauge indices, $A,B$ : $SO(8)$ indices) with $\Delta=2$ and $I_3=1$ for example.

When we add the Wilson line $W_{\Box}$, the Schur index computed from the $q$-deformed Yang-Mills formula is given by
\begin{align}
\nonumber & \calI^{\rm Schur}_{\tiny SU(2)\;N_f=4 \mbox{ SQCD w/. } W_{\Box}}(a,b,c,d)= q^{1/2} \left[
\chi(\rmbf{2},\rmbf{2},\rmbf{1},\rmbf{1})
+
\chi(\rmbf{1},\rmbf{1},\rmbf{2},\rmbf{2})
\right]+q^{3/2}
\left[
\chi(\rmbf{2},\rmbf{2},\rmbf{1},\rmbf{1})\right. \\
\nonumber & \quad
+
\chi(\rmbf{1},\rmbf{1},\rmbf{2},\rmbf{2})
+
\chi(\rmbf{2},\rmbf{2},\rmbf{3},\rmbf{3})
+
\chi(\rmbf{3},\rmbf{3},\rmbf{2},\rmbf{2})
+
\chi(\rmbf{3},\rmbf{1},\rmbf{2},\rmbf{2})
+
\chi(\rmbf{1},\rmbf{3},\rmbf{2},\rmbf{2})
+
\chi(\rmbf{2},\rmbf{2},\rmbf{3},\rmbf{1})
\\
\nonumber & \quad +\left.
\chi(\rmbf{2},\rmbf{2},\rmbf{1},\rmbf{3})
+
\chi(\rmbf{4},\rmbf{2},\rmbf{1},\rmbf{1})
+
\chi(\rmbf{2},\rmbf{4},\rmbf{1},\rmbf{1})
+
\chi(\rmbf{1},\rmbf{1},\rmbf{4},\rmbf{2})
+
\chi(\rmbf{1},\rmbf{1},\rmbf{2},\rmbf{4}) \right]
+
\calO(q^{5/2}) \\
&\quad=
q^{1/2} \chi_{\rmbf{8}_v}(a,b,c,d)%weight=[1,0,0,0]
+
q^{3/2} \chi_{\rmbf{160}_v}(a,b,c,d)%weight=[1,1,0,0]
+
\calO(q^{5/2})
\end{align}
which matches with the known result in \cite{CGS1606} for example.
\!\footnote{Of course, this is obviously true once the expression of the free bi-fundamental hypermultiplets is given by the correlator of the 2D $q$-deformed Yang-Mills theory. The insertion of the corresponding character and the integration over the gauge group are same operations.}

Next, let us see the dual 't Hooft loop operators.
Although its field theoretical definition is not known yet, from the geometrical point of view, this acts on the above Wilson line expression simply as the permutation of two $[1^2]$-type punctures.
In particular, this permutation is equivalent to the triality action of $SO(8)$, that is to say, the exchange of the simple roots $\alpha_1$ and $\alpha_4$. Therefore, its Schur index expression is
\begin{align}
\nonumber \calI^{\rm Schur}_{\tiny SU(2)\;N_f=4 \mbox{ SQCD w/. } T^{\rm dual}_{\Box}}(a,b,c,d)&=
q^{1/2} \chi_{\rmbf{8}_s}
+
q^{3/2} \chi_{\rmbf{160}_s} \\
&+
q^{5/2} \left[ \chi_{\rmbf{8}_s}+\chi_{\rmbf{56}_s}+\chi_{\rmbf{160}_s}+\chi_{\rmbf{1400}_s} \right]
+
\calO(q^{7/2}).
\end{align}

According to \cite{CGS1606} for example, we can interpret the Schur index as the count of local operators playing the role of line changing operators between the fundamental Wilson line and this 't Hooft line.

Finally, let us evaluate the two loops coexisting case on the right side in Fig.~\ref{fig:loops in SQCD for A1}.
By representing the four punctured sphere as the disk by removing one point in the sphere, we can geometrically compute the OPE of the Wilson loop and the dual 't Hooft loop with the crossing resolutions (see \cite{AGGTV09,TW1504} for example) as follows:
\begin{align}
\nonumber
\raisebox{-0.47\height}{\tikzsetnextfilename{-figure-computation-su2SQCD_exps1}
\begin{tikzpicture}
\def\ra{0.5}
\def\rb{0.35}
\def\rx{1.3}
\def\ry{1.0}
\coordinate (O) at (0,0);
\coordinate (A) at (-\ra,\rb);
\coordinate (B) at (-\ra,-\rb);
\coordinate (C) at (\ra,-\rb);
\coordinate (D) at (\ra,\rb);
\puncture[above left]{(A)}{\scriptsize A}
\puncture[below left]{(B)}{\scriptsize B}
\puncture[below right]{(C)}{\scriptsize C}
\puncture[above right]{(D)}{\scriptsize D}
\draw[thick] (O) circle [x radius=\rx, y radius=\ry];
\draw[red,ultra thick] ($(O)+(0,-\ry)$) -- ($(O)+(0,\ry)$);
\draw[white,line width=3.5pt] ($(B)!0.5!(C)$) circle [x radius=1.3*\ra, y radius=\rb];
\draw[red,ultra thick] ($(B)!0.5!(C)$) circle [x radius=1.3*\ra, y radius=\rb];
\end{tikzpicture}}
&=
\raisebox{-0.47\height}{\tikzsetnextfilename{-figure-computation-su2SQCD_exps2_1}
\begin{tikzpicture}
\def\ra{0.5}
\def\rb{0.35}
\def\rx{1.3}
\def\ry{1.0}
\def\rc{0.3}
\def\rd{0.3}
\coordinate (O) at (0,0);
\coordinate (A) at (-\ra,\rb);
\coordinate (B) at (-\ra,-\rb);
\coordinate (C) at (\ra,-\rb);
\coordinate (D) at (\ra,\rb);
\coordinate (S) at (0,-\ry);
\coordinate (T) at (0,\ry);
\coordinate (V) at ($(O)+(-80:0.5*\ry)$);
\coordinate (P) at ($(C)+(-70:\rd)$);
\coordinate (Q) at ($(C)+(0:\rd)$);
\coordinate (R) at ($(C)+(70:\rd)$);
\coordinate (U) at ($(O)+(80:\rd)$);
\simplepunc{(A)}
\simplepunc{(B)}
\simplepunc{(C)}
\simplepunc{(D)}
\draw[thick] (O) circle [x radius=\rx, y radius=\ry];
\draw[red,ultra thick] plot [smooth,tension=0.7] coordinates {(S) (V) (P) (Q) (R) (U) (T)};
\draw[red,ultra thick] (B) circle [x radius=\rc, y radius=\rc];
\end{tikzpicture}}
+
\raisebox{-0.47\height}{\tikzsetnextfilename{-figure-computation-su2SQCD_exps2_2}
\begin{tikzpicture}
\def\ra{0.5}
\def\rb{0.35}
\def\rx{1.3}
\def\ry{1.0}
\def\rc{0.3}
\def\rd{0.3}
\coordinate (O) at (0,0);
\coordinate (A) at (-\ra,\rb);
\coordinate (B) at (-\ra,-\rb);
\coordinate (C) at (\ra,-\rb);
\coordinate (D) at (\ra,\rb);
\coordinate (S) at (0,-\ry);
\coordinate (T) at (0,\ry);
\coordinate (V) at ($(O)+(-100:0.5*\ry)$);
\coordinate (P) at ($(B)+(-110:\rd)$);
\coordinate (Q) at ($(B)+(180:\rd)$);
\coordinate (R) at ($(B)+(110:\rd)$);
\coordinate (U) at ($(O)+(100:\rd)$);
\simplepunc{(A)}
\simplepunc{(B)}
\simplepunc{(C)}
\simplepunc{(D)}
\draw[thick] (O) circle [x radius=\rx, y radius=\ry];
\draw[red,ultra thick] plot [smooth,tension=0.7] coordinates {(S) (V) (P) (Q) (R) (U) (T)};
\draw[red,ultra thick] (C) circle [x radius=\rc, y radius=\rc];
\end{tikzpicture}}
\\
& -q^{1/2}
\raisebox{-0.47\height}{\tikzsetnextfilename{-figure-computation-su2SQCD_exps2_3}
\begin{tikzpicture}
\def\ra{0.5}
\def\rb{0.35}
\def\rx{1.3}
\def\ry{1.0}
\def\rc{0.3}
\def\rd{0.6}
\coordinate (O) at (0,0);
\coordinate (A) at (-\ra,\rb);
\coordinate (B) at (-\ra,-\rb);
\coordinate (C) at (\ra,-\rb);
\coordinate (D) at (\ra,\rb);
\coordinate (S) at (0,-\ry);
\coordinate (T) at (0,\ry);
\coordinate (P) at ($(O)+(-100:\rd)$);
\coordinate (Q) at ($(B)+(-100:\rc)$);
\coordinate (R) at ($(B)+(180:\rc)$);
\coordinate (U) at ($(B)+(90:\rc)$);
\coordinate (V) at ($(C)+(-90:\rc)$);
\coordinate (W) at ($(C)+(0:\rc)$);
\coordinate (X) at ($(C)+(100:\rc)$);
\coordinate (Y) at ($(O)+(80:\rd)$);
\simplepunc{(A)}
\simplepunc{(B)}
\simplepunc{(C)}
\simplepunc{(D)}
\draw[thick] (O) circle [x radius=\rx, y radius=\ry];
\draw[red,ultra thick] plot [smooth,tension=1.0] coordinates {(S) (P) (Q) (R) (U) ($(B)!0.5!(C)$) (V) (W) (X) (Y) (T)};
\end{tikzpicture}}
-q^{-1/2}
\raisebox{-0.47\height}{\tikzsetnextfilename{-figure-computation-su2SQCD_exps2_4}
\begin{tikzpicture}
\def\ra{0.5}
\def\rb{0.35}
\def\rx{1.3}
\def\ry{1.0}
\def\rc{0.3}
\def\rd{0.6}
\coordinate (O) at (0,0);
\coordinate (A) at (-\ra,\rb);
\coordinate (B) at (-\ra,-\rb);
\coordinate (C) at (\ra,-\rb);
\coordinate (D) at (\ra,\rb);
\coordinate (S) at (0,-\ry);
\coordinate (T) at (0,\ry);
\coordinate (P) at ($(O)+(-80:\rd)$);
\coordinate (Q) at ($(C)+(-80:\rc)$);
\coordinate (R) at ($(C)+(0:\rc)$);
\coordinate (U) at ($(C)+(90:\rc)$);
\coordinate (V) at ($(B)+(-90:\rc)$);
\coordinate (W) at ($(B)+(-180:\rc)$);
\coordinate (X) at ($(B)+(80:\rc)$);
\coordinate (Y) at ($(O)+(100:\rd)$);
\simplepunc{(A)}
\simplepunc{(B)}
\simplepunc{(C)}
\simplepunc{(D)}
\draw[thick] (O) circle [x radius=\rx, y radius=\ry];
\draw[red,ultra thick] plot [smooth,tension=1.0] coordinates {(S) (P) (Q) (R) (U) ($(B)!0.5!(C)$) (V) (W) (X) (Y) (T)};
\end{tikzpicture}} \\
\nonumber &=
\raisebox{-0.47\height}{\tikzsetnextfilename{-figure-computation-su2SQCD_exps3_1}
\begin{tikzpicture}
\def\ra{0.5}
\def\rb{0.35}
\def\rx{1.3}
\def\ry{1.0}
\def\rc{0.3}
\coordinate (O) at (0,0);
\coordinate (A) at (-\ra,\rb);
\coordinate (B) at (-\ra,-\rb);
\coordinate (C) at (\ra,-\rb);
\coordinate (D) at (\ra,\rb);
\simplepunc{(A)}
\simplepunc{(B)}
\simplepunc{(C)}
\simplepunc{(D)}
\draw[thick] (O) circle [x radius=\rx, y radius=\ry];
\draw[red,ultra thick] (D) circle [x radius=\rc, y radius=\rc];
\draw[red,ultra thick] (B) circle [x radius=\rc, y radius=\rc];
\end{tikzpicture}}
+
\raisebox{-0.47\height}{\tikzsetnextfilename{-figure-computation-su2SQCD_exps3_2}
\begin{tikzpicture}
\def\ra{0.5}
\def\rb{0.35}
\def\rx{1.3}
\def\ry{1.0}
\def\rc{0.3}
\coordinate (O) at (0,0);
\coordinate (A) at (-\ra,\rb);
\coordinate (B) at (-\ra,-\rb);
\coordinate (C) at (\ra,-\rb);
\coordinate (D) at (\ra,\rb);
\simplepunc{(A)}
\simplepunc{(B)}
\simplepunc{(C)}
\simplepunc{(D)}
\draw[thick] (O) circle [x radius=\rx, y radius=\ry];
\draw[red,ultra thick] (A) circle [x radius=\rc, y radius=\rc];
\draw[red,ultra thick] (C) circle [x radius=\rc, y radius=\rc];
\end{tikzpicture}}
\\
& -q^{1/2}
\raisebox{-0.47\height}{\tikzsetnextfilename{-figure-computation-su2SQCD_exps3_3}
\begin{tikzpicture}
\def\ra{0.5}
\def\rb{0.35}
\def\rx{1.3}
\def\ry{1.0}
\def\rc{0.25}
\def\rd{0.35}
\coordinate (O) at (0,0);
\coordinate (A) at (-\ra,\rb);
\coordinate (B) at (-\ra,-\rb);
\coordinate (C) at (\ra,-\rb);
\coordinate (D) at (\ra,\rb);
\simplepunc{(A)}
\simplepunc{(B)}
\simplepunc{(C)}
\simplepunc{(D)}
\draw[thick] (O) circle [x radius=\rx, y radius=\ry];
\draw[red,ultra thick] plot [smooth,tension=0.6] coordinates {($(B)+(-135:\rd)$) ($(B)+(150:\rc)$) ($(D)+(120:\rc)$) ($(D)+(45:\rd)$) ($(D)+(-30:\rc)$) ($(B)+(-60:\rc)$) ($(B)+(-135:\rd)$)};
\end{tikzpicture}}
-q^{-1/2}
\raisebox{-0.47\height}{\tikzsetnextfilename{-figure-computation-su2SQCD_exps3_4}
\begin{tikzpicture}
\def\ra{0.5}
\def\rb{0.35}
\def\rx{1.3}
\def\ry{1.0}
\def\rc{0.25}
\def\rd{0.35}
\coordinate (O) at (0,0);
\coordinate (A) at (-\ra,\rb);
\coordinate (B) at (-\ra,-\rb);
\coordinate (C) at (\ra,-\rb);
\coordinate (D) at (\ra,\rb);
\simplepunc{(A)}
\simplepunc{(B)}
\simplepunc{(C)}
\simplepunc{(D)}
\draw[thick] (O) circle [x radius=\rx, y radius=\ry];
\draw[red,ultra thick] plot [smooth,tension=0.6] coordinates {($(C)+(-45:\rd)$) ($(C)+(-130:\rc)$) ($(A)+(210:\rc)$) ($(A)+(135:\rd)$) ($(A)+(60:\rc)$) ($(C)+(30:\rc)$) ($(C)+(-45:\rd)$)};
\end{tikzpicture}}
\end{align}

The first two terms give
\begin{align}
\chi_{\rmbf{2}}(b)\chi_{\rmbf{2}}(d)
+
\chi_{\rmbf{2}}(a)\chi_{\rmbf{2}}(c)
=
\chi^{SO(8)}_{\rmbf{8}_c}(a,b,c,d)
\end{align}
and the latter two have the same Schur index ( different at the 4D line operators level ) up to the prefactors $q^{\pm 1/2}$. The final expression is
\begin{align}
\nonumber & \calI^{\rm Schur}_{\tiny SU(2)\;N_f=4 \mbox{ SQCD w/. } T^{\rm dual}_{\Box} \circ W_{\Box}}(a,b,c,d)
=
\calI^{\rm Schur}_{\tiny SU(2)\;N_f=4 \mbox{ SQCD w/. } W_{\Box} \circ T^{\rm dual}_{\Box}}(a,b,c,d)
\\
&= \chi_{\rmbf{8}_c}\calI^{\rm Schur}_{\phi}-[2]_q\calI^{\rm Schur}_{WT_{\Box}}=
q\chi_{\rmbf{56}_c}+q^2\left[\chi_{\rmbf{8}_c}+\chi_{\rmbf{840}_c}\right]+\calO(q^3).
\end{align}

Since we have no duality frame where both loops are simultaneously magnetically neutral, it is difficult to interpret this result based on the Lagrangian description.

\subsubsection*{$SU(3)$ $N_f=6$ SCQCD case}

\begin{figure}[th]
\centering
\begin{tabular}{rcl}
\begin{minipage}{0.3\hsize}
\includegraphics[width=35mm]{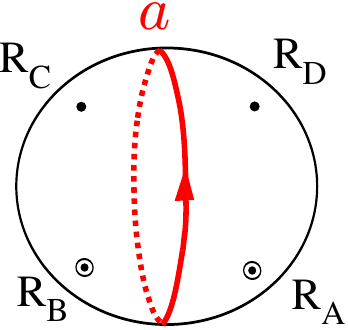}
\end{minipage}
&
\begin{minipage}{0.3\hsize}
\includegraphics[width=35mm]{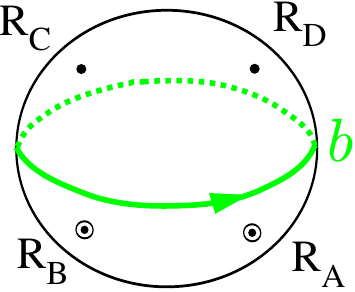}
\end{minipage}
&
\begin{minipage}{0.3\hsize}
\includegraphics[width=35mm]{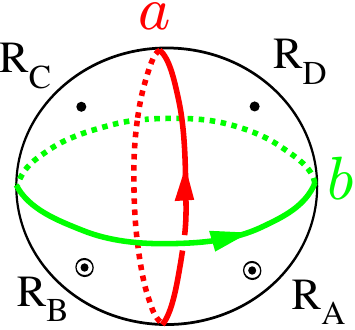}
\end{minipage}
\end{tabular}
\caption{For the higher rank cases, Wilson loop in $\wedge^a \Box$ (Left), A 't Hooft loop labelled by $\wedge^b \Box$ (Middle) and both Wilson and 't Hooft loops (Right). Notice that the ordering of the insertions of two loops is irrelevant in this case.}
\label{fig:loops in SQCD for higher rank}
\end{figure}

The $SU(3)$ $N_f=6$ SQCD is given by $T^\calS[C(2\cdot[1^3],2\cdot[2,1])]$ as shown in Fig.~\ref{fig:loops in SQCD for higher rank}. That is to say, the four punctured sphere has two full punctures ($[1^3]$-type) and two simple punctures ($[21]$-type).

In the absence of line defects,
\begin{align}
\nonumber & \calI^{\rm Schur}_{\tiny SU(3)\; N_f=6\;\mbox{ SQCD }}(a,b,c,d)= 1 + \;q \; (2+c^{-1}d^{-1}\chi_{\rmbf{\ovl{3}}}(a)\chi_{\rmbf{\ovl{3}}}(b)+\chi_{\rmbf{8}}(a)\\ 
\nonumber & \qquad +cd\chi_{\rmbf{3}}(a)\chi_{\rmbf{3}}(b)+\chi_{\rmbf{8}}(b))+ \;q^{3/2} \;(
c^{-2}d\chi_{\rmbf{3}}(a)\chi_{\rmbf{3}}(b)+cd^{-2}\chi_{\rmbf{3}}(a)\chi_{\rmbf{3}}(b)\\
& \qquad +c^{-1}d^{2}\chi_{\rmbf{\ovl{3}}}(a)\chi_{\rmbf{\ovl{3}}}(b)+c^{2}d^{-1}\chi_{\rmbf{\ovl{3}}}(a)\chi_{\rmbf{\ovl{3}}}(b)+c^{-3}+c^{3}+d^{3}+d^{-3})+\calO(q^2) \\
& \qquad =1+q(\chi^{U(6)}_{\Adj}+1)+q^{3/2}(\chi^{U(6)}_{\wedge^3\rmbf{6}}+\chi^{U(6)}_{\wedge^3\rmbf{\ovl{6}}})+\calO(q^2)
\end{align}
where $\chi^{U(6)}_{\rmbf{6}}=\chi^{U(6)}_{[10000]_0}=c\chi_{\rmbf{3}}(b)+d^{-1}\chi_{\rmbf{\ovl{3}}}(a)$ and $\chi^{U(6)}_{\rmbf{\ovl{6}}}=\chi^{U(6)}_{[00001]_{-1}}=d\chi_{\rmbf{3}}(a)+c^{-1}\chi_{\rmbf{\ovl{3}}}(b)$.

The Schur indices with the fundamental Wilson loop $W_{\Box}$ (Left in Fig.~\ref{fig:loops in SQCD for higher rank}) and the dual 't Hooft loop $T^{\rm dual}_{\Box}$ (Right in Fig.~\ref{fig:loops in SQCD for higher rank}) are given as
\begin{align}
\nonumber & \calI^{\rm Schur}_{\tiny SU(3)\; N_f=6\mbox{ SQCD w/. }W_{\Box}}(a,b,c,d)
=
\;q^{1/2}\; \left(
c\chi_{\rmbf{3}}(b)+d^{-1}\chi_{\rmbf{\ovl{3}}}(a)
\right) \\
\nonumber & +\;q\; \left(
c^{-1}d\chi_{\rmbf{3}}(a)\chi_{\rmbf{\ovl{3}}}(b)+c^{-2}\chi_{\rmbf{3}}(b)+d^{2}\chi_{\rmbf{\ovl{3}}}(a)
\right)+\;q^{3/2}\;\left(
2d^{-1}\chi_{\rmbf{\ovl{3}}}(a)\chi_{\rmbf{8}}(b)+c\chi_{\rmbf{15}'}(b) \right. \\
\nonumber & \quad +3c\chi_{\rmbf{3}}(b)+d^{-1}\chi_{\rmbf{\ovl{15}}'}(a)+c^{-1}d^{-2}\chi_{\rmbf{\ovl{6}}}(a)\chi_{\rmbf{\ovl{3}}}(b)+3d^{-1}\chi_{\rmbf{\ovl{3}}}(a)+c^{2}d\chi_{\rmbf{3}}(a)\chi_{\rmbf{\ovl{3}}}(b)\\
& \left.\quad +2c\chi_{\rmbf{8}}(a)\chi_{\rmbf{3}}(b)+c\chi_{\rmbf{\ovl{6}}}(b)+c^{2}d\chi_{\rmbf{3}}(a)\chi_{\rmbf{6}}(b)+d^{-1}\chi_{\rmbf{6}}(a)+c^{-1}d^{-2}\chi_{\rmbf{3}}(a)\chi_{\rmbf{\ovl{3}}}(b)
\right)+\calO(q^2) \\
& =\;q^{1/2}\; \chi^{U(6)}_{\rmbf{6}}+\;q\;\chi^{U(6)}_{\rmbf{\wedge^2\ovl{6}}}+\;q^{3/2}\; \left[ \chi^{U(6)}_{[20001]_{-1}}+\chi^{U(6)}_{[01001]_{-1}}+\chi^{U(6)}_{\rmbf{6}} \right]+\calO(q^2)
\end{align}
and
\begin{align}
\nonumber & \calI^{\rm Schur}_{\tiny SU(3)\; N_f=6\mbox{ SQCD w/. }T^{\rm dual}_{\Box}}(a,b,c,d)
=
cd+\;q^{1/2}\; \left( cd^{-2}+c^{-2}d \right)+\\
\nonumber & +\;q\;\left( c^{-1}d^{-1}\chi_{\rmbf{3}}(a)\chi_{\rmbf{3}}(b)+cd\chi_{\rmbf{8}}(b)+cd\chi_{\rmbf{8}}(a)+3cd+c^{2}d^{2}\chi_{\rmbf{3}}(a)\chi_{\rmbf{3}}(b)+c^{-2}d^{-2} \right.\\
\nonumber & \left.\quad +2\chi_{\rmbf{\ovl{3}}}(a)\chi_{\rmbf{\ovl{3}}}(b)\right)+\;q^{3/2}\; \left( cd^{4}+2c^{2}d^{-1}\chi_{\rmbf{3}}(a)\chi_{\rmbf{3}}(b)+c^{4}d+cd^{-2}\chi_{\rmbf{8}}(b) \right.\\
\nonumber & \quad +d^{3}\chi_{\rmbf{\ovl{3}}}(a)\chi_{\rmbf{\ovl{3}}}(b)+c^{-2}d\chi_{\rmbf{8}}(b)+c^{-3}\chi_{\rmbf{\ovl{3}}}(a)\chi_{\rmbf{\ovl{3}}}(b)+2c^{-1}d^{2}\chi_{\rmbf{3}}(a)\chi_{\rmbf{3}}(b)+3c^{-2}d\\
& \left.\quad +c^{-2}d\chi_{\rmbf{8}}(a)+c^{3}\chi_{\rmbf{\ovl{3}}}(a)\chi_{\rmbf{\ovl{3}}}(b)+3cd^{-2}+cd^{-2}\chi_{\rmbf{8}}(a)+d^{-3}\chi_{\rmbf{\ovl{3}}}(a)\chi_{\rmbf{\ovl{3}}}(b)\right)
+\calO(q^2) \\
\nonumber & =cd+\;q^{1/2}\; cd(c^{-3}+d^{-3})+\;q\;
cd \left[ 1+\chi^{U(6)}_{\Adj}+c^{-3}\chi^{U(6)}_{\wedge^3\rmbf{6}} \right] \\
& \quad+\;q^{3/2}\; cd\left[ (c^{-3}+d^{-3})(\chi^{U(6)}_{\Adj}+1)+\chi^{U(6)}_{\wedge^3\rmbf{6}}+\chi^{U(6)}_{\wedge^3\rmbf{\ovl{6}}} \right] +\calO(q^2)
\end{align}
respectively.
Notice that $\chi^{U(6)}_{[0\ldots\underset{k}{1}\ldots]_{p}}:=(\chi^{U(6)}_{\wedge^6\rmbf{6}})^p\chi^{U(6)}_{\wedge^k\rmbf{6}}$ and $\chi^{U(6)}_{\wedge^6\rmbf{6}}=c^3d^{-3}$.

When we add both fundamental Wilson loop and the dual 't Hooft loop as shown in Fig.~\ref{fig:loops in SQCD for higher rank} (Right) with $a=b=1$, the answer is
\begin{align}
\nonumber & \calI^{\rm Schur}_{\tiny SU(3)\; N_f=6\mbox{ SQCD w/. }T^{\rm dual}_{\Box} \circ W_{\Box}}(a,b,c,d)= q^{1/2} \; (c\chi_{\rmbf{\ovl{3}}}(a)+c^{2}d\chi_{\rmbf{3}}(b)) \\ 
\nonumber & + \;q \; (c^{-2}\chi_{\rmbf{\ovl{3}}}(a)+cd^{-3}\chi_{\rmbf{\ovl{3}}}(a)+c^{2}d^{-2}\chi_{\rmbf{3}}(b)+2c^{-1}d\chi_{\rmbf{3}}(b)+cd^{3}\chi_{\rmbf{\ovl{3}}}(a)+d^{2}\chi_{\rmbf{3}}(a)\chi_{\rmbf{\ovl{3}}}(b)) \\ 
\nonumber & +\;q^{3/2}\;
(3d^{-1}\chi_{\rmbf{3}}(a)\chi_{\rmbf{\ovl{3}}}(b)+2c^{-1}d^{-2}\chi_{\rmbf{3}}(b)+c^{3}d^{2}\chi_{\rmbf{3}}(a)\chi_{\rmbf{6}}(b)+c^{-3}d^{2}\chi_{\rmbf{3}}(a)\chi_{\rmbf{\ovl{3}}}(b)+c^{-4}d\chi_{\rmbf{3}}(b)\\
\nonumber & \quad
+5c\chi_{\rmbf{\ovl{3}}}(a)+3c\chi_{\rmbf{\ovl{3}}}(a)\chi_{\rmbf{8}}(b)+c^{2}d\chi_{\rmbf{\ovl{6}}}(b)+d^{-1}\chi_{\rmbf{3}}(a)\chi_{\rmbf{6}}(b)+c^{-2}d^{3}\chi_{\rmbf{\ovl{3}}}(a)+c^{3}d^{2}\chi_{\rmbf{3}}(a)\chi_{\rmbf{\ovl{3}}}(b)\\
\nonumber & \quad
+2c^{2}d\chi_{\rmbf{8}}(a)\chi_{\rmbf{3}}(b)+c^{-1}d^{-2}\chi_{\rmbf{8}}(a)\chi_{\rmbf{3}}(b)+c^{2}d\chi_{\rmbf{15}'}(b)+2d^{-1}\chi_{\rmbf{\ovl{6}}}(a)\chi_{\rmbf{\ovl{3}}}(b)+4c^{2}d\chi_{\rmbf{3}}(b)\\
& \quad +c\chi_{\rmbf{\ovl{15}'}}(a)+c\chi_{\rmbf{6}}(a)+c^{-2}d^{-3}\chi_{\rmbf{\ovl{3}}}(a))+\calO(q^{2}). \\
\nonumber &=q^{1/2}cd\chi^{U(6)}_{\rmbf{6}}+q\left[cd\chi^{U(6)}_{\rmbf{\wedge^2\ovl{6}}}+cd(c^{-3}+d^{-3})\chi^{U(6)}_{\rmbf{6}}\right] \\
&+q^{3/2}\; (c^{-2}d)\left[ d^3\chi^{U(6)}_{[20001]_0}+d^3\chi^{U(6)}_{[01001]_0}+\chi^{U(6)}_{[10100]_0}+(1+c^3d^{-3})\chi^{U(6)}_{\wedge^4 \rmbf{6}}+c^3\chi^{U(6)}_{\rmbf{6}}
\right]+\calO(q^2)
\label{eq:SU(3) SQCD with W and T loops}
\end{align}
where we use the Boltzmann weight result for each crossing given in \cite{Watanabe1603}.
However, we can derive this result in a different way as shown in Sec.~\ref{subsec:Application}.

\begin{comment}

\subsubsection{$\SUSYN{2^\ast}$ gauge theory}

\end{comment}

%\end{document}

%\end{comment}
\section{New kinds of skein relations}
\label{sec:new kinds of skein relations}

\begin{figure}[htb]
\centering
\begin{tikzpicture}
\punctureindigonPP{$c$}{$a$}{$b$}{above}{$Y$}{0.6}{0.08}{1.2}
\end{tikzpicture}
\caption{The local digon network with a puncture $Y$ inside it.}
\label{fig:digon with puncture inside}
\end{figure}
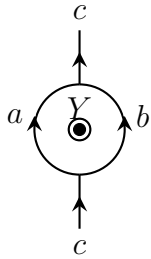

In this section, we show several new kinds of skein relations based on the computations which the formula in ~\eqref{eq:formula for qYM correlator with elementary pants network} gives.
They are new in the sense that both codimension two defects (punctures) and codimension four defects (networks) are included in the relations and, in addition, non-trivially related.
In particular, we focus on the skein relation what we call digon type skein relations.
They are associated to one puncture as shown in Fig.~\ref{fig:digon with puncture inside}.

The strategy to find the skein relations is to compute the $q$-deformed Yang-Mills correlators with every elementary pants network of the theory $T^\calS[C([1^N],[1^N],Y)]$ for each puncture $Y$.
%\!\footnote{The difference between the Schur indices and $q$-deformed Yang-Mills correlators is just overall factors given in Sec.~\ref{}. This does not affect the skein relations.}
We analyze the punctures other than the full, simple (instead given in Sec.~\ref{subsec:Simple punctures}) and null/trivial punctures in $A_3=\Liesu(4)$ (Sec.~\ref{subsec:A3 case}), $A_4=\Liesu(5)$ (Sec.~\ref{subsec:A4 case}) and $A_5=\Liesu(6)$ (Sec.~\ref{subsec:A5 case}) cases.
From many examples, In Sec.\ref{subsec:General punctures}, we give a unified formula to generate all the skein relations.
From the argument in the first part of Sec.~\ref{sec:Schur indices with line defects},
we have a natural conjecture that the number of the linearly independent skein relations (See the explanation later) is equal to $d_C^{[1^N]}-d_C^Y$. See ~\eqref{eq:total Coulomb dim formula for puncture}.

Notice that we assume that all the skein relations admit the mirror operation defined as
\begin{align}
& M \;:\quad
\raisebox{-0.47\height}{
\begin{tikzpicture}
\punctureindigonPP{$c$}{$a$}{$b$}{above}{$Y,x$}{0.6}{0.08}{1.2}
\end{tikzpicture}
}
\longleftrightarrow
\raisebox{-0.47\height}{
\begin{tikzpicture}
\punctureindigonPP{$c$}{$b$}{$a$}{above}{$Y,x^\ast$}{0.6}{0.08}{1.2}
\end{tikzpicture}
} \\
& M \;:\qqqquad\qquad
x \longleftrightarrow x^{\ast}=x^{-1} \\
& M \;:\quad
\;\raisebox{-0.38\height}{
\begin{tikzpicture}
\linenearpuncture{$c$}{left}{$Y$}{1.32}{0.45}{0.08}
\end{tikzpicture}
}\quad
\longleftrightarrow
\;\raisebox{-0.38\height}{
\begin{tikzpicture}
\linenearpuncture{$c$}{right}{$Y$}{1.32}{-0.45}{0.08}
\end{tikzpicture}
}
\end{align}
and some skein relations are invariant and others gives the new relations. This operation is clearly involution and no more relations appear.
We can check that the whole skein relations are closed under this operation in all the example below just by computing them.
Hereafter, when we exhibit a skein relation, we abbreviate its mirror operated relation.

\subsubsection*{Class S skein relations}

We recall the concept of class S skein relations.
This is the same as those used in the theory of knots in mathematics. However, in our set-up, we have no three geometry as long as we add surface operators or consider the multiple lines.
For concrete examples, see the following sections or \cite{TW1504}.

Let $W_{\Gamma}(\{ a \};q)$ be the expectation value of the Wilson operator associated with a network $\Gamma$ on the punctured Riemann surface. $\{ a \}$ are all fugacities for the flavor symmetries or holonomies around punctures of $C$.

And let us consider two sub graphs $\gamma_A$ and $\gamma_B$ which allow the common punctures inside them. For any pair of two graphs $\Gamma_A$ and $\Gamma_B$ which include $\gamma_A$ and $\gamma_B$ respectively but are same on removing these sub graphs,
when the equality shown just below always holds true, we identify $\gamma_A$ and $\gamma_B$ and write this as $\gamma_A \sim \gamma_B$. The equality is
\begin{align}
W_{\Gamma_A}(\{ a,c \};q)=F_{\gamma_A \to \gamma_B}(c;q)W_{\Gamma_B}(\{ a,c \};q)
\end{align}
where $F_{\gamma_A \to \gamma_B}(c;q)$ is a function of $q$ and $\{c\}$ which are holonomies around the punctures included in $\gamma_A$ or $\gamma_B$, independent of holonomies around other punctures and also determined by $\gamma_A$ and $\gamma_B$ only.
Now we have the equivalence relations $\sim$ and refer to them as the ``(class S) skein relations" hereafter.

Finally, we make a few comments.
$\gamma_B$ or $\Gamma_B$ may consist of sum of several networks and $W$ is the homomorphism from the networks to the functions of $q$ and all holonomies.
Related to this, when we present the skein relations, we must take some basis. For example, multiplying both hand sides by the some common factors gives the same relation. When there are two independent relations, we can take some linear combination of two but the new relation is not linearly independent. This point will be important later in Sec.~\ref{subsec:General punctures}.
%\!\footnote{In mathematical language, the networks form the ``module" over $\bbZ[\{a\},q^{1/2},q^{-1/2}]$ divided by the skein relations and $a_1a_2\cdot a_N=1$. In our presentation of skein relations, we must take some basis of that.}

In all examples we know, $F(c;q)$ is a polynomial of $q^{\pm \frac{1}{2}}$ and $c$. In addition, we can take the basis such that $F(c;q)=F(c;q^{-1})$ and the invariance under the permutations of $c$ in any fixed (simple) flavor symmetry always hold true. The former property comes from the symmetry (an assumption, however) of the 2D $q$-deformed Yang-Mills theory. Actually, we can compute the skein relations in the 2D topological $q$-deformed Yang-Mills theory not in the Schur indices because they differ by just the overall factors. Furthermore, in many cases, this relation is enough local and independent of the choice of $C$.

The final comment is related to the class S set-up.
Under the parameter identification $q=e^{2\pi i b^2}$ \cite{DGG1108,TW1504} where $b$ is a physical parameter in the Liouville/Toda CFT, the skein relations are common both in the CFTs and in the 2D $q$-deformed Yang-Mills theory.
This is because the skein relations are expected to be the local relations about codimension four defects in the 6D $\SUSYN{(2,0)}$ SCFTs and to be independent of the four dimensional global background geometries, namely, the choice of $S^4_b$ or $S^1 \times_q S^3$.

\subsection{Simple punctures}
\label{subsec:Simple punctures}

This puncture corresponds to $Y=[N-1,1]$-type and the original $U(1)^N/U(1)\subset SU(N)$ fugacity is expressed as $c_{[N-1,1]}=(q^{\tfrac{N-2}{2}}c,q^{\tfrac{N-4}{2}}c,\ldots,q^{\tfrac{-N+4}{2}}c,q^{\tfrac{-N+2}{2}}c,c^{-(N-1)})$ where $c$ is the $U(1)$ fugacity.
In this case, we have already seen the examples in Sec.~\ref{subsec:Free hypermultiplets} and the conjecture ~\eqref{eq:flavor Wilson of bi-fundamental type} for general rank.
We can read off the following relations from them.

\begin{align}
\raisebox{-0.47\height}{
\begin{tikzpicture}
\punctureindigonPP{$a+b$}{$a$}{$b$}{above}{$Y$}{0.6}{0.08}{1.2}
\end{tikzpicture}
}
=
c^{-a}\left[\begin{array}{c} a+b-1 \\ a \end{array}\right]_q
\;\raisebox{-0.38\height}{
\begin{tikzpicture}
\linenearpuncture{$a+b$}{left}{$Y$}{1.32}{0.45}{0.08}
\end{tikzpicture}
}\quad
+
c^{b}\left[\begin{array}{c} a+b-1 \\ b \end{array}\right]_q
\;\raisebox{-0.38\height}{
\begin{tikzpicture}
\linenearpuncture{$a+b$}{right}{$Y$}{1.32}{-0.45}{0.08}
\end{tikzpicture}
}\quad
\label{eq:digon skein simple puncture}
\end{align}
where $1\le a,b$ and $a+b \le N-1$.

%\end{comment}
\subsection{$A_3$ case}
\label{subsec:A3 case}

Let us see the non-trivial skein relations in the $A_3=\Liesu(4)$ case.
There are five types of regular punctures, two of which are subjects. See the Table ~\ref{tab:A3 examples of leading exponent}.

\begin{table}[htb]
\centering
\begin{tabular}{|c|c|c|c|c|}
\hline
$[1^4]$ & $[21^2]$ & $[2^2]$ & $[31]$ & $[4]$ \\
\hline
[0, 0, 0] & [-1, -1, -1] & [-1, -2, -1] & [-2, -2, -2] & [-3, -4, -3] \\
\hline
[1, 2, 3] & [1, 2, 2] & [1, 1, 2] & [1, 1, 1] & [0, 0, 0] \\
\hline
\end{tabular}
\caption{The list for $\Lieg=\Liesu(4)$ ($A_3$) : the puncture type $Y$, $2\rho^{\alpha}_{L,Y}$ and the numbers of Coulomb branch operators with scaling dimensions $2,3,4$ in this order.}
\label{tab:A3 examples of leading exponent}
\end{table}

First of all, we list up all the good theories in the sense discussed in Sec.~\ref{subsec:Good theories for q-series expansion}.

\paragraph{rank 3}

\begin{align}
T^{(3)}_{A_3}=T^\calS[C([1^4],[1^4],[1^4])]
\end{align}

\paragraph{rank 2}

\begin{align}
T^{(2)}_{A_3}=T^\calS[C([1^4],[1^4],[21^2])]
\end{align}

\paragraph{rank 1}

\begin{align}
T^{(1-1)}_{A_3}=T^\calS[C([1^4],[1^4],[2^2])] \qquad T^{(1-2)}_{A_3}=T^\calS[C([1^4],[21^2],[21^2])]
\end{align}

\paragraph{rank 0}

\begin{align}
F^{\rm (bf)}_{A_3}=T^\calS[C([1^4],[1^4],[31])] \qquad F^{\rm (as)}_{A_3}=T^\calS[C([1^4],[21^2],[2^2])]
\end{align}

\subsubsection*{$\bf Y=[2^2]$}

The flavor symmetry associated with this puncture $Y=[2^2]$ is $SU(2)$.
The associated fugacity is given by $c_{[2^2]}=(q^{1/2}c,q^{-1/2}c,q^{1/2}c^{-1},q^{-1/2}c^{-1})$.
It is expected that there are 2 independent skein relations.

\paragraph{i}
\begin{align}
\raisebox{-0.47\height}{\tikzsetnextfilename{-figure-A3-digon321}
\begin{tikzpicture}
\punctureindigonNP{$3$}{$2$}{$1$}{above}{$Y$}{0.6}{0.08}{1.2}
\end{tikzpicture}
}
=
\;\raisebox{-0.38\height}{\tikzsetnextfilename{-figure-A3-line3passright}
\begin{tikzpicture}
\linenearpuncture{$3$}{left}{$Y$}{1.32}{0.45}{0.08}
\end{tikzpicture}
}\quad
+
\chi_{\rmbf{2}}^{SU(2)}(c)
\;\raisebox{-0.38\height}{\tikzsetnextfilename{-figure-A3-line3passleft}
\begin{tikzpicture}
\linenearpuncture{$3$}{right}{$Y$}{1.32}{-0.45}{0.08}
\end{tikzpicture}
}\quad
\end{align}
and its mirror operated relation.

\subsubsection*{$\bf Y=[21^2]$}

The flavor symmetry associated with this puncture $Y=[21^2]$ is $S(U(1)\times U(2)) \simeq U(1) \times SU(2)$.
The associated fugacity is given by $c_{[21^2]}=(q^{1/2}c_1,q^{-1/2}c_1,c_2,c_3)$.
In particular, we introduce the natural recombined fugacities $c:=c_1$ and $c':=c_2/{(c_2c_3)^{1/2}}=c_1c_2$.
It is expected that there is 1 independent skein relation.

\paragraph{i}
\begin{align}
c^{1/2}
\;\raisebox{-0.47\height}{\tikzsetnextfilename{-figure-A3-digon321}
\begin{tikzpicture}
\punctureindigonNP{$3$}{$2$}{$1$}{above}{$Y$}{0.6}{0.08}{1.2}
\end{tikzpicture}
}
-
c^{-1/2}
\;\raisebox{-0.47\height}{\tikzsetnextfilename{-figure-A3-digon312}
\begin{tikzpicture}
\punctureindigonPN{$3$}{$1$}{$2$}{above}{$Y$}{0.6}{0.08}{1.2}
\end{tikzpicture}
}
=
c^{3/2}
\;\raisebox{-0.38\height}{\tikzsetnextfilename{-figure-A3-line3passleft}
\begin{tikzpicture}
\linenearpuncture{$3$}{right}{$Y$}{1.32}{-0.45}{0.08}
\end{tikzpicture}
}\quad
-
c^{-3/2}
\;\raisebox{-0.38\height}{\tikzsetnextfilename{-figure-A3-line3passright}
\begin{tikzpicture}
\linenearpuncture{$3$}{left}{$Y$}{1.32}{0.45}{0.08}
\end{tikzpicture}
}\quad
\label{eq:new skein:[211]}
\end{align}

\subsubsection*{Check for $T^{(1-2)}_{A_3}$}

The above skein relations give some constraints on the elementary pants networks of $T^{(1-2)}_{A_3}=T^\calS[C([1^4],[21^2],[21^2])]$.
The explicit flavor symmetry is given by $SU(4) \times U(1)_B \times SU(2)_B \times U(1)_C \times SU(2)_C$ and we write the associated fugacities as $a_{[1^4]}=(a_1,a_2,a_3,a_4)$ ($a_1a_2a_3a_4=1$), $b_{[21^2]}=(q^{1/2}b,q^{1/2}b^{-1},b'/b,1/bb')$ and $c_{[21^2]}=(q^{1/2}c,q^{1/2}c^{-1},c'/c,1/cc')$.

The relation ~\eqref{eq:new skein:[211]} leads to the following equalities :
\begin{align}
\nonumber & \qquad \calI^{\rm Schur}_{T^{(1-2)}_{A_3}\mbox{ w/. }\wp(1,2,1)}(a,b',c') \\
\nonumber &=
c^{-1}\;\calI^{\rm Schur}_{T^{(1-2)}_{A_3}\mbox{ w/. }\wp(1,1,2)}(a,b',c')
+
\left( [2]_qb^{-1}c+bc\chi^{SU(2)_B}_{\rmbf{2}}(b')-c^{-2}\chi_{\rmbf{4}}^{SU(4)}(a) \right)
\calI^{\rm Schur}_{T^{(1-2)}_{A_3}}(a,b',c') \\
&=
b\;\calI^{\rm Schur}_{T^{(1-2)}_{A_3}\mbox{ w/. }\wp(2,1,1)}(a,b',c')
+
\left( [2]_qb^{-1}c+b^{-1}c^{-1}\chi^{SU(2)_C}_{\rmbf{2}}(c')-b^{2}\chi_{\rmbf{\ovl{4}}}^{SU(4)}(a) \right)
\calI^{\rm Schur}_{T^{(1-2)}_{A_3}}(a,b',c').
\end{align}
%See Appendix~\ref{app-subsec:T^{(1-2)}_{A_3}} for the $q$-deformed Yang-Mills expression up to the $q^{3/2}$-order.

In this case, three elementary pants networks are equivalent up to flavor loops and there is only one independent elementary pants network which coincides with the rank one.

\subsubsection*{Check for $F^{\rm (as)}_{A_3}$}

Let us focus on the rank 0 free SCFT $F^{\rm (as)}_{A_3}=T^\calS[C([1^4],[21^2],[2^2])]$.
We can show that all the elementary pants networks reduce to the flavor Wilson lines by applying the above skein relations and they reproduce the result ~\eqref{eq:flavor Wilson for A3 as-type free hypers}.

%\end{comment}
\subsection{$A_4$ case}
\label{subsec:A4 case}

The next case we consider is the $A_4=\Liesu(5)$ case.
There are seven types of regular punctures, four of which are subjects. See the Table ~\ref{tab:A4 examples of leading exponent}.

\begin{center}
\begin{table}[htb]
\begin{tabular}{|c|c|c|c|c|c|c|}
\hline
$[1^5]$ & $[21^3]$ & $[2^21]$ & $[31^2]$ & $[32]$ & $[41]$ & $[5]$ \\
\hline
[0,0,0,0] & [-1,-1,-1,-1] & [-1,-2,-2,-1] & [-2,-2,-2,-2] & [-2,-3,-3,-2] & [-3,-4,-4,-3] & [-4,-6,-6,-4] \\
\hline
[1, 2, 3, 4] & [1, 2, 3, 3] & [1, 2, 2, 3] & [1, 2, 2, 2] & [1, 1, 2, 2] & [1, 1, 1, 1] & [0, 0, 0, 0] \\
\hline
\end{tabular}
\caption{The list for $\Lieg=\Liesu(5)$ ($A_4$) : the puncture type $Y$, $2\rho^{\alpha}_{L,Y}$ and the numbers of Coulomb branch operators with scaling dimensions $2,3,4,5$ in this order.}
\label{tab:A4 examples of leading exponent}
\end{table}
\end{center}

\begin{comment}

We list up all the good theories.

\paragraph{rank 6}

\begin{align}
T^{(6)}_{A_4}=T^\calS[C([1^5],[1^5],[1^5])]
\end{align}

\paragraph{rank 5}

\begin{align}
T^{(5)}_{A_4}=T^\calS[C([1^5],[1^5],[21^3])]
\end{align}

\paragraph{rank 4}

\begin{align}
T^{(4-1)}_{A_4}=T^\calS[C([1^5],[1^5],[2^21])]
\qquad
T^{(4-2)}_{A_4}=T^\calS[C([1^5],[21^3],[21^3])]
\end{align}

\paragraph{rank 3}

\begin{align}
T^{(3-1)}_{A_4}=T^\calS[C([1^5],[1^5],[31^3])]
\qquad
T^{(3-2)}_{A_4}=T^\calS[C([1^5],[21^3],[2^21])] \\
T^{(3-3)}_{A_4}=T^\calS[C([21^3],[21^3],[21^3])]
\end{align}

\paragraph{rank 2}

\begin{align}
T^{(2-1)}_{A_4}=T^\calS[C([1^5],[1^5],[32])]
\qquad
T^{(2-2)}_{A_4}=T^\calS[C([1^5],[21^3],[31^2])] \\
T^{(2-3)}_{A_4}=T^\calS[C([21^3],[21^3],[2^21])]
\qquad
T^{(2-4)}_{A_4}=T^\calS[C([1^5],[2^21],[2^21])]
\end{align}

\paragraph{rank 1}

\begin{align}
T^{(1-1)}_{A_4}=T^\calS[C([1^5],[21^3],[32])] \qquad T^{(1-2)}_{A_4}=T^\calS[C([1^5],[2^21],[31^2])]\\ T^{(1-3)}_{A_4}=T^\calS[C([21^3],[2^21],[2^21])]
\end{align}

\paragraph{rank 0}

\begin{align}
F^{\rm (bf)}_{A_4}=T^\calS[C([1^5],[1^5],[41])] \qquad F^{\rm (as)}_{A_4}=T^\calS[C([1^5],[2^21],[32])]
\end{align}

\end{comment}

\subsubsection*{$\bf Y=[32]$}

The flavor symmetry associated with this puncture $Y=[32]$ is $U(1)$.
The associated fugacity is given by $c_{[32]}=(qc_1,c_1,q^{-1}c_1,q^{1/2}c_2,q^{-1/2}c_2)$.
In particular, we introduce the natural recombined fugacities $c:=c_1$. % and $c':=c_2/{(c_2c_3)^{1/2}}=c_1^{3/2}c_2$.
It is expected that there are 4 independent skein relations.

\paragraph{i-1\& i-2}
\begin{align}
\;\raisebox{-0.47\height}{\tikzsetnextfilename{-figure-A4-digon422}
\begin{tikzpicture}
\punctureindigonPP{$4$}{$2$}{$2$}{above}{$Y$}{0.6}{0.08}{1.2}
\end{tikzpicture}
}
=
\left( c^2+[2]_q c^{-1/2} \right)
\;\raisebox{-0.38\height}{\tikzsetnextfilename{-figure-A4-line4passright}
\begin{tikzpicture}
\linenearpuncture{$4$}{left}{$Y$}{1.32}{0.45}{0.08}
\end{tikzpicture}
}
+
\left( c^{-2}+[2]_q c^{1/2} \right)
\;\raisebox{-0.38\height}{\tikzsetnextfilename{-figure-A4-line4passleft}
\begin{tikzpicture}
\linenearpuncture{$4$}{right}{$Y$}{1.32}{-0.45}{0.08}
\end{tikzpicture}
}\quad.
\end{align}

\begin{align}
&
\raisebox{-0.47\height}{\tikzsetnextfilename{-figure-A4-digon431}
\begin{tikzpicture}
\punctureindigonPP{$4$}{$3$}{$1$}{above}{$Y$}{0.6}{0.08}{1.2}
\end{tikzpicture}
}
=
\left( c^{-3/2}+[2]_q c \right)
\;\raisebox{-0.38\height}{\tikzsetnextfilename{-figure-A4-line4passleft}
\begin{tikzpicture}
\linenearpuncture{$4$}{right}{$Y$}{1.32}{-0.45}{0.08}
\end{tikzpicture}
}\quad
+
c^{-1/2}
\;\raisebox{-0.38\height}{\tikzsetnextfilename{-figure-A4-line4passright}
\begin{tikzpicture}
\linenearpuncture{$4$}{left}{$Y$}{1.32}{0.45}{0.08}
\end{tikzpicture}
}\quad
\end{align}
and its mirror operated relation.

\paragraph{ii}
\begin{align}
c^{-1/2}
\raisebox{-0.47\height}{\tikzsetnextfilename{-figure-A4-digon312}
\begin{tikzpicture}
\punctureindigonPP{$3$}{$1$}{$2$}{above}{$Y$}{0.6}{0.08}{1.2}
\end{tikzpicture}
}
-
c^{1/2}
\;\raisebox{-0.47\height}{\tikzsetnextfilename{-figure-A4-digon321}
\begin{tikzpicture}
\punctureindigonPP{$3$}{$2$}{$1$}{above}{$Y$}{0.6}{0.08}{1.2}
\end{tikzpicture}
}
=
c^{-3/2}
\;\raisebox{-0.38\height}{\tikzsetnextfilename{-figure-A4-line3passright}
\begin{tikzpicture}
\linenearpuncture{$3$}{left}{$Y$}{1.32}{0.45}{0.08}
\end{tikzpicture}
}\quad
-
c^{3/2}
\;\raisebox{-0.38\height}{\tikzsetnextfilename{-figure-A4-line3passleft}
\begin{tikzpicture}
\linenearpuncture{$3$}{right}{$Y$}{1.32}{-0.45}{0.08}
\end{tikzpicture}
}\quad.
\label{eq:skein [32]-type class ii}
\end{align}

\begin{comment}
\paragraph{i}
\begin{align}
c^{-1}
\;\raisebox{-0.47\height}{\input{figure/A4/digon413.tikz}}
-
c
\;\raisebox{-0.47\height}{\input{figure/A4/digon431.tikz}}
=
[2]_q \left[
c^{-2}
\;\raisebox{-0.38\height}{\input{figure/A4/line4passright.tikz}}\quad
-
c^2
\;\raisebox{-0.38\height}{\input{figure/A4/line4passleft.tikz}}\quad \right]
\end{align}
and
\begin{align}
&
c^{-3/2}
\raisebox{-0.47\height}{\input{figure/A4/digon431.tikz}}
+
c^{3/2}
\;\raisebox{-0.47\height}{\input{figure/A4/digon413.tikz}}
-
\raisebox{-0.47\height}{\input{figure/A4/digon422.tikz}}=
c^{3}
\;\raisebox{-0.38\height}{\input{figure/A4/line4passright.tikz}}\quad
+
c^{-3}
\;\raisebox{-0.38\height}{\input{figure/A4/line4passleft.tikz}}\quad.
\end{align}

\paragraph{ii}
\begin{align}
c^{-1/2}
\raisebox{-0.47\height}{\input{figure/A4/digon312.tikz}}
-
c^{1/2}
\;\raisebox{-0.47\height}{\input{figure/A4/digon321.tikz}}
=
c^{-3/2}
\;\raisebox{-0.38\height}{\input{figure/A4/line3passright.tikz}}\quad
-
c^{3/2}
\;\raisebox{-0.38\height}{\input{figure/A4/line3passleft.tikz}}\quad.
\end{align}

\paragraph{iii}
\begin{align}
\raisebox{-0.47\height}{\input{figure/A4/digon211.tikz}}
=
c^{-1}
\;\raisebox{-0.38\height}{\input{figure/A4/line2passright.tikz}}\quad
+
c
\;\raisebox{-0.38\height}{\input{figure/A4/line2passleft.tikz}}\quad.
\end{align}
\end{comment}

\subsubsection*{$\bf Y=[31^2]$}

The flavor symmetry associated with this puncture $Y=[31^2]$ is $S(U(1)\times U(2)) \simeq U(1) \times SU(2)$.
The associated fugacity is given by $c_{[31^2]}=(qc_1,c_1,q^{-1}c_1,c_2,c_3)$.
In particular, we introduce the natural recombined fugacities $c:=c_1$. % and $c':=c_2/{(c_2c_3)^{1/2}}=c_1^{3/2}c_2$.
It is expected that there are 3 independent skein relations.

Notice that the mirror operated relation for the second one in the category \textbf{i} is linearly dependent on the first and the second.

\paragraph{i}
\begin{align}
c
\;\raisebox{-0.47\height}{\tikzsetnextfilename{-figure-A4-digon431}
\begin{tikzpicture}
\punctureindigonPP{$4$}{$3$}{$1$}{above}{$Y$}{0.6}{0.08}{1.2}
\end{tikzpicture}
}
-
c^{-1}
\raisebox{-0.47\height}{\tikzsetnextfilename{-figure-A4-digon413}
\begin{tikzpicture}
\punctureindigonPP{$4$}{$1$}{$3$}{above}{$Y$}{0.6}{0.08}{1.2}
\end{tikzpicture}
}
=
[2]_q\; \left[
c^{2}
\;\raisebox{-0.38\height}{\tikzsetnextfilename{-figure-A4-line4passleft}
\begin{tikzpicture}
\linenearpuncture{$4$}{right}{$Y$}{1.32}{-0.45}{0.08}
\end{tikzpicture}
}\quad
-
c^{-2}
\;\raisebox{-0.38\height}{\tikzsetnextfilename{-figure-A4-line4passright}
\begin{tikzpicture}
\linenearpuncture{$4$}{left}{$Y$}{1.32}{0.45}{0.08}
\end{tikzpicture}
}\quad
\right]
\end{align}
and
\begin{align}
\;\raisebox{-0.47\height}{\tikzsetnextfilename{-figure-A4-digon422}
\begin{tikzpicture}
\punctureindigonPP{$4$}{$2$}{$2$}{above}{$Y$}{0.6}{0.08}{1.2}
\end{tikzpicture}
}
-
[2]_q c^{-1}
\raisebox{-0.47\height}{\tikzsetnextfilename{-figure-A4-digon413}
\begin{tikzpicture}
\punctureindigonPP{$4$}{$1$}{$3$}{above}{$Y$}{0.6}{0.08}{1.2}
\end{tikzpicture}
}
=
c^2
\;\raisebox{-0.38\height}{\tikzsetnextfilename{-figure-A4-line4passleft}
\begin{tikzpicture}
\linenearpuncture{$4$}{right}{$Y$}{1.32}{-0.45}{0.08}
\end{tikzpicture}
}\quad
-[3]_q c^{-2}
\;\raisebox{-0.38\height}{\tikzsetnextfilename{-figure-A4-line4passright}
\begin{tikzpicture}
\linenearpuncture{$4$}{left}{$Y$}{1.32}{0.45}{0.08}
\end{tikzpicture}
}\quad.
\end{align}

\paragraph{ii}
We have the same skein relation ~\eqref{eq:skein [32]-type class ii}.

\subsubsection*{$\bf Y=[2^21]$}

The flavor symmetry associated with this puncture $Y=[2^21]$ is $S(U(2)\times U(1)) \simeq U(1) \times SU(2)$.
The associated fugacity is given by $c_{[2^21]}=(q^{1/2}c_1,q^{-1/2}c_1,q^{1/2}c_2,q^{-1/2}c_2,c_3)$.
In particular, we introduce the natural recombined fugacities $c:=(c_1c_2)^{1/2}$ and $c'_i:=c_{i}/c$ for $i=1,2$.
It is expected that there are 2 independent skein relations.

\paragraph{i}
\begin{align}
\raisebox{-0.47\height}{\tikzsetnextfilename{-figure-A4-digon422}
\begin{tikzpicture}
\punctureindigonPP{$4$}{$2$}{$2$}{above}{$Y$}{0.6}{0.08}{1.2}
\end{tikzpicture}
}
-
c\chi^{SU(2)}_{\rmbf{2}}(c')
\raisebox{-0.47\height}{\tikzsetnextfilename{-figure-A4-digon431}
\begin{tikzpicture}
\punctureindigonPP{$4$}{$3$}{$1$}{above}{$Y$}{0.6}{0.08}{1.2}
\end{tikzpicture}
}
=
c^{-2}
\;\raisebox{-0.38\height}{\tikzsetnextfilename{-figure-A4-line4passright}
\begin{tikzpicture}
\linenearpuncture{$4$}{left}{$Y$}{1.32}{0.45}{0.08}
\end{tikzpicture}
}\quad
-
c^{2}\chi^{SU(2)}_{\rmbf{3}}(c')
\;\raisebox{-0.38\height}{\tikzsetnextfilename{-figure-A4-line4passleft}
\begin{tikzpicture}
\linenearpuncture{$4$}{right}{$Y$}{1.32}{-0.45}{0.08}
\end{tikzpicture}
}\quad
\end{align}
and
\begin{align}
c
\;\raisebox{-0.47\height}{\tikzsetnextfilename{-figure-A4-digon431}
\begin{tikzpicture}
\punctureindigonPP{$4$}{$3$}{$1$}{above}{$Y$}{0.6}{0.08}{1.2}
\end{tikzpicture}
}
-
c^{-1}
\raisebox{-0.47\height}{\tikzsetnextfilename{-figure-A4-digon413}
\begin{tikzpicture}
\punctureindigonPP{$4$}{$1$}{$3$}{above}{$Y$}{0.6}{0.08}{1.2}
\end{tikzpicture}
}
=
c^{2}\chi^{SU(2)}_{\rmbf{2}}(c')
\;\raisebox{-0.38\height}{\tikzsetnextfilename{-figure-A4-line4passleft}
\begin{tikzpicture}
\linenearpuncture{$4$}{right}{$Y$}{1.32}{-0.45}{0.08}
\end{tikzpicture}
}\quad
-
c^{-2}\chi^{SU(2)}_{\rmbf{2}}(c')
\;\raisebox{-0.38\height}{\tikzsetnextfilename{-figure-A4-line4passright}
\begin{tikzpicture}
\linenearpuncture{$4$}{left}{$Y$}{1.32}{0.45}{0.08}
\end{tikzpicture}
}\quad.
\end{align}

\subsubsection*{$\bf Y=[21^3]$}

The flavor symmetry associated with this puncture $Y=[21^3]$ is $S(U(1)\times U(3)) \simeq U(1) \times SU(3)$.
The associated fugacity is given by $c_{[21^3]}=(q^{1/2}c_1,q^{-1/2}c_1,c_2,c_3,c_4)$.
In particular, we introduce the natural recombined fugacities $c:=c_1$.% and $c'_i:=c_{i+1}/{(c_2c_3c_4)^{1/3}}=c_1^{2/3}c_2$ for $i=1,2,3$.
It is expected that there is 1 independent skein relation.

\paragraph{i}
\begin{align}
&
c
\raisebox{-0.47\height}{\tikzsetnextfilename{-figure-A4-digon431}
\begin{tikzpicture}
\punctureindigonPP{$4$}{$3$}{$1$}{above}{$Y$}{0.6}{0.08}{1.2}
\end{tikzpicture}
}
+
c^{-1}
\;\raisebox{-0.47\height}{\tikzsetnextfilename{-figure-A4-digon413}
\begin{tikzpicture}
\punctureindigonPP{$4$}{$1$}{$3$}{above}{$Y$}{0.6}{0.08}{1.2}
\end{tikzpicture}
}
-
\raisebox{-0.47\height}{\tikzsetnextfilename{-figure-A4-digon422}
\begin{tikzpicture}
\punctureindigonPP{$4$}{$2$}{$2$}{above}{$Y$}{0.6}{0.08}{1.2}
\end{tikzpicture}
}=
c^{2}
\;\raisebox{-0.38\height}{\tikzsetnextfilename{-figure-A4-line4passleft}
\begin{tikzpicture}
\linenearpuncture{$4$}{right}{$Y$}{1.32}{-0.45}{0.08}
\end{tikzpicture}
}\quad
+
c^{-2}
\;\raisebox{-0.38\height}{\tikzsetnextfilename{-figure-A4-line4passright}
\begin{tikzpicture}
\linenearpuncture{$4$}{left}{$Y$}{1.32}{0.45}{0.08}
\end{tikzpicture}
}\quad.
\end{align}

%\end{comment}
\subsection{$A_5$ case}
\label{subsec:A5 case}

The final example is $A_5=\Liesu(6)$.
There are eleven types of regular punctures, eight of which are subjects. See the Table ~\ref{tab:A5 examples of leading exponent}.

\begin{center}
\begin{table}[htb]
\centering
\begin{tabular}{|c|c|c|c|c|c|}
\hline
$[1^6]$ & $[21^4]$ & $[2^21^2]$ & $[2^3]$ \\
\hline
[0, 0, 0, 0, 0] &
[-1, -1, -1, -1, -1] &
[-1, -2, -2, -2, -1] &
[-1, -2, -3, -2, -1] \\
\hline
[1, 2, 3, 4, 5] &
[1, 2, 3, 4, 4] &
[1, 2, 3, 3, 4] &
[1, 2, 2, 3, 4] \\
\hline
\hline
$[31^3]$ & $[321]$ & $[3^2]$ & $[41^2]$ \\
\hline
[-2, -2, -2, -2, -2] &
[-2, -3, -3, -3, -2] &
[-2, -4, -4, -4, -2] &
[-3, -4, -4, -4, -3] \\
\hline
[1, 2, 3, 3, 3] &
[1, 2, 2, 3, 3] &
[1, 1, 2, 2, 3] &
[1, 2, 2, 2, 2] \\
\hline
\hline
$[42]$ & $[51]$ & $[6]$ & \\
\hline
[-3, -4, -5, -4, -3] &
[-4, -6, -6, -6, -4] &
[-5, -8, -9, -8, -5] & \\
\hline
[1, 1, 2, 2, 2] &
[1, 1, 1, 1, 1] &
[0, 0, 0, 0, 0] & \\
\hline
\end{tabular}
\caption{The list for $\Lieg=\Liesu(6)$ ($A_5$) : the puncture type $Y$, $2\rho^{\alpha}_{L,Y}$ and the numbers of Coulomb branch operators with scaling dimensions $2,3,4,5,6$ in this order.}
\label{tab:A5 examples of leading exponent}
\end{table}
\end{center}

\begin{comment}
rank 6

\begin{align}
T^{(6)}_{A_5}=T^\calS[C([1^5],[1^5],[1^5])]
\end{align}

rank 5

\begin{align}
T^{(5)}_{A_5}=T^\calS[C([1^5],[1^5],[21^3])]
\end{align}

rank 4

\begin{align}
T^{(4-1)}_{A_5}=T^\calS[C([1^5],[1^5],[2^21])]
\qquad
T^{(4-2)}_{A_5}=T^\calS[C([1^5],[21^3],[21^3])]
\end{align}

rank 3

\begin{align}
T^{(3-1)}_{A_5}=T^\calS[C([1^5],[1^5],[31^3])]
\qquad
T^{(3-2)}_{A_5}=T^\calS[C([1^5],[21^3],[2^21])] \\
T^{(3-3)}_{A_5}=T^\calS[C([21^3],[21^3],[21^3])]
\end{align}

rank 2

\begin{align}
T^{(2-1)}_{A_5}=T^\calS[C([1^5],[1^5],[32])]
\qquad
T^{(2-2)}_{A_5}=T^\calS[C([1^5],[21^3],[31^2])] \\
T^{(2-3)}_{A_5}=T^\calS[C([21^3],[21^3],[2^21])]
\qquad
T^{(2-4)}_{A_5}=T^\calS[C([1^5],[2^21],[2^21])]
\end{align}

rank 1

\begin{align}
T^{(1-1)}_{A_5}=T^\calS[C([1^5],[21^3],[32])] \qquad T^{(1-2)}_{A_5}=T^\calS[C([1^5],[2^21],[31^2])]\\ T^{(1-3)}_{A_5}=T^\calS[C([21^3],[2^21],[2^21])]
\end{align}

rank 0 SCFT

\begin{align}
F^{\rm (bf)}_{A_5}=T^\calS[C([1^6],[1^6],[41])] \qquad F^{\rm (as)}_{A_5}=T^\calS[C([1^6],[321],[3^2])] \\
F^{\rm (ex1)}_{A_5}=T^\calS[C([1^6],[2^3],[42])] \qquad F^{\rm (ex2)}_{A_5}=T^\calS[C([21^4],[2^3],[3^2])]
\end{align}

\end{comment}

\subsubsection*{$\bf Y=[42]$}

The flavor symmetry associated with this puncture $Y=[42]$ is $S(U(1)\times U(1)) \simeq U(1)$.
The associated fugacity is given by $c_{[42]}=(q^{3/2}c_1,q^{1/2}c_1,q^{-1/2}c_1,q^{-3/2}c_1,q^{1/2}c_2,q^{-1/2}c_2)$.
In particular, we introduce $c:=c_1$.
It is expected that there are 7 independent skein relations.

\paragraph{i-1 \& i-2}
\begin{align}
\raisebox{-0.47\height}{\tikzsetnextfilename{-figure-A5-digon514}
\begin{tikzpicture}
\punctureindigonPP{$5$}{$1$}{$4$}{above}{$Y$}{0.6}{0.08}{1.2}
\end{tikzpicture}
}
=
([3]_q c^{-1} + c^2)
\;\raisebox{-0.38\height}{\tikzsetnextfilename{-figure-A5-line5passright}
\begin{tikzpicture}
\linenearpuncture{$5$}{left}{$Y$}{1.32}{0.45}{0.08}
\end{tikzpicture}
}\quad
+
c
\;\raisebox{-0.38\height}{\tikzsetnextfilename{-figure-A5-line5passleft}
\begin{tikzpicture}
\linenearpuncture{$5$}{right}{$Y$}{1.32}{-0.45}{0.08}
\end{tikzpicture}
}\quad
\end{align}

\begin{align}
\raisebox{-0.47\height}{\tikzsetnextfilename{-figure-A5-digon523}
\begin{tikzpicture}
\punctureindigonPN{$5$}{$2$}{$3$}{above}{$Y$}{0.6}{0.08}{1.2}
\end{tikzpicture}
}
=
[3]_q (c + c^{-2})
\;\raisebox{-0.38\height}{\tikzsetnextfilename{-figure-A5-line5passright}
\begin{tikzpicture}
\linenearpuncture{$5$}{left}{$Y$}{1.32}{0.45}{0.08}
\end{tikzpicture}
}\quad
+
([3]_q + c^3)
\;\raisebox{-0.38\height}{\tikzsetnextfilename{-figure-A5-line5passleft}
\begin{tikzpicture}
\linenearpuncture{$5$}{right}{$Y$}{1.32}{-0.45}{0.08}
\end{tikzpicture}
}\quad
\end{align}

\paragraph{ii}
\begin{align}
&\raisebox{-0.47\height}{\tikzsetnextfilename{-figure-A5-digon422}
\begin{tikzpicture}
\punctureindigonPP{$4$}{$2$}{$2$}{above}{$Y$}{0.6}{0.08}{1.2}
\end{tikzpicture}
}
-
[2]_qc
\;\raisebox{-0.47\height}{\tikzsetnextfilename{-figure-A5-digon431}
\begin{tikzpicture}
\punctureindigonNP{$4$}{$3$}{$1$}{above}{$Y$}{0.6}{0.08}{1.2}
\end{tikzpicture}
}
=
c^{-2}
\;\raisebox{-0.38\height}{\tikzsetnextfilename{-figure-A5-line4passright}
\begin{tikzpicture}
\linenearpuncture{$4$}{left}{$Y$}{1.32}{0.45}{0.08}
\end{tikzpicture}
}\quad
-
[3]_q \; c^2
\;\raisebox{-0.38\height}{\tikzsetnextfilename{-figure-A5-line4passleft}
\begin{tikzpicture}
\linenearpuncture{$4$}{right}{$Y$}{1.32}{-0.45}{0.08}
\end{tikzpicture}
}\quad.
\label{eq:skein [42]-type class ii}
\end{align}
and its mirror operated relation.

\paragraph{iii}
\begin{align}
c
\;\raisebox{-0.47\height}{\tikzsetnextfilename{-figure-A5-digon321}
\begin{tikzpicture}
\punctureindigonPPintoN{$3$}{$2$}{$1$}{above}{$Y$}{0.6}{0.08}{1.2}
\end{tikzpicture}
}
-
\raisebox{-0.47\height}{\tikzsetnextfilename{-figure-A5-digon312}
\begin{tikzpicture}
\punctureindigonPPintoN{$3$}{$1$}{$2$}{above}{$Y$}{0.6}{0.08}{1.2}
\end{tikzpicture}
}
=
c^{2}
\;\raisebox{-0.38\height}{\tikzsetnextfilename{-figure-A5-line3passleft}
\begin{tikzpicture}
\linewithnoarrownearpuncture{$3$}{right}{$Y$}{1.32}{-0.45}{0.08}
\end{tikzpicture}
}
-
c^{-1}
\;\raisebox{-0.38\height}{\tikzsetnextfilename{-figure-A5-line3passright}
\begin{tikzpicture}
\linewithnoarrownearpuncture{$3$}{left}{$Y$}{1.32}{0.45}{0.08}
\end{tikzpicture}
}\quad.
\label{eq:skein [42]-type class iii}
\end{align}

\subsubsection*{$\bf Y=[41^2]$}

The flavor symmetry associated with this puncture $Y=[41^2]$ is $S(U(1)\times U(2)) \simeq U(1) \times SU(2)$.
The associated fugacity is given by $c_{[41^2]}=(q^{3/2}c_1,q^{1/2}c_1,q^{-1/2}c_1,q^{-3/2}c_1,c_2,c_3)$.
In particular, we introduce $c:=c_1$.
It is expected that there are 6 independent skein relations.

\paragraph{i}
\begin{align}
c^{-3/2}
\;\raisebox{-0.47\height}{\tikzsetnextfilename{-figure-A5-digon514}
\begin{tikzpicture}
\punctureindigonPP{$5$}{$1$}{$4$}{above}{$Y$}{0.6}{0.08}{1.2}
\end{tikzpicture}
}
-
c^{3/2}
\raisebox{-0.47\height}{\tikzsetnextfilename{-figure-A5-digon541}
\begin{tikzpicture}
\punctureindigonPP{$5$}{$4$}{$1$}{above}{$Y$}{0.6}{0.08}{1.2}
\end{tikzpicture}
}
=
[3]_q\; \left[
c^{-5/2}
\;\raisebox{-0.38\height}{\tikzsetnextfilename{-figure-A5-line5passright}
\begin{tikzpicture}
\linenearpuncture{$5$}{left}{$Y$}{1.32}{0.45}{0.08}
\end{tikzpicture}
}\quad
-
c^{5/2}
\;\raisebox{-0.38\height}{\tikzsetnextfilename{-figure-A5-line5passleft}
\begin{tikzpicture}
\linenearpuncture{$5$}{right}{$Y$}{1.32}{-0.45}{0.08}
\end{tikzpicture}
}\quad
\right]
\end{align}
and
\begin{align}
c^{3}
\;\raisebox{-0.47\height}{\tikzsetnextfilename{-figure-A5-digon532}
\begin{tikzpicture}
\punctureindigonNP{$5$}{$3$}{$2$}{above}{$Y$}{0.6}{0.08}{1.2}
\end{tikzpicture}
}
-
[3]_qc^{4}
\;\raisebox{-0.47\height}{\tikzsetnextfilename{-figure-A5-digon541}
\begin{tikzpicture}
\punctureindigonPP{$5$}{$4$}{$1$}{above}{$Y$}{0.6}{0.08}{1.2}
\end{tikzpicture}
}
=
\;\raisebox{-0.38\height}{\tikzsetnextfilename{-figure-A5-line5passright}
\begin{tikzpicture}
\linenearpuncture{$5$}{left}{$Y$}{1.32}{0.45}{0.08}
\end{tikzpicture}
}\quad
-
c^{5}
\;\raisebox{-0.38\height}{\tikzsetnextfilename{-figure-A5-line5passleft}
\begin{tikzpicture}
\linenearpuncture{$5$}{right}{$Y$}{1.32}{-0.45}{0.08}
\end{tikzpicture}
}\quad.
\end{align}

\paragraph{ii}
We have the same skein relation ~\eqref{eq:skein [42]-type class ii}.

\paragraph{iii}
We have the same skein relation ~\eqref{eq:skein [42]-type class iii}.

\subsubsection*{$\bf Y=[3^2]$}

The flavor symmetry associated with this puncture $Y=[3^2]$ is $S(U(2)) \simeq SU(2)$.
The associated fugacity is given by $c_{[3^2]}=(qc,c,q^{-1}c,qc^{-1},c^{-1},qc^{-1})$.
It is expected that there are 6 independent skein relations.

\paragraph{i}
\begin{align}
\raisebox{-0.47\height}{\tikzsetnextfilename{-figure-A5-digon541}
\begin{tikzpicture}
\punctureindigonPP{$5$}{$4$}{$1$}{above}{$Y$}{0.6}{0.08}{1.2}
\end{tikzpicture}
}
=
\;\raisebox{-0.38\height}{\tikzsetnextfilename{-figure-A5-line5passright}
\begin{tikzpicture}
\linenearpuncture{$5$}{left}{$Y$}{1.32}{0.45}{0.08}
\end{tikzpicture}
}\quad
+
[2]_q \chi_{\rmbf{2}}^{SU(2)}(c)
\;\raisebox{-0.38\height}{\tikzsetnextfilename{-figure-A5-line5passleft}
\begin{tikzpicture}
\linenearpuncture{$5$}{right}{$Y$}{1.32}{-0.45}{0.08}
\end{tikzpicture}
}\quad
\end{align}
and
\begin{align}
\raisebox{-0.47\height}{\tikzsetnextfilename{-figure-A5-digon523}
\begin{tikzpicture}
\punctureindigonPN{$5$}{$2$}{$3$}{above}{$Y$}{0.6}{0.08}{1.2}
\end{tikzpicture}
}
=
([3]_q + \chi_{\rmbf{3}}^{SU(2)}(c))
\;\raisebox{-0.38\height}{\tikzsetnextfilename{-figure-A5-line5passright}
\begin{tikzpicture}
\linenearpuncture{$5$}{left}{$Y$}{1.32}{0.45}{0.08}
\end{tikzpicture}
}\quad
+
[2]_q \chi_{\rmbf{2}}^{SU(2)}(c)
\;\raisebox{-0.38\height}{\tikzsetnextfilename{-figure-A5-line5passleft}
\begin{tikzpicture}
\linenearpuncture{$5$}{right}{$Y$}{1.32}{-0.45}{0.08}
\end{tikzpicture}
}\quad
\end{align}
and their mirror operated relations.

\paragraph{ii}
\begin{align}
\chi^{SU(2)}_{\rmbf{2}}(c)
\;\raisebox{-0.47\height}{\tikzsetnextfilename{-figure-A5-digon431}
\begin{tikzpicture}
\punctureindigonNP{$4$}{$3$}{$1$}{above}{$Y$}{0.6}{0.08}{1.2}
\end{tikzpicture}
}
-
\raisebox{-0.47\height}{\tikzsetnextfilename{-figure-A5-digon422}
\begin{tikzpicture}
\punctureindigonPP{$4$}{$2$}{$2$}{above}{$Y$}{0.6}{0.08}{1.2}
\end{tikzpicture}
}
=
\chi_{\rmbf{3}}^{SU(2)}(c)
\;\raisebox{-0.38\height}{\tikzsetnextfilename{-figure-A5-line4passleft}
\begin{tikzpicture}
\linenearpuncture{$4$}{right}{$Y$}{1.32}{-0.45}{0.08}
\end{tikzpicture}
}\quad
-
\;\raisebox{-0.38\height}{\tikzsetnextfilename{-figure-A5-line4passright}
\begin{tikzpicture}
\linenearpuncture{$4$}{left}{$Y$}{1.32}{0.45}{0.08}
\end{tikzpicture}
}\quad
\end{align}
and its mirror operated relation.

\begin{comment}
<ab>=digon(a,b) with Y inside it
[3]\chi^{SU(2)}_3(c)<05>-[2]\chi_2<14>+<23>+<41>-[2]\chi_2<50>=0
[4]\chi_4<05>-[3]\chi_3<14>+[2]\chi_2<23>-<32>-<50>=0
\end{comment}

\subsubsection*{$\bf Y=[321]$}

The flavor symmetry associated with this puncture $Y=[321]$ is $S(U(1) \times U(1) \times U(1)) \simeq U(1)_1 \times U(1)_2$.
The associated fugacity is given by $c_{[321]}=(qc_1,c_1,q^{-1}c_1,q^{1/2}c_2,q^{-1/2}c_2,c_1^{-3}c_2^{-2})$.
It is expected that there are 4 independent skein relations.

\paragraph{i-1}
\begin{align}
\nonumber & [2]_qc_1^{3/2}
\;\raisebox{-0.47\height}{\tikzsetnextfilename{-figure-A5-digon541}
\begin{tikzpicture}
\punctureindigonPP{$5$}{$4$}{$1$}{above}{$Y$}{0.6}{0.08}{1.2}
\end{tikzpicture}
}
-
c_1^{1/2}
\;\raisebox{-0.47\height}{\tikzsetnextfilename{-figure-A5-digon532}
\begin{tikzpicture}
\punctureindigonNP{$5$}{$3$}{$2$}{above}{$Y$}{0.6}{0.08}{1.2}
\end{tikzpicture}
}
+
c_1^{-3/2}
\;\raisebox{-0.47\height}{\tikzsetnextfilename{-figure-A5-digon514}
\begin{tikzpicture}
\punctureindigonPP{$5$}{$1$}{$4$}{above}{$Y$}{0.6}{0.08}{1.2}
\end{tikzpicture}
} \\
&=
[3]_qc_1^{5/2}
\;\raisebox{-0.38\height}{\tikzsetnextfilename{-figure-A5-line5passleft}
\begin{tikzpicture}
\linenearpuncture{$5$}{right}{$Y$}{1.32}{-0.45}{0.08}
\end{tikzpicture}
}\quad
+
[2]_qc_1^{-5/2}
\;\raisebox{-0.38\height}{\tikzsetnextfilename{-figure-A5-line5passright}
\begin{tikzpicture}
\linenearpuncture{$5$}{left}{$Y$}{1.32}{0.45}{0.08}
\end{tikzpicture}
}\quad
\end{align}
and its mirror operated relation.

\paragraph{i-2}
\begin{align}
\nonumber & c_2^{3/2}
\;\raisebox{-0.47\height}{\tikzsetnextfilename{-figure-A5-digon541}
\begin{tikzpicture}
\punctureindigonPP{$5$}{$4$}{$1$}{above}{$Y$}{0.6}{0.08}{1.2}
\end{tikzpicture}
}
-
c_2^{1/2}
\;\raisebox{-0.47\height}{\tikzsetnextfilename{-figure-A5-digon532}
\begin{tikzpicture}
\punctureindigonNP{$5$}{$3$}{$2$}{above}{$Y$}{0.6}{0.08}{1.2}
\end{tikzpicture}
}
+
c_2^{-1/2}
\;\raisebox{-0.47\height}{\tikzsetnextfilename{-figure-A5-digon523}
\begin{tikzpicture}
\punctureindigonPN{$5$}{$2$}{$3$}{above}{$Y$}{0.6}{0.08}{1.2}
\end{tikzpicture}
}
-
c_2^{-3/2}
\;\raisebox{-0.47\height}{\tikzsetnextfilename{-figure-A5-digon514}
\begin{tikzpicture}
\punctureindigonPP{$5$}{$1$}{$4$}{above}{$Y$}{0.6}{0.08}{1.2}
\end{tikzpicture}
} \\
&=
c_2^{5/2}
\;\raisebox{-0.38\height}{\tikzsetnextfilename{-figure-A5-line5passleft}
\begin{tikzpicture}
\linenearpuncture{$5$}{right}{$Y$}{1.32}{-0.45}{0.08}
\end{tikzpicture}
}\quad
-
c_2^{-5/2}
\;\raisebox{-0.38\height}{\tikzsetnextfilename{-figure-A5-line5passright}
\begin{tikzpicture}
\linenearpuncture{$5$}{left}{$Y$}{1.32}{0.45}{0.08}
\end{tikzpicture}
}\quad.
\end{align}

\paragraph{ii}
\begin{align}
c_1^{-1}
\;\raisebox{-0.47\height}{\tikzsetnextfilename{-figure-A5-digon413}
\begin{tikzpicture}
\punctureindigonPN{$4$}{$1$}{$3$}{above}{$Y$}{0.6}{0.08}{1.2}
\end{tikzpicture}
}
+
c_1
\;\raisebox{-0.47\height}{\tikzsetnextfilename{-figure-A5-digon431}
\begin{tikzpicture}
\punctureindigonNP{$4$}{$3$}{$1$}{above}{$Y$}{0.6}{0.08}{1.2}
\end{tikzpicture}
}
-
\;\raisebox{-0.47\height}{\tikzsetnextfilename{-figure-A5-digon422}
\begin{tikzpicture}
\punctureindigonPP{$4$}{$2$}{$2$}{above}{$Y$}{0.6}{0.08}{1.2}
\end{tikzpicture}
}
=
c_1^{-2}
\;\raisebox{-0.38\height}{\tikzsetnextfilename{-figure-A5-line4passright}
\begin{tikzpicture}
\linenearpuncture{$4$}{left}{$Y$}{1.32}{0.45}{0.08}
\end{tikzpicture}
}\quad
+
c_1^2
\;\raisebox{-0.38\height}{\tikzsetnextfilename{-figure-A5-line4passleft}
\begin{tikzpicture}
\linenearpuncture{$4$}{right}{$Y$}{1.32}{-0.45}{0.08}
\end{tikzpicture}
}\quad.
\label{eq:skein [321]-type class ii}
\end{align}

\subsubsection*{$\bf Y=[31^3]$}

The flavor symmetry associated with this puncture $Y=[31^3]$ is $S(U(1) \times U(3)) \simeq U(1) \times SU(3)$.
The associated fugacity is given by $c_{[31^3]}=(qc_1,c_1,q^{-1}c_1,c_2,c_3,c_4)$.
In particular, we introduce the natural recombined fugacities $c:=c_1$.
It is expected that there are 3 independent skein relations.

\paragraph{i}
\begin{align}
\nonumber & c^{-3/2}
\;\raisebox{-0.47\height}{\tikzsetnextfilename{-figure-A5-digon514}
\begin{tikzpicture}
\punctureindigonPP{$5$}{$1$}{$4$}{above}{$Y$}{0.6}{0.08}{1.2}
\end{tikzpicture}
}
+
[2]_qc^{3/2}
\;\raisebox{-0.47\height}{\tikzsetnextfilename{-figure-A5-digon541}
\begin{tikzpicture}
\punctureindigonPP{$5$}{$4$}{$1$}{above}{$Y$}{0.6}{0.08}{1.2}
\end{tikzpicture}
}
-
c^{1/2}
\;\raisebox{-0.47\height}{\tikzsetnextfilename{-figure-A5-digon532}
\begin{tikzpicture}
\punctureindigonNP{$5$}{$3$}{$2$}{above}{$Y$}{0.6}{0.08}{1.2}
\end{tikzpicture}
} \\
&=
[3]_qc^{5/2}
\;\raisebox{-0.38\height}{\tikzsetnextfilename{-figure-A5-line5passleft}
\begin{tikzpicture}
\linenearpuncture{$5$}{right}{$Y$}{1.32}{-0.45}{0.08}
\end{tikzpicture}
}\quad
+
[2]_qc^{-5/2}
\;\raisebox{-0.38\height}{\tikzsetnextfilename{-figure-A5-line5passright}
\begin{tikzpicture}
\linenearpuncture{$5$}{left}{$Y$}{1.32}{0.45}{0.08}
\end{tikzpicture}
}\quad
\end{align}
and its mirror operated relation.

\paragraph{ii}
We have the same skein relation ~\eqref{eq:skein [321]-type class ii} with the replacement of $c_1$ by $c$.

\subsubsection*{$\bf Y=[2^3]$}

The flavor symmetry associated with this puncture $Y=[2^3]$ is $S(U(3)) \simeq SU(3)$.
The associated fugacity is given by $c_{[2^3]}=(q^{1/2}c_1,q^{-1/2}c_1,q^{1/2}c_2,q^{-1/2}c_2,q^{1/2}c_3,q^{-1/2}c_3)$ ($c_1c_2c_3=1$).
It is expected that there are 3 independent skein relations.

\paragraph{i}
\begin{align}
\raisebox{-0.47\height}{\tikzsetnextfilename{-figure-A5-digon532}
\begin{tikzpicture}
\punctureindigonNP{$5$}{$3$}{$2$}{above}{$Y$}{0.6}{0.08}{1.2}
\end{tikzpicture}
}
-
\chi_{\rmbf{3}}^{SU(3)}(c)
\;\raisebox{-0.47\height}{\tikzsetnextfilename{-figure-A5-digon541}
\begin{tikzpicture}
\punctureindigonPP{$5$}{$4$}{$1$}{above}{$Y$}{0.6}{0.08}{1.2}
\end{tikzpicture}
}
=
\;\raisebox{-0.38\height}{\tikzsetnextfilename{-figure-A5-line5passright}
\begin{tikzpicture}
\linenearpuncture{$5$}{left}{$Y$}{1.32}{0.45}{0.08}
\end{tikzpicture}
}\quad
-
\chi_{\rmbf{6}}^{SU(3)}(c)
\;\raisebox{-0.38\height}{\tikzsetnextfilename{-figure-A5-line5passleft}
\begin{tikzpicture}
\linenearpuncture{$5$}{right}{$Y$}{1.32}{-0.45}{0.08}
\end{tikzpicture}
}\quad
\label{eq:skein [222]-type class i-1}
\end{align}
and its mirror operated relation.

\begin{align}
\raisebox{-0.47\height}{\tikzsetnextfilename{-figure-A5-digon514}
\begin{tikzpicture}
\punctureindigonPP{$5$}{$1$}{$4$}{above}{$Y$}{0.6}{0.08}{1.2}
\end{tikzpicture}
}
-
\raisebox{-0.47\height}{\tikzsetnextfilename{-figure-A5-digon541}
\begin{tikzpicture}
\punctureindigonPP{$5$}{$4$}{$1$}{above}{$Y$}{0.6}{0.08}{1.2}
\end{tikzpicture}
}
=
\chi_{\rmbf{\ovl{3}}}^{SU(3)}(c)
\;\raisebox{-0.38\height}{\tikzsetnextfilename{-figure-A5-line5passright}
\begin{tikzpicture}
\linenearpuncture{$5$}{left}{$Y$}{1.32}{0.45}{0.08}
\end{tikzpicture}
}\quad
-
\chi_{\rmbf{3}}^{SU(3)}(c)
\;\raisebox{-0.38\height}{\tikzsetnextfilename{-figure-A5-line5passleft}
\begin{tikzpicture}
\linenearpuncture{$5$}{right}{$Y$}{1.32}{-0.45}{0.08}
\end{tikzpicture}
}\quad.
\label{eq:skein [222]-type class i-2}
\end{align}

\subsubsection*{$\bf Y=[2^21^2]$}

The flavor symmetry associated with this puncture $Y=[2^21^2]$ is $S(U(2)\times U(2)) \simeq U(1) \times SU(2)_1 \times SU(2)_2$.
The associated fugacity is given by $c_{[2^21^2]}=(q^{1/2}c_1,q^{-1/2}c_1,q^{1/2}c_2,q^{-1/2}c_2,c_3,c_4)$.
In particular, we introduce the natural recombined fugacities $c:=(c_1c_2)^{1/2}$ and $c'_i:=c_i/c$ (or $c'_1=(c_1/c_2)^{1/2}$ and $c'_2=(c_2/c_1)^{1/2}$).
It is expected that there are 2 independent skein relations.

\paragraph{i}
\begin{align}
\nonumber & c^{-2}
\;\raisebox{-0.47\height}{\tikzsetnextfilename{-figure-A5-digon514}
\begin{tikzpicture}
\punctureindigonPP{$5$}{$1$}{$4$}{above}{$Y$}{0.6}{0.08}{1.2}
\end{tikzpicture}
}
+
c\chi^{SU(2)}_{\rmbf{2}}(c')
\;\raisebox{-0.47\height}{\tikzsetnextfilename{-figure-A5-digon541}
\begin{tikzpicture}
\punctureindigonPP{$5$}{$4$}{$1$}{above}{$Y$}{0.6}{0.08}{1.2}
\end{tikzpicture}
}
-
\;\raisebox{-0.47\height}{\tikzsetnextfilename{-figure-A5-digon532}
\begin{tikzpicture}
\punctureindigonNP{$5$}{$3$}{$2$}{above}{$Y$}{0.6}{0.08}{1.2}
\end{tikzpicture}
}\\
&=
c^{-3}\chi_{\rmbf{2}}^{SU(2)}(c')
\;\raisebox{-0.38\height}{\tikzsetnextfilename{-figure-A5-line5passright}
\begin{tikzpicture}
\linenearpuncture{$5$}{left}{$Y$}{1.32}{0.45}{0.08}
\end{tikzpicture}
}\quad
+
c^2\chi_{\rmbf{3}}^{SU(2)}(c')
\;\raisebox{-0.38\height}{\tikzsetnextfilename{-figure-A5-line5passleft}
\begin{tikzpicture}
\linenearpuncture{$5$}{right}{$Y$}{1.32}{-0.45}{0.08}
\end{tikzpicture}
}\quad
\end{align}
and its mirror operated relation.

\subsubsection*{$\bf Y=[21^4]$}

The flavor symmetry associated with this puncture $Y=[21^4]$ is $S(U(1)\times U(4)) \simeq U(1) \times SU(4)$.
The associated fugacity is given by $c_{[21^4]}=(q^{1/2}c_1,q^{-1/2}c_1,c_2,c_3,c_4,c_5)$.
In particular, we introduce the natural recombined fugacities $c:=c_1$.% and $c'_i:=c_{i+1}c^{1/2}$ for $i=1,2,3,4$.
It is expected that there is 1 independent skein relation.

\paragraph{i}
\begin{align}
\nonumber & c^{3/2}
\;\raisebox{-0.47\height}{\tikzsetnextfilename{-figure-A5-digon541}
\begin{tikzpicture}
\punctureindigonPP{$5$}{$4$}{$1$}{above}{$Y$}{0.6}{0.08}{1.2}
\end{tikzpicture}
}
-
c^{-3/2}
\;\raisebox{-0.47\height}{\tikzsetnextfilename{-figure-A5-digon514}
\begin{tikzpicture}
\punctureindigonPP{$5$}{$1$}{$4$}{above}{$Y$}{0.6}{0.08}{1.2}
\end{tikzpicture}
}
+
c^{-1/2}
\;\raisebox{-0.47\height}{\tikzsetnextfilename{-figure-A5-digon523}
\begin{tikzpicture}
\punctureindigonPN{$5$}{$2$}{$3$}{above}{$Y$}{0.6}{0.08}{1.2}
\end{tikzpicture}
}
-
c^{1/2}
\;\raisebox{-0.47\height}{\tikzsetnextfilename{-figure-A5-digon532}
\begin{tikzpicture}
\punctureindigonNP{$5$}{$3$}{$2$}{above}{$Y$}{0.6}{0.08}{1.2}
\end{tikzpicture}
}\\
&=
c^{5/2}
\;\raisebox{-0.38\height}{\tikzsetnextfilename{-figure-A5-line5passleft}
\begin{tikzpicture}
\linenearpuncture{$5$}{right}{$Y$}{1.32}{-0.45}{0.08}
\end{tikzpicture}
}\quad
-
c^{-5/2}
\;\raisebox{-0.38\height}{\tikzsetnextfilename{-figure-A5-line5passright}
\begin{tikzpicture}
\linenearpuncture{$5$}{left}{$Y$}{1.32}{0.45}{0.08}
\end{tikzpicture}
}\quad.
\end{align}

\subsubsection*{Reproduction of previous results on flavor Wilson loops}

Using the above skein relations, we can reproduce ~\eqref{eq:flavor Wilson for A5 ex2-type free hypers},~\eqref{eq:flavor Wilson for A5 ex1-type free hypers} and ~\eqref{eq:flavor Wilson for rank 1 E8 SCFT}.

%\subsection{L-shaped punctures}

\subsection{General punctures}
\label{subsec:General punctures}

At this stage, we write down a formula to unify the above skein relations. We expect that this formula is valid for every puncture in each rank in the $A=\Liesu$ type.

\paragraph{Notations and symbols}
Let $Y$ be $[n_1,n_2,\ldots,n_k]=[m_1^{d_1}m_2^{d_2}\cdots m_t^{d_t}]$-type.
Then, we have a Young diagram whose $i$-th column is given by $n_i$.
To each box $y=\Box_{(a_y,h_y)} \in Y$ ($a_y$ and $h_y$ specify the position of a box $y$ from the left and the below, respectively ), we assign infinitely many digon type skein relations as we explain later.
They are not linearly independent each other and, indeed, there are only finite independent skein relations.

\begin{table}[htb]
\begin{tabular}{cc}
\begin{minipage}{0.2\hsize}
\centering
$Y=[32^21]$ \\
$d=\{1,2,1\}$ \\
\def\YsetH#1{\YsetSR{0.75}{1em}{#1}}
\def\Ya{\YsetH{\Large $a$}}
\def\Yb{\YsetH{\Large $b$}}
\def\Yc{\YsetH{\Large $c$}}
\def\Yd{\YsetH{\Large $d$}}
\def\Ye{\YsetH{\Large $e$}}
\def\Yf{\YsetH{\Large $f$}}
\def\Yg{\YsetH{\Large $g$}}
\def\Yh{\YsetH{\Large $h$}}
\rotateL{\young(\Ya\Yb\Yc,\Yd\Ye,\Yf\Yg,\Yh)}
\end{minipage}
&
\begin{minipage}{0.8\hsize}
\centering
\begin{tabular}{|c|cccccccc|}
\hline
$y_{(a,h)}$ & $a_{(1,1)}$ & $b_{(1,2)}$ & $c_{(1,3)}$ & $d_{(2,1)}$ & $e_{(2,2)}$ & $f_{(3,1)}$ & $g_{(3,2)}$ & $h_{(4,1)}$ \\
\hline\hline
$f_y$ & $1$ & $1$ & $1$ & $2$ & $2$ & $2$ & $2$ & $3$ \\
$g_y$ & $1$ & $1$ & $1$ & $1$ & $1$ & $2$ & $2$ & $1$ \\
\hline
$\ell_y$ & $3$ & $2$ & $1$ & $2$ & $1$ & $2$ & $1$ & $1$ \\
$c_y$ & $8$ & $7$ & $6$ & $8$ & $7$ & $8$ & $7$ & $8$ \\
\hline
\end{tabular}
\end{minipage}
\end{tabular}
\caption{The example of the associated values for the $[3221]=[32^21]$-type puncture in $A_7=\Liesu(8)$. The last line in the right table represents the number of independent skein relations. When $c_y=8$, the relations are trivial in the sense of the simplicity shown in ~\eqref{eq:cases digon reduction to just loop}. Otherwise, they are non-trivial ones. $\ell_y$ is the number of linearly independent skein relations. In this case, there are five non-trivial skein relations. Notice that $d_C^{[1^8]} -d_C^Y=5$. See also the later explanations.}
\label{table:puncture type to associated values}
\end{table}

To state the rule generating skein relations, let us introduce symbols $f_y=f(\Box_{(a,h)}):=\{ n \in \{1,2,\ldots,t\} \mbox{ s.t. } \sum_{s=1}^{n-1}d_s < a \le \sum_{s=1}^{n}d_s \}$, $g_y=g(\Box_{(a,h)}):=a-\sum_{s=1}^{f_y-1} d_s$ and $\ell_y:=m_{f_y}-h+1=n_i-h+1$.
This assignment corresponds to the $g_y$-th weight of the $f_y$-th flavor symmetry $U(d_{f_y})$ and the $\ell_y$-th weight of the $m_{f_y}$-dimensional irreducible representation of $SU(2)_R^{IR}$.
In addition, we use $c_y:=N+1-h_y$.
See Table.~\ref{table:puncture type to associated values} for an example.

Using the same notations introduced above, the fugacity $x_Y$ associated with the puncture $Y$ is given by $\{q^{\frac{m_{f_y}+1}{2}-\ell_y}x^{U(d_{f_y})_{f_y}}_{g_y}=q^{\frac{m_{f_y}+1}{2}-\ell_y}x^{SU(d_{f_y})_{f_y}}_{g_y}x^{U(1)_{f_y}} \}_{y \in Y}$.
$x^{SU(d_s)_{d_s}}_{i=1,2,\ldots,N}$ is the fugacity of $SU(d_s)$ \ie belongs to the Cartan of it and satisfies $\prod_{i=1}^N x^{SU(d_s)_{d_s}}_{i}=1$.
$x^{U(1)_{s}}$ is that of the $s$-th $U(1)$ for $s=1,2,\ldots,t$ but there is a constraint $\prod_{s=1}^{t} (x^{U(1)_{s}})^{m_sd_s}=1$ coming from the fact that $x_Y$ is the fugacity of $SU(N)$.

\begin{table}[htb]
\begin{tabular}{cc}
\begin{minipage}{0.35\hsize}
\centering
\def\YsetH#1{\YsetSR{0.8}{1.15em}{#1}}
\def\Ya{\YsetH{\footnotesize $q^{-1}z_1$}}
\def\Yb{\YsetH{\footnotesize $\;\;\;z_1$}}
\def\Yc{\YsetH{\footnotesize $\;qz_1$}}
\def\Yd{\YsetH{\footnotesize $q^{-\frac{1}{2}}z_2$}}
\def\Ye{\YsetH{\footnotesize $\;q^{\frac{1}{2}}z_2$}}
\def\Yf{\YsetH{\footnotesize $q^{-\frac{1}{2}}z_3$}}
\def\Yg{\YsetH{\footnotesize $\;q^{\frac{1}{2}}z_3$}}
\def\Yh{\YsetH{\footnotesize $\;\;\;z_4$}}
\scalebox{1.4}{\rotateL{\LARGE\young(\Ya\Yb\Yc,\Yd\Ye,\Yf\Yg,\Yh)}}
\end{minipage}
&
\begin{minipage}{0.65\hsize}
\centering
$\rmbf{8} \longrightarrow \rmbf{3}\otimes \bbC^{U(1)_1}_{1} \oplus \rmbf{2}\otimes \rmbf{2}^{U(2)_2}_{1} \oplus \rmbf{1}\otimes \bbC^{U(1)_3}_{1}$ \\
under $SU(2)_R \times S(U(1)_1 \times U(2)_2 \times U(1)_3)$
\end{minipage}
\end{tabular}
\caption{For $Y=[32^21]$, the $SU(8)$ fugacity $x_{[32^21]}$ consists of the above ones. Here we introduce the simplified notations as $z_1=x^{U(1)_1}_1=c_1$,$z_2=x^{U(2)_2}_1=c_2c$,$z_3=x^{U(2)_2}_2=c_2c^{-1}$ and $z_4=x^{U(1)_3}_1=c_3$ where $z_1^3z_2^2z_3^2z_4=c_1^3c_2^4c_3=1$. On the right hand side, the decomposition of the define representation ($\Box=\rmbf{8}=R(\omega_1)$) is shown.}
\label{table:fugacity assignment}
\end{table}

Next, we assign the $U(N)$ character for any element of the $\Liesu(N)$ weight lattice.
Using the Weyl's character formula, we can define it as
\begin{align}
\wtl{\chi}^{U(d)}_{\lambda}(x):=
\dfrac{\displaystyle \sum_{w\in \calW} \sigma_w x^{w(\rho+\lambda)}}{\displaystyle \sum_{w\in \calW} \sigma_w x^{w(\rho)}}
=
(x_0)^{|\lambda|}\;
\dfrac{\displaystyle \sum_{w\in \calW} \sigma_w X^{w(\rho+\lambda)}}{\displaystyle \sum_{w\in \calW} \sigma_w X^{w(\rho)}}
\end{align}
where $x=x^{U(N)}=(x_0X_1,x_0X_2,\ldots,x_0X_N)=x_0X^{SU(N)}$ with the constraint $\prod_{i=1}^N X_i=1$, $\calW$ is the Weyl group of $\Liesu(N)$ which is equivalent to the order $N$ permutation group and $\sigma_w=(-1)^{\ell(w)}$. $\ell(w)$ is the length of $w$ \ie the minimum number of generators of $\calW$ to generate $w$.

Let us see some examples.
In the case $\Lieu(2)$,
\begin{align}
\wtl{\chi}_{(n-1)\omega_1}=x_0^{n-1} \chi^{SU(2)}_{\rmbf{n}}
\qquad
\wtl{\chi}_{-\omega_1}=0
\qquad
\wtl{\chi}_{(-n-1)\omega_1}=-x_0^{-n-1}\chi^{SU(2)}_{\rmbf{n}}
\end{align}
for $n=1,2,\ldots$. In the case $\Lieu(3)$,
\begin{align}
\wtl{\chi}_{n\omega_1+m\omega_2}=x_0^{n+m}\times
\begin{cases}
\chi_{R(n\omega_1+m\omega_2)} \quad& \mathbf{I}\;:\; n,m \ge 0\\
-\chi_{R((-n-2)\omega_1+(n+m+1)\omega_2)} \quad& \mathbf{II}\;:\; n \le -2,n+m\ge -1 \\
\chi_{R(m\omega_1+(-n-m-3)\omega_2)} \quad& \mathbf{III}\;:\; m \ge 0,n+m\le -3 \\
-\chi_{R((-m-2)\omega_1+(-n-2)\omega_2)} \quad& \mathbf{IV}\;:\; n,m \le -2\\
\chi_{R((-n-m-3)\omega_1+n\omega_2)} \quad& \mathbf{V}\;:\; n \ge 0,n+m\le -3 \\
-\chi_{R((n+m+1)\omega_1+(-m-2)\omega_2)} \quad& \mathbf{VI}\;:\; m \le -2,n+m\ge -1 \\
0 \quad& \mbox{ otherwise}.
\end{cases}
\end{align}

\begin{figure}[htb]
\centering
\tikzsetnextfilename{-figure-A2_extended_weights.tikz}
\begin{tikzpicture}
\def\l{0.5}
\def\ll{3^0.5*\l}
\def\rad{0.05}

\coordinate (O) at (0,0);

\draw[black,thin,->] ($(60:\ll)+(30:-3.3*\ll)$) --++ (30:5.8*\ll) node[above right]{$n$};
\draw[black,thin,->] ($(60:\ll)+(90:-6.6*\l)$) --++ (90:10.3*\l) node[above]{$m$};
\draw[black,thin] ($(60:\ll)+(-30:-4.4*\l)$) --++ (-30:8.8*\l);

\fill[black] (O) circle (\rad);
\foreach \t in {0,60,...,300}{
\draw[dotted,thick] (\t:\ll) --++(\t-30:3.3*\l);
\draw[dotted,thick] (\t:\ll) --++(\t+30:3.3*\l);
\foreach \n in {1,...,3}{\fill[red] (\t:\n *\ll) circle (\rad);}
\foreach \n in {1,...,4}{\fill[black] (\t+30:\n*\l) circle (\rad);}
\fill[red] ($(\t:\ll)+(\t+30:3*\l)$) circle (\rad);
\fill[red] ($(\t:\ll)+(\t-30:3*\l)$) circle (\rad);
}

\foreach \t in {0,120,240}{
\starpoint{($(\t+60:\ll)+(\t+30:\l)$)}{green}{1.1*\rad}{90}
\starpoint{($(\t+60:2*\ll)+(\t+30:\l)$)}{green}{1.1*\rad}{90}
\starpoint{($(\t+60:\ll)+(\t+90:2*\l)$)}{green}{1.1*\rad}{90}
\starpoint{($(\t+120:\ll)+(\t+150:\l)$)}{green}{1.1*\rad}{90}
\starpoint{($(\t+120:2*\ll)+(\t+150:\l)$)}{green}{1.1*\rad}{90}
\starpoint{($(\t+120:\ll)+(\t+90:2*\l)$)}{green}{1.1*\rad}{90}
\trianglepoint{($(\t+60:\ll)+(\t+90:\l)$)}{blue}{1.1*\rad}{90}
\trianglepoint{($(\t+60:2*\ll)+(\t+90:\l)$)}{blue}{1.1*\rad}{90}
\trianglepoint{($(\t+60:\ll)+(\t+30:2*\l)$)}{blue}{1.1*\rad}{90}
\trianglepoint{($(\t+120:\ll)+(\t+90:\l)$)}{blue}{1.1*\rad}{90}
\trianglepoint{($(\t+120:2*\ll)+(\t+90:\l)$)}{blue}{1.1*\rad}{90}
\trianglepoint{($(\t+120:\ll)+(\t+150:2*\l)$)}{blue}{1.1*\rad}{90}
}

\node at (45:3.1*\ll) {$\bf I$};
\node at (135:3.1*\ll) {$\bf II$};
\node at (190:3.3*\ll) {$\bf III$};
\node at (230:3.3*\ll) {$\bf IV$};
\node at (310:3.3*\ll) {$\bf V$};
\node at (-10:3.3*\ll) {$\bf VI$};

\node at ($(60:\ll)+(30:\l)$) [below,green]{$\rmbf{3}$};
\node at ($(60:\ll)+(90:\l)$) [left,blue]{$\rmbf{\ovl{3}}$};
\node at ($(60:1.8*\ll)$) [above right,red]{$\rmbf{8}$};

\def\l{0.7}
\def\rad{0.07}
\def\h{0.25}
\coordinate (O) at (9,1);
\coordinate (P) at ($(O)+(120:\h)$);
\coordinate (Q) at ($(O)+(-60:\h)$);

\draw[black,thin,->] ($(O)+(30:-6.5*\l)$) --++ (30:10*\l) node[above right]{$n$};

\foreach \n in {-2,-1,0,1}{
\fill[red] ($(O)+(30:3*\n*\l)$) circle (\rad);
}
\foreach \n in {-2,-1}{
\fill[black] ($(O)+(30:\n*\l)$) circle (\rad);
}
\foreach \n in {-2,0}{
\starpoint{($(O)+(30:3*\n*\l+\l)$)}{green}{1.1*\rad}{90}
\trianglepoint{($(O)+(30:3*\n*\l+2*\l)$)}{blue}{1.1*\rad}{90}
}

\node at (O) [below right,red]{$\rmbf{1}$};
\node at ($(O)+(30:\l)$) [below right,green]{$\rmbf{3}$};
\node at ($(O)+(30:2*\l)$) [below right,blue]{$\rmbf{6}$};
\node at ($(O)+(30:3*\l)$) [below right,red]{$\rmbf{10}$};
\node at ($(O)+(30:-3*\l)$) [below right,red]{$\rmbf{1}$};
\node at ($(O)+(30:-4*\l)$) [below right,blue]{$\rmbf{\ovl{3}}$};
\node at ($(O)+(30:-5*\l)$) [below right,green]{$\rmbf{\ovl{6}}$};
\node at ($(O)+(30:-6*\l)$) [below right,red]{$\rmbf{\ovl{10}}$};

\draw[dotted,purple,ultra thick] ($(P)+(120:0.04)+(30:2.5*\l)$) --++ (-60:2*\h) --++ (210:6*\l) --++ (120:2*\h) --++ (30:6*\l) node[above left]{$(C)$};

\draw[dotted,orange,ultra thick] ($(P)+(120:0.08)+(30:0.5*\l)$) --++ (-60:2*\h) --++ (210:6*\l) --++ (120:2*\h) --++ (30:6*\l) node[above left]{$(B)$};

\draw[dotted,cyan,ultra thick] ($(P)+(120:0.12)+(30:1.5*\l)$) --++ (-60:2*\h) --++ (210:6*\l) --++ (120:2*\h) --++ (30:6*\l) node[above left]{$(A)$};
\end{tikzpicture}
\caption{The $\Liesu(3)$ weight lattice (Left). $\color{black} \bullet$ gives $0$. $\color{red} \bullet$, $\color{green} \star$ and \protect\scalebox{0.7}{$\color{blue} \blacktriangle$} respectively correspond to the charges $0,1$ and $2$ under the center symmetry $\bbZ_3$. On the walls, namely, the boundaries of the Weyl chambers, the characters vanish. The prefactors of the skein relations ~\eqref{eq:skein [222]-type class i-1}, its mirror and ~\eqref{eq:skein [222]-type class i-2} correspond to the dotted boxes $(C)$,$(A)$ and $(B)$ respectively (Right).}
\label{fig:A2_extended_weights}
\end{figure}
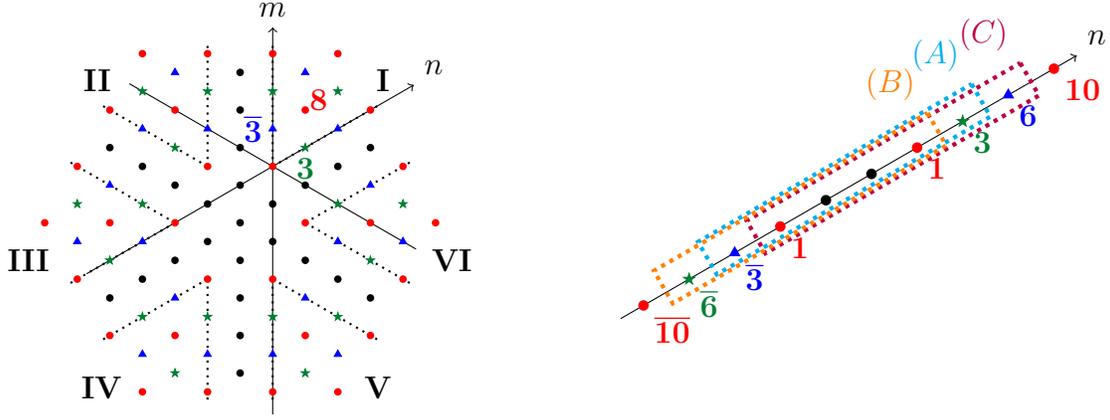
%$\scalebox{0.8}{\color{blue} \blacktriangle}$

\paragraph{Conjectural unified form of skein relations}
Now, the skein relations are given by
\begin{align}
DSR^{Y}_{\{b\},\beta} \;:\;
\sum_{p=0}^{c_y}(-1)^p \prod_{i=1}^{\ell_y-1}[p+b_i]_q\wtl{\chi}^{U(d_{f_y})}_{p\omega_1+\omega_{g_y}+\beta}(x)
\raisebox{-0.47\height}{\tikzsetnextfilename{-figure-general_digon}
\begin{tikzpicture}
\punctureindigonPP{$c_y$}{$p$}{$c_y-p$}{above}{$Y$}{0.6}{0.08}{1.2}
\end{tikzpicture}
}=0
\end{align}
where $b_1,b_2,\ldots,b_{\ell_y-1}$ are arbitrary integers and $\beta$ is an arbitrary element of the $\Liesu(N)$ root lattice. We may absorb $\omega_{g_y}$ to $\beta$ and instead $\beta$ is any element of weight lattice, where two distinct $y$ with same $f_y$ and $h_y$ give the same skein relations.
$\prod_{i=0}^{1}[p+b_1]_q=:1$ when $\ell_y=1$ and $\omega_0:=0$.
Notice also that
\begin{align}
\raisebox{-0.47\height}{\tikzsetnextfilename{-figure-digon_with_zero_L}
\begin{tikzpicture}
\punctureindigonPP{$p$}{$0$}{$p$}{above}{$Y$}{0.6}{0.08}{1.2}
\end{tikzpicture}
}=\raisebox{-0.38\height}{\tikzsetnextfilename{-figure-linep_passright}
\begin{tikzpicture}
\linenearpuncture{$p$}{left}{$Y$}{1.32}{0.45}{0.08}
\end{tikzpicture}
}
\qquad\qquad
\raisebox{-0.47\height}{\tikzsetnextfilename{-figure-digon_with_zero_R}
\begin{tikzpicture}
\punctureindigonPP{$p$}{$p$}{$0$}{above}{$Y$}{0.6}{0.08}{1.2}
\end{tikzpicture}
}=\raisebox{-0.38\height}{\tikzsetnextfilename{-figure-linep_passleft}
\begin{tikzpicture}
\linenearpuncture{$p$}{right}{$Y$}{1.32}{-0.45}{0.08}
\end{tikzpicture}
}.
\label{eq:cases digon reduction to just loop}
\end{align}

\paragraph{Linearly independent choice}
Although we generate infinitely many skein relations for digons with generic puncture inside it, there are linearly dependent.
Let us remark on the choices of linearly independent skein relations.

Since $[k]_q[2]_q=[K+1]_q+[k-1]_q$ for any integer $k$, the three relations $DSR^{Y}_{\{b\},\beta}$,$DSR^{Y}_{\{b'\},\beta}$ and $DSR^{Y}_{\{b''\},\beta}$ where $b_j=b'_j+1=b''_j$ for some $j$ and $b_i=b_i'=b''_j$ for all $i( \neq j)$ are not linearly independent. Then we can restrict $b$ form $\bbZ^{\ell-1}$ onto $\{1,2\}^{\ell-1}$ for example.
Furthermore, there is the permutation symmetry acting on $b_i$.
Combining two properties, we expect that the representative skein relations are labelled by the set $\{1,2\}^{\ell-1}$ divided by the permutation.
The complete invariant on this set is the summation of $\{b\}$ and we see that there are $\ell$ relations at least to generate all $b$.
In addition, we can find that they are linearly independent in general cases.
In the same way, we see that $\wtl{\chi}^{U(d)}_{p\omega_1+\omega_g}$ with $g=0,1,2,\ldots,d-1$ give the linearly independent skein relations and generate all the relations with any $\beta$.

Notice also that, when $h=1$, we also have the trivial skein relations :
\begin{align}
\raisebox{-0.47\height}{\tikzsetnextfilename{-figure-digon_equal_to_loop.tikz}
\begin{tikzpicture}
\punctureindigonPP{$N$}{$N-a$}{$a$}{above}{$Y$}{0.6}{0.08}{1.2}
\end{tikzpicture}
}
=
\raisebox{-0.45\height}{\tikzsetnextfilename{-figure-loop_winding_puncture_on_C}
\begin{tikzpicture}
\def\r{0.08}
\def\rad{0.6}
\def\l{1.0}
\coordinate (C) at (0,0);
\coordinate (P) at ($(C)+(0:\rad)$);

\draw[fill=white] (C) circle (1.8*\r) node[above]{$Y$};
\draw[black,thick] (C) circle (1.8*\r);
\draw[fill=black] (C) circle (\r);

\draw[thick] (C) circle (\rad);
\circlearrow{(C)}{\rad}{180}{0.3}{left}{$a$}
\end{tikzpicture}
}
=
\chi^{SU(N)}_{\wedge^a \Box=R(\omega_a)}(x_Y)
\raisebox{-0.05\height}{\tikzsetnextfilename{-figure-puncture_on_C}
\begin{tikzpicture}
\def\r{0.08}
\def\rad{0.45}
\def\l{1.0}
\coordinate (C) at (0,0);
\coordinate (P) at ($(C)+(0:\rad)$);

\draw[fill=white] (P) circle (1.8*\r) node[above]{$Y$};
\draw[black,thick] (P) circle (1.8*\r);
\draw[fill=black] (P) circle (\r);
\end{tikzpicture}
}.
\end{align}

In summary, each $y$ with $h_y>1$ gives $\ell_y=n_{a_y}-h_y+1$ non-trivial ones which are linearly independent each other.
Then, the total number of non-trivial skein relations is given by
\begin{align}
\sum_y (n_{a_y}-h_y+1)=\dfrac{1}{2} \sum_{a=1} n_a(n_a-1)=\dfrac{1}{2} \left( \sum_{a=1} n_a^2 - N \right)
\end{align}
and this is exactly equals to $d_C^{[1^N]}-d_C^Y$ which is the difference between the complex dimension of the 4D local Coulomb branch contributed by the puncture $Y$ and that by the full puncture. See ~\eqref{eq:total Coulomb dim formula for puncture}.

Let us make two comments.
First, only the $[k,N-k]$-type punctures ($N>k \ge N/2$) allow the flavor Wilson lines decomposition for any $(1,p,N-1-p)$-type elementary pants networks in the 4D good theory $T^\calS[C(Y_1,Y_2,[k,N-k])]$ for any $Y_1$ and $Y_2$.
Second, it is expected that these skein relations are valid for the Liouville-Toda CFT.
This is because the skein relations are determined locally around the line operators and then independent of the choice of the 4-manifolds on which the SCFTs are defined.
If we consider the $S^4$ partition functions instead of the superconformal indices, the corresponding 2D theories are considered to be the Liouville-Toda CFT \cite{AGT09,Wyllard09}.

\subsection{Application}
\label{subsec:Application}

We see one simple application of the above skein relations.
They allow us to compute the $q$-deformed Yang-Mills correlator with a complicated network operator easier.
Let us rewrite the network on the right hand in Fig.~\ref{fig:loops in SQCD for higher rank}.
By using the crossing resolution type skein relation (See \cite{TW1504} for example)
\begin{equation}
\raisebox{-0.5\height}{\tikzsetnextfilename{-figure-R}
\begin{tikzpicture}
\def\l{1.4}
\coordinate (UL) at (0,0);
\coordinate (DL) at ($(UL)+(0,-\l)$);
\coordinate (UR) at ($(UL)+(\l,0)$);
\coordinate (DR) at ($(UR)+(0,-\l)$);

\draw[end arrow] (DR) node[below]{$1$} -- (UL);
\draw[line width=5.0pt,white] (DL) -- (UR);
\draw[end arrow] (DL) node[below]{$1$} -- (UR);
\end{tikzpicture}
}
=
q^{\frac{1}{2N}} \raisebox{-0.4\height}{\tikzsetnextfilename{-figure-Q}
\begin{tikzpicture}
\def\l{1.4}
\coordinate (UL) at (0,0);
\coordinate (DL) at ($(UL)+(0,-\l)$);
\coordinate (UR) at ($(UL)+(\l,0)$);
\coordinate (DR) at ($(UR)+(0,-\l)$);
\coordinate (UM) at ($(UL)+(\l/2,-\l/4)$);
\coordinate (DM) at ($(DL)+(\l/2,\l/4)$);

\draw[mid arrow] (DR) node[below]{$1$} -- (DM);
\draw[mid arrow] (DL) node[below]{$1$} -- (DM);
\draw[mid arrow] (DM) -- node[right]{$2$} (UM);
\draw[mid arrow] (UM) -- (UL) node[above]{$1$};
\draw[mid arrow] (UM) -- (UR) node[above]{$1$};
\end{tikzpicture}
}
-
q^{\frac{1}{2N}-\frac{1}{2}} \raisebox{-0.5\height}{\tikzsetnextfilename{-figure-Id}
\begin{tikzpicture}
\def\l{1.4}
\coordinate (UL) at (0,0);
\coordinate (DL) at ($(UL)+(0,-\l)$);
\coordinate (UR) at ($(UL)+(\l,0)$);
\coordinate (DR) at ($(UR)+(0,-\l)$);

\draw[end arrow] (DR) node[below]{$1$} -- (UR);
\draw[end arrow] (DL) node[below]{$1$} -- (UL);
\end{tikzpicture}
}
\end{equation}
and then applying ~\eqref{eq:digon skein simple puncture} several times, we have
\begin{align}
\nonumber &
\raisebox{-0.47\height}{\tikzsetnextfilename{-figure-computation-su3SQCD_exps_0}
\begin{tikzpicture}
\def\ra{0.4}
\def\rb{0.28}
\def\rx{1.0}
\def\ry{0.8}
\def\rc{0.2}
\def\rd{0.3}
\coordinate (O) at (0,0);
\coordinate (A) at (-\ra,\rb);
\coordinate (B) at (-\ra,-\rb);
\coordinate (C) at (\ra,-\rb);
\coordinate (D) at (\ra,\rb);
\simplepunc{(A)}
\fullpunc{(B)}
\fullpunc{(C)}
\simplepunc{(D)}
\draw[thick] (O) circle [x radius=\rx, y radius=\ry];
\draw[green,ultra thick,mid-end arrow] ($(O)+(-\rx,0)$) -- ($(O)+(\rx,0)$);
\draw[red,ultra thick,mid-end arrow] ($(O)+(0,-\ry)$) -- ($(O)+(0,\ry)$);
\end{tikzpicture}}
=
cd[N-2]_q
\raisebox{-0.47\height}{\tikzsetnextfilename{-figure-computation-su3SQCD_exps_1}
\begin{tikzpicture}
\def\ra{0.4}
\def\rb{0.28}
\def\rx{1.0}
\def\ry{0.8}
\def\rc{0.2}
\def\rd{0.3}
\coordinate (O) at (0,0);
\coordinate (A) at (-\ra,\rb);
\coordinate (B) at (-\ra,-\rb);
\coordinate (C) at (\ra,-\rb);
\coordinate (D) at (\ra,\rb);
\simplepunc{(A)}
\fullpunc{(B)}
\fullpunc{(C)}
\simplepunc{(D)}
\draw[thick] (O) circle [x radius=\rx, y radius=\ry];
\draw[red,ultra thick,mid arrow] ($(O)+(0,-\ry)$) -- ($(O)+(0,\ry)$);
\end{tikzpicture}}
+d^{-1}
\raisebox{-0.47\height}{\tikzsetnextfilename{-figure-computation-su3SQCD_exps_2}
\begin{tikzpicture}
\def\ra{0.4}
\def\rb{0.28}
\def\rx{1.0}
\def\ry{0.8}
\def\rc{0.2}
\def\rd{0.3}
\coordinate (O) at (0,0);
\coordinate (A) at (-\ra,\rb);
\coordinate (B) at (-\ra,-\rb);
\coordinate (C) at (\ra,-\rb);
\coordinate (D) at (\ra,\rb);
\coordinate (P) at ($(O)+(0,-0.1*\ry)$);
\coordinate (Q) at ($(O)+(0,0.7*\ry)$);
\simplepunc{(A)}
\fullpunc{(B)}
\fullpunc{(C)}
\simplepunc{(D)}
\draw[thick] (O) circle [x radius=\rx, y radius=\ry];
\draw[red,ultra thick,mid arrow] ($(O)+(0,-\ry)$) -- (P);
\draw[red,ultra thick,mid-end arrow] (Q) -- ($(O)+(0,\ry)$);
\draw[red,ultra thick,start-mid arrow] plot [smooth,tension=0.6] coordinates {(Q) ($(A)+(120:0.2)$) ($(A)+(180:0.3)$) ($(A)+(240:0.2)$) (P)};
\draw[red,ultra thick,start-mid arrow] plot [smooth,tension=0.6] coordinates {(P) ($(D)+(-60:0.2)$) ($(D)+(0:0.3)$) ($(D)+(60:0.2)$) (Q)};
\node at ($(D)+(0:0.05)$) [red,below right]{$2$};
\end{tikzpicture}}
-[2]_qc^{-(N-2)}
\raisebox{-0.47\height}{\tikzsetnextfilename{-figure-computation-su3SQCD_exps_3}
\begin{tikzpicture}
\def\ra{0.4}
\def\rb{0.28}
\def\rx{1.0}
\def\ry{0.8}
\def\rc{0.2}
\def\rd{0.3}
\coordinate (O) at (0,0);
\coordinate (A) at (-\ra,\rb);
\coordinate (B) at (-\ra,-\rb);
\coordinate (C) at (\ra,-\rb);
\coordinate (D) at (\ra,\rb);
\simplepunc{(A)}
\fullpunc{(B)}
\fullpunc{(C)}
\simplepunc{(D)}
\draw[thick] (O) circle [x radius=\rx, y radius=\ry];
\draw[red,ultra thick,mid-end arrow] plot [smooth,tension=0.6] coordinates {($(C)+(-45:\rd)$) ($(C)+(-130:\rc)$) ($(A)+(210:\rc)$) ($(A)+(135:\rd)$) ($(A)+(60:\rc)$) ($(C)+(30:\rc)$) ($(C)+(-45:\rd)$)};
\end{tikzpicture}} \\
&
+
\left[ c^{-(N-2)}d\chi_{\Box}(b)+c^{-(N-1)}\chi_{\ovl{\Box}}(a)-c[N-3]_q \chi_{\ovl{\Box}}(a) \right]
\raisebox{-0.47\height}{\tikzsetnextfilename{-figure-computation-su3SQCD_exps_4}
\begin{tikzpicture}
\def\ra{0.4}
\def\rb{0.28}
\def\rx{1.0}
\def\ry{0.8}
\def\rc{0.2}
\def\rd{0.3}
\coordinate (O) at (0,0);
\coordinate (A) at (-\ra,\rb);
\coordinate (B) at (-\ra,-\rb);
\coordinate (C) at (\ra,-\rb);
\coordinate (D) at (\ra,\rb);
\simplepunc{(A)}
\fullpunc{(B)}
\fullpunc{(C)}
\simplepunc{(D)}
\draw[thick] (O) circle [x radius=\rx, y radius=\ry];
\end{tikzpicture}}
\end{align}
where we have used $\chi_\Box(c_{[N-1,1]})=[N-1]_q c+c^{-(N-1)}$ and the topological property of correlators allowing the punctures to move around continuously.
\!\footnote{Notice that there is another crossing at the point we have removed for the disk representation.}
In particular, in $N=3$ case, they reproduce the result ~\eqref{eq:SU(3) SQCD with W and T loops} again.

\paragraph{Acknowledgements}
The author is supported by the Advanced Leading Graduate Course for Photon Science, one of the Program for Leading Graduate Schools lead by Japan Society for the Promotion of Science, MEXT and also the World Premier International Research Center Initiative (WPI), Kavli IPMU, the University of Tokyo.

\appendix

%\end{comment}

\section{Lie algebra convention}
\label{app:Lie algebra convention}

First of all, $\Pi(R)$, $\calP^\Lieg_+$ and $\rho$ denote weights of the representation $R$, the set of dominant weights and the Weyl vector $\displaystyle =\sum_{a=1}^{\rk \Lieg} \omega_a=\dfrac{1}{2}\sum_{\alpha \in \Delta^+_\Lieg} \alpha$ respectively.
And $R(\lambda)$ denotes the irreducible representation associated with a dominant weight $\lambda$, $\lambda_R$ does the dominant weight to $R$ conversely.

In this paper, we use several conventions on representations.
To express the irreducible representation, for example, we use the Dynkin weights (labels) $[0,\ldots,\underset{k}{1},\ldots,0]$ or $(0\cdots0\underset{k}{1}0\cdots 0)$, weight vector $R(\omega_k)$ and the dimension $\wedge^k \rmbf{N}$ as necessary.
For $\Lieg=\Liesu(N)$ case, we also use the Young diagram $\wedge^{k} \Box$.

\subsection{$A$-type Lie algebra convention}
\label{app:type A convention}

Let $\{ \alpha_a \}_{a=1,2,\ldots,\rk \Lieg=N-1}$ be a set of chosen positive simple roots.
Notice all choices are equivalent under the Weyl reflection actions.
$\omega_{\alpha=1,2,...,N-1}$ are fundamental weights, $h_{i=1,2,\ldots,N}$ are weights in $\Pi(R(\omega_1)=\Box)$.
\footnote{The perfect order of the indices of $h_i$ is determined by the partial order in the weight lattice.}
Using the orthonormal basis $e_{i=1,2,\ldots,N}$ in $\bbR^N$ satisfying $(e_i,e_j)=\delta_{i,j}$, there are relations among them as $h_a=\omega_a-\omega_{a-1}$ where $\omega_N=\omega_0=0$, $\alpha_a=e_a-e_{a+1}$ and $\displaystyle h_i=e_i - \tfrac{1}{N} \sum_{i=1}^N e_i$.

The Dynkin labels are given by $\lambda_{a}:=(\lambda,\alpha_a)$.
In other words, $\lambda=\sum_{a=1}^{N-1}\lambda_{a}\omega_a$. Then, the dominant weight set is given by
\begin{align}
\calP^{\Liesu(N)}_+=\{ \lambda \;|\; \lambda_a \in \bbZ_{\ge 0} \mbox{ for }\forall\; a \}.
\end{align}
Notice that $\rho_a=1$ for all $a$.
Notice also that $\lambda_{\hat{i}}:=(\lambda_R,h_i-h_N)$ means the number of the boxes in the $i$-th row of the corresponding Young tableau of the irreducible representation $R$ ans satisfies $\lambda_i=\lambda_{\hat{i}}-\lambda_{\hat{i+1}}$.
We also define a symbol $\displaystyle |\lambda|:=\sum_{i=1}^{N} \lambda_{\hat{i}}$.

The quantum dimension of $R(\lambda)$ is given by
\begin{align}
\dim R=\prod_{\alpha \in \Delta^+_\Lieg} \dfrac{[(\lambda_R+\rho,\alpha)]_q}{[(\rho,\alpha)]_q}.
\label{eq:definition of quantum dimension}
\end{align}

\section{The formula for Coulomb and Higgs branch dimension associated to regular punctures}

\begin{figure}[htb]
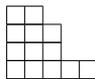

\centering
{\tiny \young(\;\;,\;\;\;,\;\;\;,\;\;\;\;\;)}
\caption{$N=13$ example : the corresponding standard partition is given by $[n]=[4,4,3,1,1]=[4^231^2]$
and its dual representation is $\{s\}=\{5,3,3,2\}=\{53^22\}$. $d_C^Y=p^Y=\{ 0,1,2,3,4,4,5,6,6,7,8,8,9 \}$.}
\label{fig:partition representation}
\end{figure}

Let $\calB_{\rm Coulomb}$ and $\calM_{\rm Higgs}$ be the 4D Coulomb branch and the 4D Higgs branch respectively. According to \cite{Gaiotto0904,NanopoulosXie09,CDT12}, the contribution from the $Y$-type puncture is given as
\begin{align}
& d_C^Y:=\frac{1}{2} \left( N^2 - \sum_{i} n_i^2 \right)
\label{eq:total Coulomb dim formula for puncture} \\
& d_H^Y:= \frac{1}{2} \left( \sum_{i} s_i^2 - N \right)
\end{align}
where $[n]$ is the partition of $N$ and $\{s\}$ is its dual representation. See Fig.~\ref{fig:partition representation}.

The total dimensions are given by
\begin{align}
& \dim_\bbC \calB_{\rm Coulomb} = \sum_{A} \dim_\bbC \calB_{\rm Coulomb}(Y_A) - \frac{1}{2}\chi_C \; \dim_\bbR \Lieg \\
& \dim_\bbH \calM_{\rm Higgs} = \sum_{A} \dim_\bbH \calM_{\rm Higgs}(Y_A) + \; \rk \Lieg
\label{eq:formulae for dimensions of Coulomb/Higgs branches}
\end{align}
where $\chi_C$ is the Euler number of the Riemann surface on ignoring the punctures.
Keep in mind that the analysis up to now are locally and do not give any global information.

However, the 4D Coulomb branch has more information about the scaling dimension of the Coulomb branch operators, namely graded by the scaling dimension.
By analyzing the Seiberg-Witten curve and its relation to Hitchin systems, we see
\begin{align}
d_{C,k}^Y=p_k^{Y}=k-\min\{ a \;|\; \sum_{b=1}^{a} s_b \ge k \} \quad \for k=2,3,\ldots,N
\label{eq:dimension formula for Coulomb branch with each scaling dimension}
\end{align}
and
\begin{align}
\dim_\bbC \calB^{(k)}_{\rm Coulomb}=\sum_{A} d_{C,k}^Y-(2k-1)
\end{align}
for $k=2,3,\ldots,N$.

\bibliographystyle{h-physrev}%alpha
\bibliography{ref}

\begin{thebibliography}{10}

\bibitem{Watanabe1603}
N.~Watanabe,
\newblock JHEP {\bf 12}, 063 (2016), 1603.02939.

\bibitem{Pestun07}
V.~Pestun,
\newblock Commun.Math.Phys. {\bf 313}, 71 (2012), 0712.2824.

\bibitem{KWY0909}
A.~Kapustin, B.~Willett, and I.~Yaakov,
\newblock JHEP {\bf 03}, 089 (2010), 0909.4559.

\bibitem{GOP11}
J.~Gomis, T.~Okuda, and V.~Pestun,
\newblock JHEP {\bf 1205}, 141 (2012), 1105.2568.

\bibitem{GKL1201}
D.~Gang, E.~Koh, and K.~Lee,
\newblock JHEP {\bf 1205}, 007 (2012), 1201.5539.

\bibitem{Okuda1412_review}
T.~Okuda,
\newblock {Line operators in supersymmetric gauge theories and the 2d-4d
  relation},
\newblock in {\em New Dualities of Supersymmetric Gauge Theories}, edited by
  J.~Teschner, pp. 195--222, 2016, 1412.7126.

\bibitem{GLMR15}
F.~Gliozzi, P.~Liendo, M.~Meineri, and A.~Rago,
\newblock JHEP {\bf 05}, 036 (2015), 1502.07217.

\bibitem{BGLM16}
M.~Billò, V.~Gonçalves, E.~Lauria, and M.~Meineri,
\newblock JHEP {\bf 04}, 091 (2016), 1601.02883.

\bibitem{Gadde1602}
A.~Gadde,
\newblock (2016), 1602.06354.

\bibitem{FFRS06}
J.~Frohlich, J.~Fuchs, I.~Runkel, and C.~Schweigert,
\newblock Nucl. Phys. {\bf B763}, 354 (2007), hep-th/0607247.

\bibitem{DGG10}
N.~Drukker, D.~Gaiotto, and J.~Gomis,
\newblock JHEP {\bf 1106}, 025 (2011), 1003.1112.

\bibitem{Witten88Jones}
E.~Witten,
\newblock Commun.Math.Phys. {\bf 121}, 351 (1989).

\bibitem{Gaiotto0904}
D.~Gaiotto,
\newblock JHEP {\bf 1208}, 034 (2012), 0904.2715.

\bibitem{GMN2}
D.~Gaiotto, G.~W. Moore, and A.~Neitzke,
\newblock (2009), 0907.3987.

\bibitem{AGT09}
L.~F. Alday, D.~Gaiotto, and Y.~Tachikawa,
\newblock Lett.Math.Phys. {\bf 91}, 167 (2010), 0906.3219.

\bibitem{Wyllard09}
N.~Wyllard,
\newblock JHEP {\bf 11}, 002 (2009), 0907.2189.

\bibitem{GPRR0910}
A.~Gadde, E.~Pomoni, L.~Rastelli, and S.~S. Razamat,
\newblock JHEP {\bf 1003}, 032 (2010), 0910.2225.

\bibitem{GRRY1104}
A.~Gadde, L.~Rastelli, S.~S. Razamat, and W.~Yan,
\newblock Phys.Rev.Lett. {\bf 106}, 241602 (2011), 1104.3850.

\bibitem{DMO09}
N.~Drukker, D.~R. Morrison, and T.~Okuda,
\newblock JHEP {\bf 0909}, 031 (2009), 0907.2593.

\bibitem{AGGTV09}
L.~F. Alday, D.~Gaiotto, S.~Gukov, Y.~Tachikawa, and H.~Verlinde,
\newblock JHEP {\bf 1001}, 113 (2010), 0909.0945.

\bibitem{DGOT09}
N.~Drukker, J.~Gomis, T.~Okuda, and J.~Teschner,
\newblock JHEP {\bf 1002}, 057 (2010), 0909.1105.

\bibitem{Passerini10}
F.~Passerini,
\newblock JHEP {\bf 1003}, 125 (2010), 1003.1151.

\bibitem{GomisLeFloch10}
J.~Gomis and B.~Le~Floch,
\newblock JHEP {\bf 1111}, 114 (2011), 1008.4139.

\bibitem{Xie1304}
D.~Xie,
\newblock (2013), 1304.2390.

\bibitem{Bullimore13}
M.~Bullimore,
\newblock (2013), 1312.5001.

\bibitem{TW1504}
Y.~Tachikawa and N.~Watanabe,
\newblock JHEP {\bf 06}, 186 (2015), 1504.00121.

\bibitem{CGT1505}
I.~Coman, M.~Gabella, and J.~Teschner,
\newblock JHEP {\bf 10}, 143 (2015), 1505.05898.

\bibitem{Gabella1603}
M.~Gabella,
\newblock (2016), 1603.05258.

\bibitem{GaiottoWitten0804}
D.~Gaiotto and E.~Witten,
\newblock J.Statist.Phys. {\bf 135}, 789 (2009), 0804.2902.

\bibitem{GRR1207}
D.~Gaiotto, L.~Rastelli, and S.~S. Razamat,
\newblock JHEP {\bf 01}, 022 (2013), 1207.3577.

\bibitem{Tachikawa1504}
Y.~Tachikawa,
\newblock PTEP {\bf 2015}, 11B102 (2015), 1504.01481.

\bibitem{CMR94}
S.~Cordes, G.~W. Moore, and S.~Ramgoolam,
\newblock Nucl. Phys. Proc. Suppl. {\bf 41}, 184 (1995), hep-th/9411210.

\bibitem{BR94}
E.~Buffenoir and P.~Roche,
\newblock Commun.Math.Phys. {\bf 170}, 669 (1995), hep-th/9405126.

\bibitem{AOSV04}
M.~Aganagic, H.~Ooguri, N.~Saulina, and C.~Vafa,
\newblock Nucl.Phys. {\bf B715}, 304 (2005), hep-th/0411280.

\bibitem{GMT11}
D.~Gaiotto, G.~W. Moore, and Y.~Tachikawa,
\newblock PTEP {\bf 2013}, 013B03 (2013), 1110.2657.

\bibitem{CGS1606}
C.~Cordova, D.~Gaiotto, and S.-H. Shao,
\newblock JHEP {\bf 11}, 106 (2016), 1606.08429.

\bibitem{GMN3}
D.~Gaiotto, G.~W. Moore, and A.~Neitzke,
\newblock (2010), 1006.0146.

\bibitem{IOT11}
Y.~Ito, T.~Okuda, and M.~Taki,
\newblock JHEP {\bf 1204}, 010 (2012), 1111.4221.

\bibitem{LieART}
R.~Feger and T.~W. Kephart,
\newblock Comput. Phys. Commun. {\bf 192}, 166 (2015), 1206.6379.

\bibitem{BLLPRR13}
C.~Beem {\em et~al.},
\newblock Commun. Math. Phys. {\bf 336}, 1359 (2015), 1312.5344.

\bibitem{CordovaShao1506}
C.~Cordova and S.-H. Shao,
\newblock JHEP {\bf 01}, 040 (2016), 1506.00265.

\bibitem{GMN5}
D.~Gaiotto, G.~W. Moore, and A.~Neitzke,
\newblock Annales Henri Poincare {\bf 14}, 1643 (2013), 1204.4824.

\bibitem{HollandsNeitzke1607}
L.~Hollands and A.~Neitzke,
\newblock (2016), 1607.01743.

\bibitem{DGG1108}
T.~Dimofte, D.~Gaiotto, and S.~Gukov,
\newblock Commun. Math. Phys. {\bf 325}, 367 (2014), 1108.4389.

\bibitem{NanopoulosXie09}
D.~Nanopoulos and D.~Xie,
\newblock JHEP {\bf 1003}, 043 (2010), 0911.1990.

\bibitem{CDT12}
O.~Chacaltana, J.~Distler, and Y.~Tachikawa,
\newblock Int.J.Mod.Phys. {\bf A28}, 1340006 (2013), 1203.2930.

\end{thebibliography}

\end{document}